\documentclass[fleqn,usenatbib]{mnras}

\usepackage{lmodern}

\usepackage[T1]{fontenc}
\usepackage{ae,aecompl}

\usepackage{graphicx}	
\usepackage{amsmath}	
\usepackage{amssymb}	

\graphicspath{ {Figures/} }
\usepackage{multirow}
\usepackage{booktabs}
\usepackage{xspace} 
\usepackage[FIGTOPCAP]{subfigure}

\def\Rcl{R_{\rm cl}}
\def\krho{k_{\rho}}
\def\Sigmacl{\Sigma_{\rm cloud} }
\def\sigmas{\sigma_{s}}
\def\MSun{M_{\odot}}
\def\Msun{M_{\odot}}
\def\phipbar{\phi_{\bar{P}}}
\def\phipcl{\phi_{P,\rm cl}}
\def\phib{\phi_{\rm B}}
\def\fbin{f_{\rm binary}}

\def\tsf{ t_* } 
\def\epsilonff{ \epsilon_{\rm ff}}

\def\fiducial{\texttt{fiducial}\xspace}

\def\segregated{\texttt{segregated}\xspace}
\def\fullbinaries{\texttt{binaries\_100}\xspace}

\defcitealias{mt03}{MT03}
\defcitealias{t13}{T13}
\defcitealias{F17}{Paper I}

\title[Gradual Star Cluster Formation]{Star Cluster Formation from Turbulent Clumps.
II.\\
Gradual Star Cluster Formation \\}

\author[Farias, Tan and Chatterjee]{
Juan P. Farias,$^{1}$\thanks{E-mail: juan.farias@chalmers.se}
Jonathan C. Tan$^{1,2}$
and Sourav Chatterjee$^{3}$
\\
$^{1}$Dept. of Space, Earth \& Environment, Chalmers University of Technology, Gothenburg, Sweden\\
$^{2}$Dept. of Astronomy, University of Virginia, Charlottesville, VA 22904, USA\\
$^{3}$Tata Institute of Fundamental Research, Homi Bhabha Road, Mumbai 400005, India\\
}

\date{Accepted XXX. Received YYY; in original form ZZZ}

\pubyear{2018}

\begin{document}
\label{firstpage}
\pagerange{\pageref{firstpage}--\pageref{lastpage}}
\maketitle

\begin{abstract}
We investigate the dynamical evolution of star clusters during their
formation, assuming that they are born from a turbulent starless clump
of a given mass that is embedded within a parent self-gravitating
molecular cloud characterized by a particular mass surface density.
In contrast to the standard practice of most $N$-body studies, we do
not assume that all stars are formed at once. Rather, we explore the
effects of different star formation rates on the global structure and
evolution of young embedded star clusters, also considering various
primordial binary fractions and mass segregation levels. Our fiducial
clumps studied in this paper have initial masses of $M_{\rm cl} =
3000\,M_\odot$, are embedded in ambient cloud environments of
$\Sigma_{\rm cloud} = 0.1$ and 1 g~cm$^{-2}$, and gradually form stars
with an overall efficiency of 50\% until the gas is exhausted. We
investigate star formation efficiencies per free-fall time in the
range $\epsilon_{\rm ff}=0.01$ to 1, and also compare to the
instantaneous case ($\epsilon_{\rm ff}=\infty$) of Paper I. We show
that most of the interesting dynamical processes that determine the
future of the cluster, happen during the early formation phase. In
particular, the ejected stellar population is sensitive to the
duration of star cluster formation: for example, clusters with longer
formation times produce more runaway stars, since these clusters
remain in a dense state for longer, thus favouring occurrence of
dynamical ejections. We also show that the presence of radial age
gradients in star clusters depends sensitively on the star formation
efficiency per free fall time, with observed values being matched best
by our slowest forming clusters with $\epsilon_{\rm ff}\lesssim0.03$.
\end{abstract}




\section{Introduction}
\begin{table*}
\centering
\caption{Parent clump parameters}

\begin{tabular*}{\textwidth}{c @{\extracolsep{\fill}}cccccccccc} \toprule
        & $\Sigmacl$ (g cm$^{-2}$) & $M_{\rm cl}$($\Msun$)& $t_{\rm ff}$ (Myr) & $\Rcl$ (pc)     & $\krho$ & $\phipcl$ & $\phipbar$
        & $\phib$ & $\sigma_{\rm cl}$ (km/s)  \\[0.1cm] \hline
  Low-$\Sigma$ Clump  & 0.1 & 3000  &0.39  & 1.154 & 1.5 & 2 & 1.31 & 2.8 & 1.71 \\ 
  High-$\Sigma$ Clump & 1   & 3000  & 0.069& 0.365 & 1.5 & 2 & 1.31 & 2.8 & 3.04 \\ \bottomrule
\end{tabular*}
\label{tab:clumps}
\end{table*}

\begin{table*}
\centering
\caption{Initial conditions for simulation sets.
For each of the sets named in the first column, 20 simulations were
performed for each of the $\epsilonff$ values listed in column 2 and for each set
of clumps parameters listed in Table~\ref{tab:clumps}. Third column shows the assumed
overall SFE. Fourth and fifth columns show the times
needed to form stars for each $\Sigma$ case, sixth
column shows the average number of stars per simulation, seventh column shows the primordial
binary fraction, eighth column shows the assumed IMF (here all taken to be from Kroupa
2001, c.f., Paper I), ninth column shows whether or not the cluster is initially mass
segregated (IMS), and tenth column shows if stellar evolution (SE) is
included in the set. }
\label{tab:ic}
\begin{tabular*}{\textwidth}{r @{\extracolsep{\fill}}ccccccccc} \toprule
        Set name & $\epsilonff$  & $\epsilon$ &  \multicolumn{2}{c}{ $t_*\,$ (Myr) }
 &$\langle N_* \rangle$      & $\fbin$ & IMF            & IMS & SE\\
 & & &Low-$\Sigma$ & High-$\Sigma$ & &  & & &\\ \hline
  \fiducial     &0.01& 0.5 & 3.35 &19.50 &$4000$&0.5 & \cite{Kroupa2001} & N &Y    \\
                &0.03& 0.5 & 1.17 &6.50  &$4000$&0.5 & \cite{Kroupa2001} & N &Y    \\
                &0.1 & 0.5 & 0.33 &1.95  &$4000$ &0.5 & \cite{Kroupa2001} & N &Y    \\
                & 1  & 0.5 & 0.03 &0.19  &$4000$&0.5 & \cite{Kroupa2001} & N &Y
                \\\hline 
  \fullbinaries &0.03 & 0.5 & 1.17 & 6.50 &$4000$ &1.0 & \cite{Kroupa2001} & N &Y   \\
  \hline                                
  \segregated   &0.03 & 0.5 &1.17&6.50&$4000$ &0.5 & \cite{Kroupa2001} & Y &Y   \\
\bottomrule
\end{tabular*}
\end{table*}

Most stars appear to form in clusters
\citep[e.g.,][]{Lada2003,Gutermut2009} and therefore understanding
star formation is in large part linked to understanding how and where
star cluster formation takes place. Observationally, it is known that
star clusters are born from dense gas clumps within giant molecular
clouds (GMC) \citep[e.g.][]{Mckee2007}. However, several aspects about
this process are still under active debate. In particular, there is no
consensus yet if star cluster formation is a relatively fast process,
i.e., happening within about one or a few free fall times,
\citep{Elmegreen2000,Elmegreen2007,Hartmann2007}, or occurs slowly
over several to many local free fall times
\citep{Tan2006,Nakamura2007,Nakamura2014}.

Turbulent motions and magnetic fields have been proposed as being
important for stabilizing the collapse of star-forming clumps and thus
regulating their star formation rates \citep[e.g.,][]{Krumholz2005}.
Although, turbulence is expected to decay in about one crossing
time (Stone et al. 1998; Mac Low et al. 1998), i.e., about two
free-fall times for a virialized clump, turbulence could be maintained
by protostellar outflow feedback from a relatively low level of star formation
\citep{Tan2006,Nakamura2007,Nakamura2014}.

Here, continuing from our initial study \citep[Paper I]{F17}, we
examine the dynamical evolution of a forming stellar population of a
star-forming clump that will yield a bound cluster, along with an
unbound and dynamically-ejected population of stars.  We use a
dedicated $N$-body code, since, unlike most other previous $N$-body
studies \citep[e.g.,][]{Bastian2006,Parker2014a,Pfalzner2015,Wang2018},
we treat the gradual formation of clusters, while still aiming to
accurately follow stellar orbits, including binary and higher-order
multiple systems. The gas is treated as an evolving background
potential and various aspects of the star formation process, including
its rate, degree of primordial binarity, primordial binary properties,
degree of primordial mass segregation within the spherical clump, are
parameterized and the effects of these choices explored. The
calculations are relatively inexpensive, which allows many
realizations to be carried out. This is necessary since stellar
dynamics is a chaotic process and the stellar initial mass function
(IMF) is sparsely sampled at the high-mass end, so we need many
realizations of clusters to find average quantities.

Our approach is complementary to other theoretical/numerical studies
that attempt to follow the detailed dynamical evolution of
star-forming gas structures with (magneto-)hydrodynamic (M)HD
codes. These studies always need to assume subgrid models for
individual star formation, e.g., often treated as sink particles. They
typically do not follow the dynamics of stellar multiples with high
accuracy. Furthermore, they are usually very computationally
expensive, which limits the number of realizations that can be
performed for a given initial condition, and so parameter space
exploration of these initial conditions (and of other parameters
associated with the sub-grid modeling) is more challenging.

This paper is organized as follows. In \S\ref{sec:ic} we describe the
theoretical framework of our star cluster formation simulations, as
well as the initial conditions and methods used in this work. In
\S\ref{sec:results} we present the results on various aspects of star
cluster evolution when changing the star cluster formation timescale
and initial conditions. Then we discuss our results in
\S\ref{sec:discussion} and summarize our conclusions in
\S\ref{sec:conclusions}.

\section{Methods}\label{sec:ic}

\subsection{Background gas model}
\label{sec:gasmodel}

We assume star clusters are formed from turbulent, magnetised,
gravitationally bound, initially starless gas clumps within
GMCs. Following \citetalias{F17}, we describe the structure of the
clump using the turbulent core/clump model of \cite[][hereafter
  \citetalias{mt03}]{mt03}, i.e., clumps are polytropic spheres
in virial and pressure equilibrium with their surroundings. The
density profile of such clumps is modeled as:
\begin{eqnarray}
        \label{eq:dens}
        \rho_{\rm cl} (r) &=& \rho_{\rm s,cl} \left(\frac{r}{\Rcl} \right)^{-\krho},
\end{eqnarray}
and the velocity dispersion profile as: 
\begin{eqnarray}
\sigma_{\rm cl}(r) &=& \sigmas \left( \frac{r}{\Rcl} \right)^{(2-\krho)/2},
\label{eq:sigmar}
\end{eqnarray}
where $\rho_{\rm s,cl}$ and $\sigmas$ are the density and velocity
dispersion at the surface of the clump, respectively, and $\Rcl$ is
the radius of the clump, i.e., the location of this surface
boundary. The values of these three characteristic quantities are
defined by the surrounding cloud's mass surface density, $\Sigmacl$,
and are given by:
\begin{eqnarray}
\sigmas &=& 5.08 \left( \frac{\phipcl \phipbar}{Ak_{P}^2 \phib^4} \right)^{1/8} 
\left(\frac{M_{\rm cl}}{3000~\Msun} \right)^{1/4} \nonumber \\
& & \times \left( \frac{\Sigmacl }{1~{\rm g~cm^{-2}}} \right)^{1/4}\:{\rm km~s^{-1}},\\
&\rightarrow & 3.04 \left(\frac{M_{\rm cl}}{3000~\Msun} \right)^{1/4}
\times \left( \frac{\Sigmacl }{1~{\rm g~cm^{-2}}} \right)^{1/4}\:{\rm km~s^{-1}},
\nonumber
\label{eq:sigmas} 
\end{eqnarray}
and:
\begin{eqnarray}
\Rcl &=& 
         0.50 \left( \frac{A}{k_{p} \phipcl \phipbar} \right)^{1/4} 
         \left( \frac{M_{\rm cl}}{3000 \Msun}  \right)^{1/2} \nonumber \\
         & & \times \left( \frac{\Sigmacl}{1 {\rm ~g~cm^{-2}}} \right)^{-1/2}\:{\rm pc},
         \label{eq:rcl} \\
         &\rightarrow&  0.365
         \left( \frac{M_{\rm cl}}{3000 \Msun}  \right)^{1/2} 
         \times \left( \frac{\Sigmacl}{1 {\rm ~g~cm^{-2}}} \right)^{-1/2}\:{\rm pc},
         \nonumber
\end{eqnarray}
where $k_p=2(\krho -1)$ is the power law exponent of the pressure
($P$) within the clump; $\phipcl$ is the ratio between the pressure at
the surface of the clump ($P_{\rm s,cl}$) and the mean pressure inside
the cloud, $\bar{P}_{\rm cloud}$; $\phipbar$ is a normalization
constant, $\sim{\cal O}(1)$, in the relation $\bar{P}_{\rm
  cloud}\equiv\phipbar G\Sigmacl^2$; $A=(3-\krho)(\krho-1)f_g
\rightarrow 3/4$; and right arrows indicate quantities obtained using
the fiducial set of constants. We assume the clump is initially
starless and therefore $f_g=1$.  We keep the same fiducial values of
$\phi_{P,\rm cl}=2$ and $\phipbar=1.32$ as in \citetalias{mt03} and
\citetalias{F17}.  Following \citetalias{F17}, we model clumps in two
different cloud environments: the high-$\Sigma$ case with
$\Sigmacl=1.0\,$g cm$^{-2}$ and the low-$\Sigma$ case with
$\Sigmacl=0.1\,$g cm$^{-2}$. The properties of the clumps are
summarized in Table~{\ref{tab:clumps}}. Then, given $\Rcl$, defined by
$\Sigmacl$, the density at the surface of the clump is:
\begin{eqnarray}
        \rho_{\rm s,cl}&=& \frac{(3-\krho)M_{\rm cl}}{4\pi\Rcl^3}.
\end{eqnarray}

\begin{figure*}
\includegraphics[width=\textwidth]{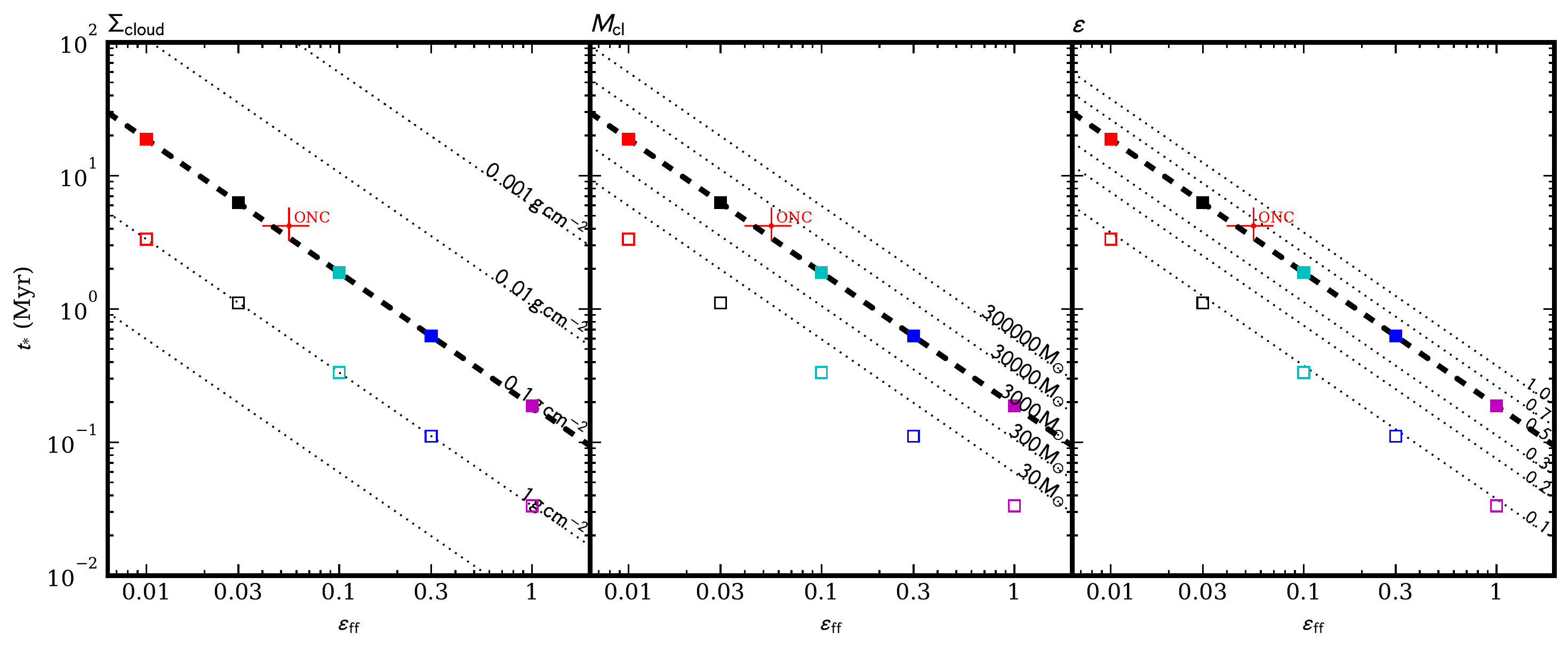}
\caption{
The star cluster formation timescale as a function of star formation
efficiency per free-fall time ($\epsilonff$) for the models
presented in this work. In the first panel we show the dependence on
different values of $\Sigmacl$. As a reference, we mark
$\Sigmacl=0.1\,$g cm$^{-2}$ with a thick dashed line, and we
show it in the second and third panels with its variation for
different values of clump mass ($M_{\rm cl}$) and overall star
formation efficiency ($\epsilon$), respectively. Square symbols show
the models for which we performed simulations in this work with a
color scheme used consistently for different values of $\epsilonff$
in the following figures. The red point with errorbars shows
the properties estimated for the ONC \citep{Dario2014}.}
        \label{fig:timescale}
\end{figure*}

In \citetalias{F17} we only simulated clusters in the ``fast formation
limit'', which means that star formation happened instantly and any
residual gas also was expelled from the system instantly. In this
paper, we now leave this approximation behind and acknowledge that
stars are born gradually. Therefore natal gas is still present while
stars are being formed. The presence of the gas, with a radial profile
described by Eq.~\ref{eq:dens}, is emulated by a time-dependent
background potential, i.e.,
\begin{eqnarray}
        \label{eq:pot}
        \Phi_{\rm gas} (r,t) &=&  \left\{
        \begin{array}{lr}
                \dfrac{ GM_{\rm cl}(t) }{(2-\krho) \Rcl } \left[ \left(
                \dfrac{r}{\Rcl} \right)^{2-\krho}\hspace{-20pt} - 3 + \krho \right] & ,r\leq \Rcl\\
                &\\
                -\dfrac{GM_{\rm cl}(t)}{r} &, r>\Rcl
        \end{array}
        \right.,
\end{eqnarray}
where $G$ is the gravitational constant and $M_{\rm cl}(t)$ the
time-dependent clump gas mass.  Note that the radius of the clump is
truncated at $R_{\rm cl}$ and therefore we do not include any effects
of additional gas mass beyond this radius.
Since the background gas is modeled with a smooth analytic potential, we
also do not include effects such as
dynamical friction caused by
sub-structured gas acting on the stars.

In this paper we assume a constant star formation rate (SFR) defined
using the {\it initial} parameters of the clump, i.e.,
\begin{eqnarray}
        \label{eq:mdot}
        \dot{M}_* &=& \frac{\epsilonff M_{\rm cl,0} }{t_{\rm ff,0}}.
\end{eqnarray}
As stars are formed, some gas is assumed to be ejected from the system instantly
according to a given local star formation efficiency, $\epsilon_*$, defined as
the ratio between the stellar mass formed and the total mass required
to form such a stellar mass.
The fiducial value we adopt for $\epsilon_*$ is 0.5. This can be
viewed as the star formation efficiency from an individual
self-gravitating core, with the remainder of the core's gas assumed to
be ejected quickly from the clump by local feedback, expected to be
dominated by protostellar outflows \citep[see, e.g.,][]{Tanaka2017}.

The time-evolution of the global gaseous mass of the clump is thus
given by:
\begin{eqnarray}
        \label{eq:mgas}
        M_{\rm cl}(t) &=& \left\{  
        \begin{array}{lr}
                M_{\rm cl,0} - \dfrac{ \dot{M}_* }{ \epsilon } t &, t \leq \tsf \\
                & \\
                0 &, t > \tsf\\
        \end{array} 
        \right.
\end{eqnarray}\\
where $t_*$ is the time at which gas is exhausted. Since we use a
constant SFR and $\epsilon$, Equation~\ref{eq:mgas} takes a linear
form and the gas exhaustion time is simply $\tsf =
(\epsilon/\epsilonff) \times t_{\rm ff,0}$.

However, the value of $t_{\rm ff,0}$, defined as
$\sqrt{3\pi/32\bar{\rho_{\rm 0}}}$, depends on $\Sigmacl$ and
$M_{\rm cl,0}$ since these values determine the size of the clump and
therefore its mean density. Using the definition of $R_{\rm cl}$
(Eq.~\ref{eq:rcl}), the initial global free fall time reduces to:
\begin{eqnarray}
        t_{\rm ff,0} &=& 0.107 \left( \frac{A}{k_{\rm p}\phi_{
        P,cl}\phi_{\bar{P} }}  \right)^{3/8}  
        \left( \frac{M_{\rm cl, 0}}{3000\,{\MSun}} \right)^{1/4}  \nonumber\\
        & & \times \left( \frac{\Sigmacl} {1\,{\rm g\,cm^{-2}}} \right)^{-3/4} \, {\rm Myr}\\
        & \rightarrow & 0.069 \left( \frac{M_{\rm cl, 0}}{3000\,{\MSun}}
        \right)^{1/4}\left( \frac{\Sigmacl} {1\,{\rm g\,cm^{-2}}} \right)^{-3/4}
        \, {\rm Myr}. \nonumber 
\end{eqnarray}

Figure~\ref{fig:timescale} shows the variation of $t_{*}$ given a
different choice of initial parameters such as $\Sigmacl$
(left) , $M_{\rm cl}$ (middle) and SFE (right) as a function of
$\epsilonff$ when keeping all other variables as constants.
High $\Sigmacl$ results in shorter free-fall times and
therefore $t_*$ is smaller for the same $\epsilonff$. We
highlight the case $\Sigmacl=0.1\rm\,g\,cm^{-2}$ (thick dashed
line) and take it as reference line in the next two panels.  The
variation with the mass of the clump is less obvious: the size of the
clump scales with mass as $R_{\rm cl} \propto M_{\rm cl}^{1/2}$ in
order to keep the clump in equilibrium. The density of the clump
scales linearly with the mass, this results in the relation
$\bar{\rho}_{0} \propto M_{\rm cl}/\Rcl^3 \propto M_{\rm
  cl}^{-1/2}$. And then $t_{\rm ff} \propto \bar{\rho_0}^{-1/2}
\propto M_{\rm cl}^{1/4}$. The third panel shows the linear relation
between $t_*$ and the SFE, which simply means that if the SFE is high,
more mass will be converted into stars and therefore more time is
needed to obtain the final stellar mass.

In the model presented here, one of the main assumptions in the star
cluster formation scheme is that the SFR is constant with time, i.e.,
parameters such as $t_{\rm ff}$ and $\epsilonff$ are defined based on
the initial state of the clump. We note, however, that as stars are
being born and the gas is being ejected/exhausted, the total
(stars$+$gas) mass density decreases (eventually by a factor of two in
the fiducial case) and so the free fall time becomes longer by a
factor of $2^{1/2}$. Thus if the instantaneous value of $\epsilonff$
were held fixed, i.e., a fixed fraction of gas converted into stars
per local total free fall time, then the SFR would taper off towards
zero \citep[see, e.g.][]{Parmentier2013}. However, given our desire to first
explore simplest models, we will defer exploring such varying SFR
cases until a future paper. We note also that observational values of
$\epsilonff$ \citep[e.g.,][]{Dario2014} are based on time averaged
SFRs.

\subsection{Gradual formation of stars}

In order to simulate the birth of stars in a cluster at different star
formation rates, we have developed a modified version of the direct
$N$-body code \texttt{Nbody6++} \citep{Aarseth2003,Wang2015}. We
introduced routines in order to add particles during run-time,
including formation of primordial binaries. The introduction of stars
is carried out as follows: (1) We create the final phase-space
distribution of stars following the model described in
\citetalias{F17}. (2) The set of particles is unsorted creating the
formation sequence.  The used unsort method ensures that the global
binary fraction is preserved at all times if no binaries are lost
during the simulation. (3) We select an initial set of $N_{\rm i} =
150$ particles from the list that are introduced in the simulation as
usual, i.e., as the initial cluster, which is a small fraction of the
eventual total. (4) For each particle in the subsequent list, the time
needed for its formation is calculated as $t_{\rm *,i} = m_{\rm
  i}/{\rm SFR}$ and the cluster is evolved until $t + t_{*,\rm i}$
when the particle is born. This step is repeated until there is no
more gas available.  Binaries are always added together, and the time
necessary for the formation of a binary is given by its total mass.

As in \citetalias{F17}, stellar mass loss from stellar evolution is
included in the simulations using the analytical models developed by
\cite{Hurley2000,Hurley2002} implemented in \texttt{Nbody6++}.
We also include artificial velocity kicks from asymmetrical supernovae
ejections. The magnitude of the velocity kicks follows a Maxwellian
velocity distribution with $\sigma = 265$ km/s based on proper motion
observations of runaway pulsars \citep{Hobbs2005} and the same
velocity kick distribution is used for neutron stars and
black-holes. With this high velocity kick distribution most
black-holes and neutron stars are ejected from the system
\citep{Pavlik2018} and therefore dynamical effects caused by these
objects are negligible.

\begin{figure*}
                \subfigure[Virial ratio] {
                \includegraphics[width=\columnwidth]{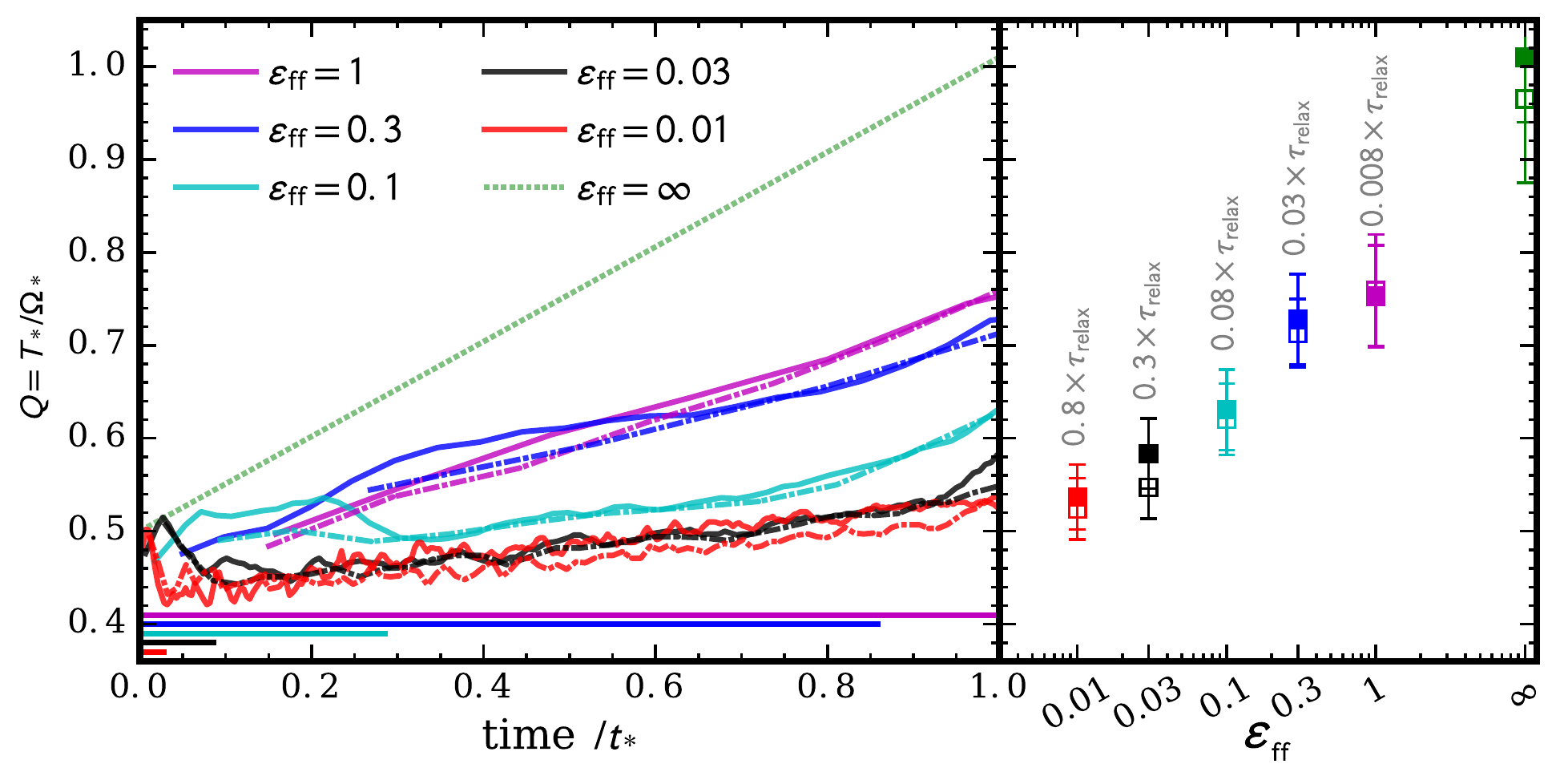}    
                }
                \subfigure[Half mass radii]{
        \includegraphics[width=\columnwidth]{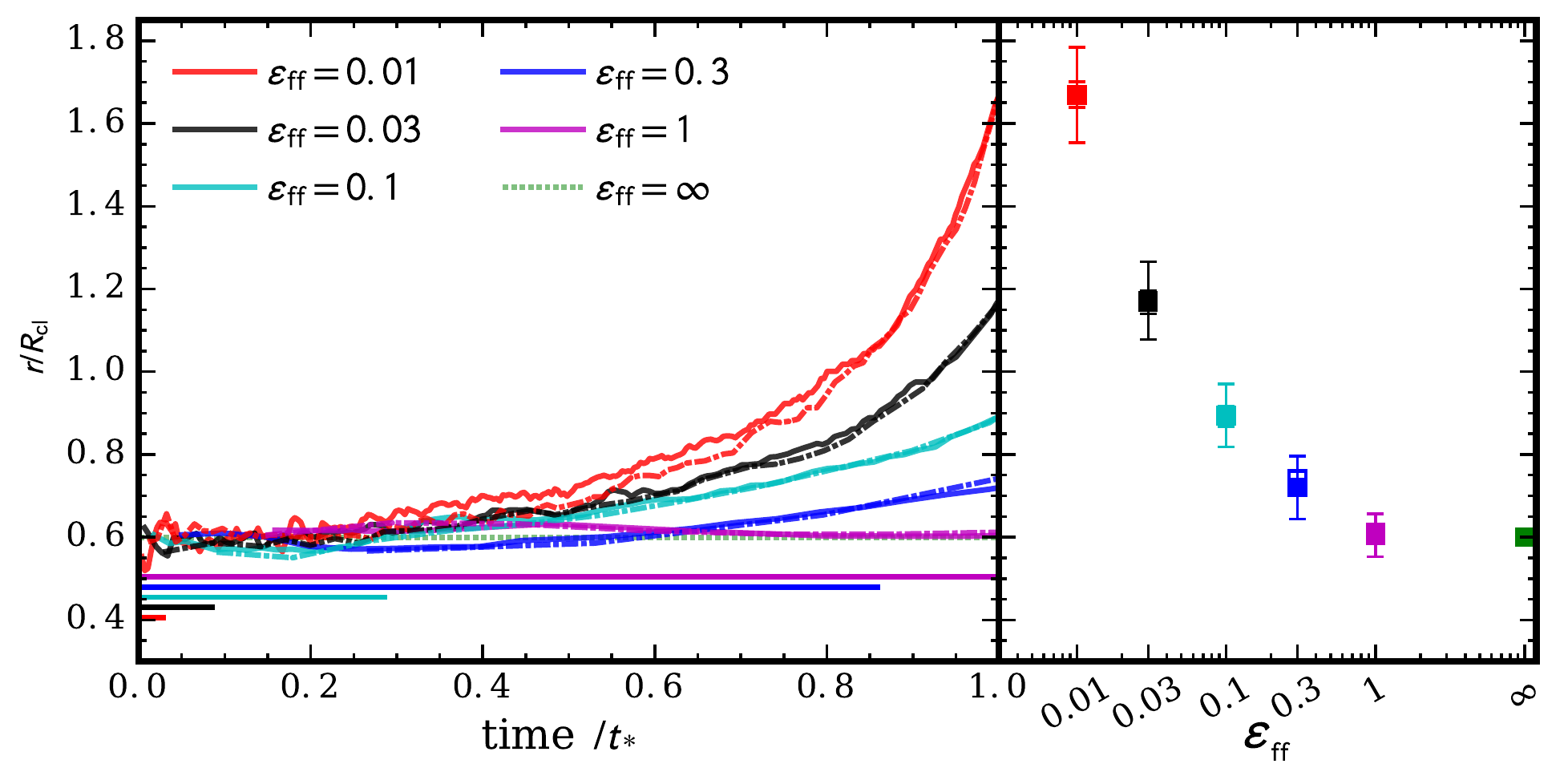}  
        }\\
        \subfigure[Velocity dispersion]{
        \includegraphics[width=\columnwidth]{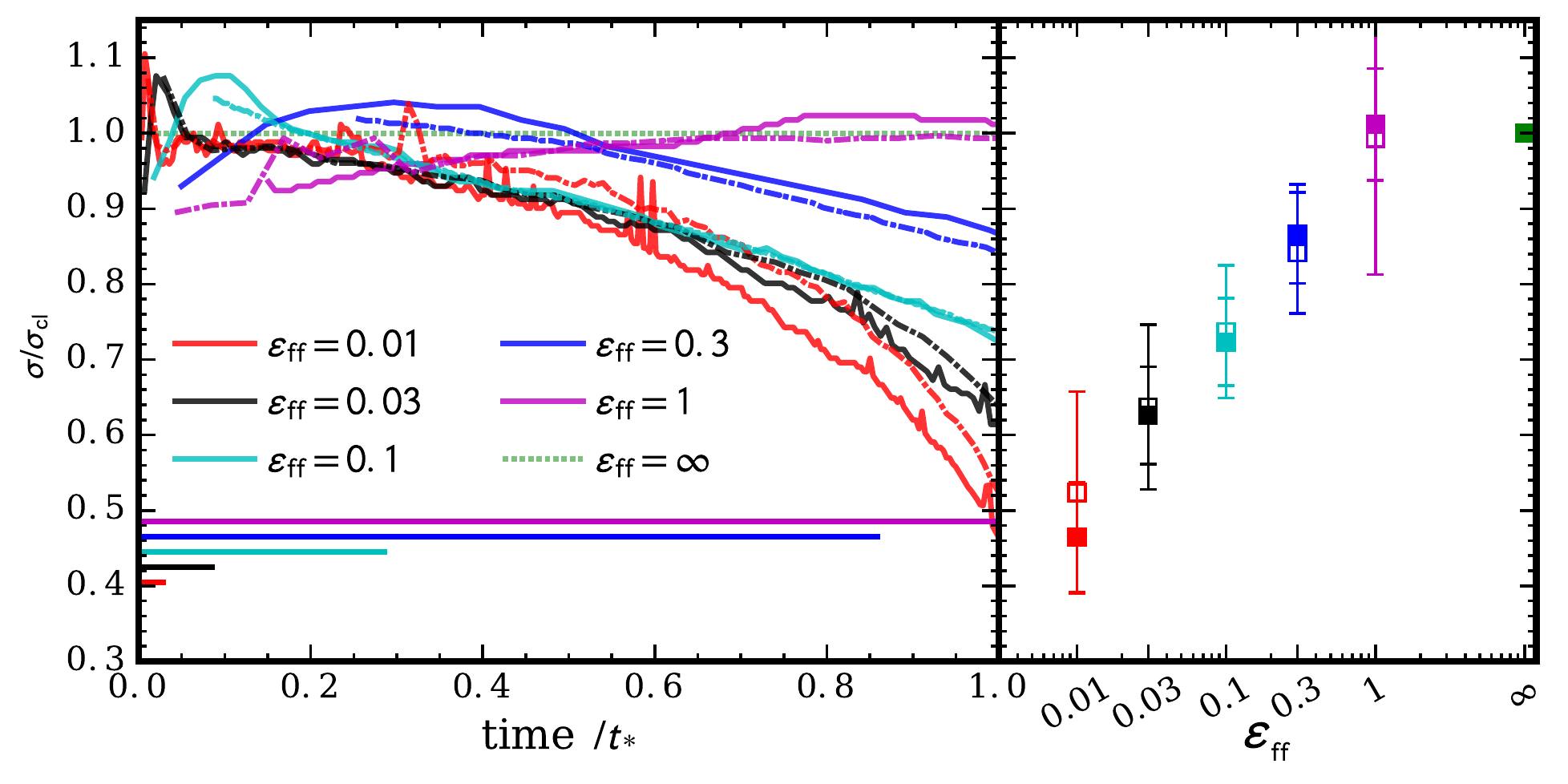}  
        }
        \subfigure[Mass segregation parameter]{
        \includegraphics[width=\columnwidth]{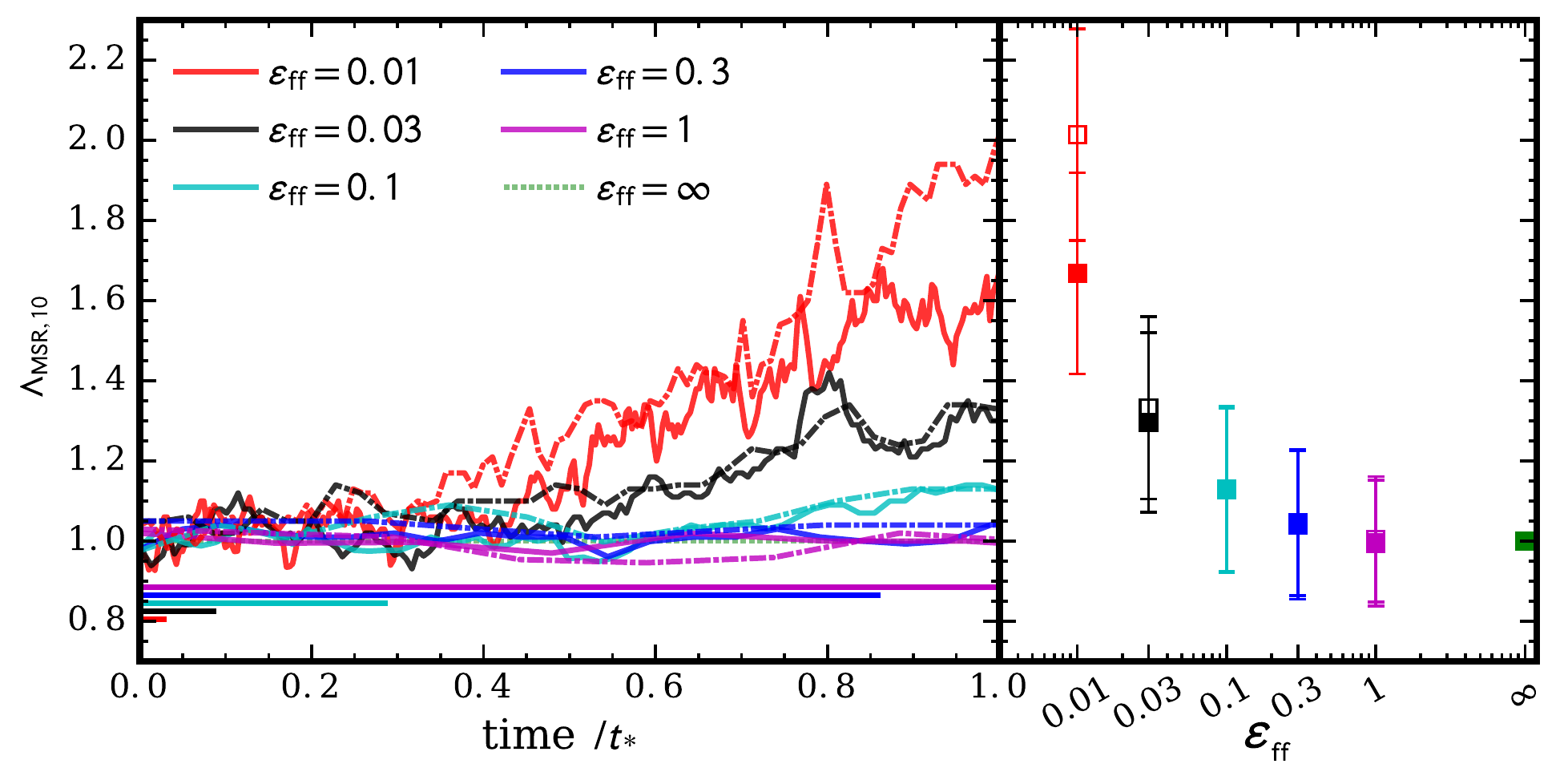}  
        }
        \caption{
Evolution of the virial ratio (a), half-mass radius (b), velocity
dispersion (c) and mass segregation parameter based on ten most
massive stars (d) during the formation of the star clusters for
different values of $\epsilonff$.  Low-$\Sigma$ case is drawn
with solid lines and High-$\Sigma$ with dot-dashed lines.  The time is
normalized by the duration of the star-forming stage, $t_*$. For
reference, for each value of $\epsilonff$, the length of a
crossing time is drawn as an horizontal line in the bottom of the
panel. In the $\epsilonff=1$ case (magenta lines),
$t_*\approx0.4t_{\rm cross}$, so this line extends beyond the range
shown in the panels. Each value at the end of the formation stage is
shown in the right respective panels as a function of $\epsilonff$.
High and Low-$\Sigma$ cases are represented with open and
filled squares respectively.  Errorbars show the standard deviation of
the mean values over the 20 realizations performed for each set.  In
panel (a), each point is labelled with the length of $t_*$ in terms of
the relaxation time of the system using the final number of stars.  }
        \label{fig:qevol}
\end{figure*}

\subsection{Overview of the Cluster Models}
\label{sec:binary}

In this work, we make exactly the same assumptions for the stellar
distribution of stars in space and in velocity structure as in
\citetalias{F17}. We assume stars are formed from the turbulent clump
of gas, and therefore they follow the same global structure given by
Equations~\ref{eq:dens} and \ref{eq:sigmar}.  We test the parent
clumps in two extreme scenarios with $\Sigmacl = 0.1$, and 1.0
g cm$^{-2}$ that we refer as high and low $\Sigma$ cases, respectively.
In this paper we focus on the differences generated when varying the
rate at which stars are formed as parametrized by $\epsilonff$,
for which the case of \citetalias{F17} can be compared as
$\epsilonff\rightarrow\infty$. We focus on the \fiducial set of
simulations defined in \citetalias{F17} (see Table~\ref{tab:ic}),
i.e., using a standard initial mass function \citep{Kroupa2001} with a
binary fraction $f_{\rm bin}$ with circular orbits.  Binary properties
are drawn from a log-normal period distribution with a mean of
$P=293.3\,$yr and standard deviation of $\sigma_{\log P} = 2.28$ (with
$P$ in days) according to observations of \cite{Raghavan2010}. The
mass ratio distribution follows the form $dN/dq\propto q^{0.7}$ as
observed in young star clusters \citep{Reggiani2011}.  We test values
of $\epsilonff=$ 0.01, 0.03, 0.1, 0.3, and 1, adopting
$\epsilonff=0.03$ as a fiducial value.

In this work, we deviate from the fiducial case only to see
differences of two extremes: the \fullbinaries case with 100\%
binaries and the \segregated case where stars are radially sorted
based on mass, which emulates an extreme case of mass segregation. However,
these cases are only tested against the fiducial value of
$\epsilonff$.

A summary of the initial conditions is shown in Table~\ref{tab:ic}, as well as the
parameters chosen for the parent clumps in Table~\ref{tab:clumps}.

\section{Results}
\label{sec:results}

\subsection{Star cluster formation stage}
\label{sec:formation}

The main difference between this and most previous $N$-body studies,
is that we recreate the formation stage of star clusters by gradually
adding stars during the simulation. Since the phase-space distribution
of the stars follows the structure of the gas, this causes that
star clusters are born with a dynamical
structure that does not represent a stellar system in dynamical
equilibrium. Even though the virial ratio of a system with SFE of
100\% would be $Q = 0.5$ (see Figure 1 of \citetalias{F17}), the
system is not in equilibrium since the velocity structure of the gas
(a velocity dispersion that increases with the central distance) is the
opposite of a dynamically stable stellar system (for which the
velocity dispersion decreases with radius).
Thus a longer formation timescale (i.e., lower $\epsilonff$)
can be crucial for the star cluster in order to rearrange internally
before residual gas is dispersed, and therefore start its gas free
evolution closer to equilibrium.

Figure~\ref{fig:qevol}a shows the evolution of $Q$, i.e., degree of
gravitational boundedness, with time during the formation stage for
the different values of $\epsilonff$ adopted in this work. The
time is normalized by $t_*$ (see Table~\ref{tab:ic}), and for
reference we also show the length of a crossing time for each value of
$\epsilonff$ in this normalization. Initially all systems start
their evolution close to $Q=0.5$ since all the mass of the clump is
present. As the cluster evolves, the stars that are present attempt to
relax to equilibrium as more stars arrive with the relatively high
velocity dispersion fixed from the initial conditions and the mass of
the gas decreases. However, the initially small stellar cluster needs
about one relaxation time ($t_{\rm relax}\sim50\, t_{\rm cross}$ in
the systems studied here) to reach equilibrium. The duration of star
cluster formation in the fast regime, $\epsilonff>0.1$, is
comparable to the crossing time of the system. In this short time the
star cluster is unable to reach equilibrium and the consequence is
that it emerges from the gas with supervirial velocities. In contrast,
in the slow regime $\epsilonff \ll 0.1$, i.e., with slow star
formation, the timescale is comparable to the relaxation time, and the
stellar cluster has plenty of time to approach towards equilibrium.
In this scenario, the formation of the star cluster is a race to reach
equilibrium, and the path to equilibrium is truncated when the gas is
gone and the cluster emerges from its natal clump with a global virial
ratio given by this truncation (see right panels of
Figure~\ref{fig:qevol}).
Here the two extreme values of $Q$ that a cluster is able to reach at
the end of its formation are $Q \simeq 0.5$ when $t_* \gtrsim t_{\rm
  relax}$ and $Q \simeq 0.75$ when $t_*$ is relatively short, with a
limit of $Q\simeq 1$ in the instantaneous formation case of Paper I.

Figure~\ref{fig:qevol}b shows that the rearrangement of the stars into
a relaxed configuration implies that the cluster expands even before
all gas is exhausted.  While the star-forming region, i.e., the clump
of gas, is forced to remain with the same size, a considerably amount
of stars start orbiting the system outside the clump. Long orbits also
means low velocities and the overall velocity dispersion of the
cluster decreases, as shown in Figure~\ref{fig:qevol}c. As shown in
these panels b and c, how much the cluster expands and the velocity
dispersion decreases is highly dependent on the assumed $\epsilonff$.
In the slowest formation scheme tested here, $\epsilonff=0.01$, the 
half mass radius of the cluster expands a factor of
2.6 and the velocity dispersion decreases by a factor 2 compared to
the initial values. However in the fastest cases $\epsilonff=1$ and
$\infty$, the initial values are not modified at
all. This result is independent of $\Sigmacl$.

We have also followed the evolution of mass segregation during the
formation stage. We quantified mass segregation using the mass
segregation ratio $\Lambda_{\rm MSR}$ introduced by
\cite{Allison2009}. $\Lambda_{\rm MSR}$ is defined as the ratio
between the length of the minimum spanning tree (MST) connecting the
$N_{\rm MST}$ most massive stars in the cluster ($l_{\rm massive})$,
which is unique) and the MST connecting $N_{\rm MST}$ randomly
selected stars ($l_{\rm normal}$). The latter is done 500 times in
order to have good estimations of the dispersion in $l_{\rm normal}$,
$\sigma_{\rm norm}$, then:
\begin{equation}
        \Lambda_{\rm MSR} \equiv \frac{\langle l_{\rm normal} \rangle }{l_{\rm massive}} \pm 
        \frac{\sigma_{\rm normal}}{l_{\rm massive}}
\end{equation}
where $\Lambda_{\rm MSR}\sim1$ means no mass segregation and
$\Lambda_{\rm MSR}>1$ indicates a mass segregated star cluster, i.e.,
massive stars are more closely distributed than other stars. For our
numerical evaluations of $\Lambda_{\rm MSR}$ we adopt $N_{\rm MST}=10$.

In our fiducial models with different values of $\epsilonff$,
the star clusters are created without primordial mass
segregation. Figure~\ref{fig:qevol}d shows how in all cases
$\Lambda_{\rm MSR}=1$ at $t=0$ and mass segregation gradually develops
as time progresses. The level of mass segregation that star clusters
are able to reach at the end of star formation thus depends on
$\epsilonff$. Mass segregation is developed dynamically as
massive stars sink to the center and kick out lower mass stars.
Numerical studies have shown that star clusters can develop
significant mass segregation on timescales $< 1$~Myr with
\citep{Dominguez2018} and without background gas
\citep{Allison2009b}. In these studies, fast segregation is normally
explained by the means of high levels of primordial substructure and
sub-virial velocities involving the merging of subclumps of stars.
Figure~\ref{fig:qevol}d shows that longer formation timescales develop
high levels of mass segregation, and this result is independent of the
initial density. The only exception appears in the case of
$\epsilonff=0.01$ (red lines) where in the low $\Sigma$ case
stellar evolution affects the result as massive stars die truncating
the growth of segregation. However, in all other cases the result is the
same, low values of $\epsilonff$ implies more crossing times in
a dense state and more time for developing mass segregation.

In agreement with previous studies, with our spherically symmetric
models that have minimal substructure, we are not able to develop very
high levels of mass segregation in $<1$~Myr, even in the high density
cases. The case where mass segregation developed most was in the
high-$\Sigma$ case with $\epsilonff=0.01$. Here star clusters
reached $\Lambda_{\rm MSR}=2$ but it takes $t_* \approx 3$ Myr to
develop it.  Future improvements of our modeling will include the
addition of primordial substructure and we will be able to study how
fast mass segregation can be developed in different density
environments and formation timescales.

\subsection{Emergence from the gas}
\label{sec:fbounds}

\begin{figure}
        \includegraphics[width=\columnwidth]{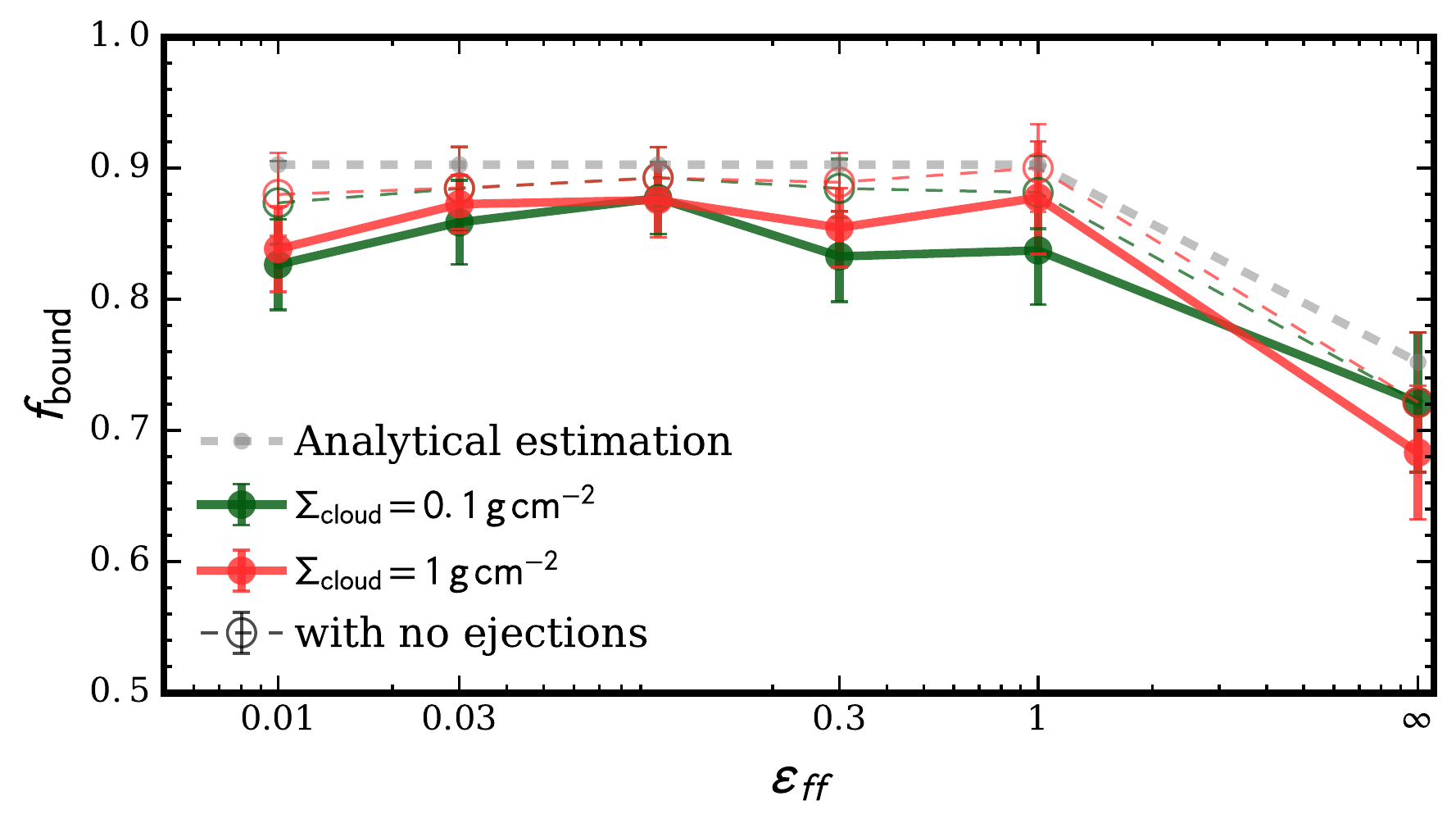}
        \caption{
Bound mass fraction measured at the time at which the gas is exhausted
($\tsf$), for simulations in the \fiducial set in the high (red) and
low (green) $\Sigma$ cases. Gray dashed line shows a semi-analytical
estimation (see \S\ref{sec:bound_model}). Each point with errorbars
represents the mean and standard deviation for 20 simulations. Thin
lines with open symbols show the resulting bound fractions, if
dynamical ejections are counted as bound, for comparison with the
analytic model.  }
        \label{fig:bounds}
\end{figure}

\begin{figure*}
\includegraphics[width=\textwidth]{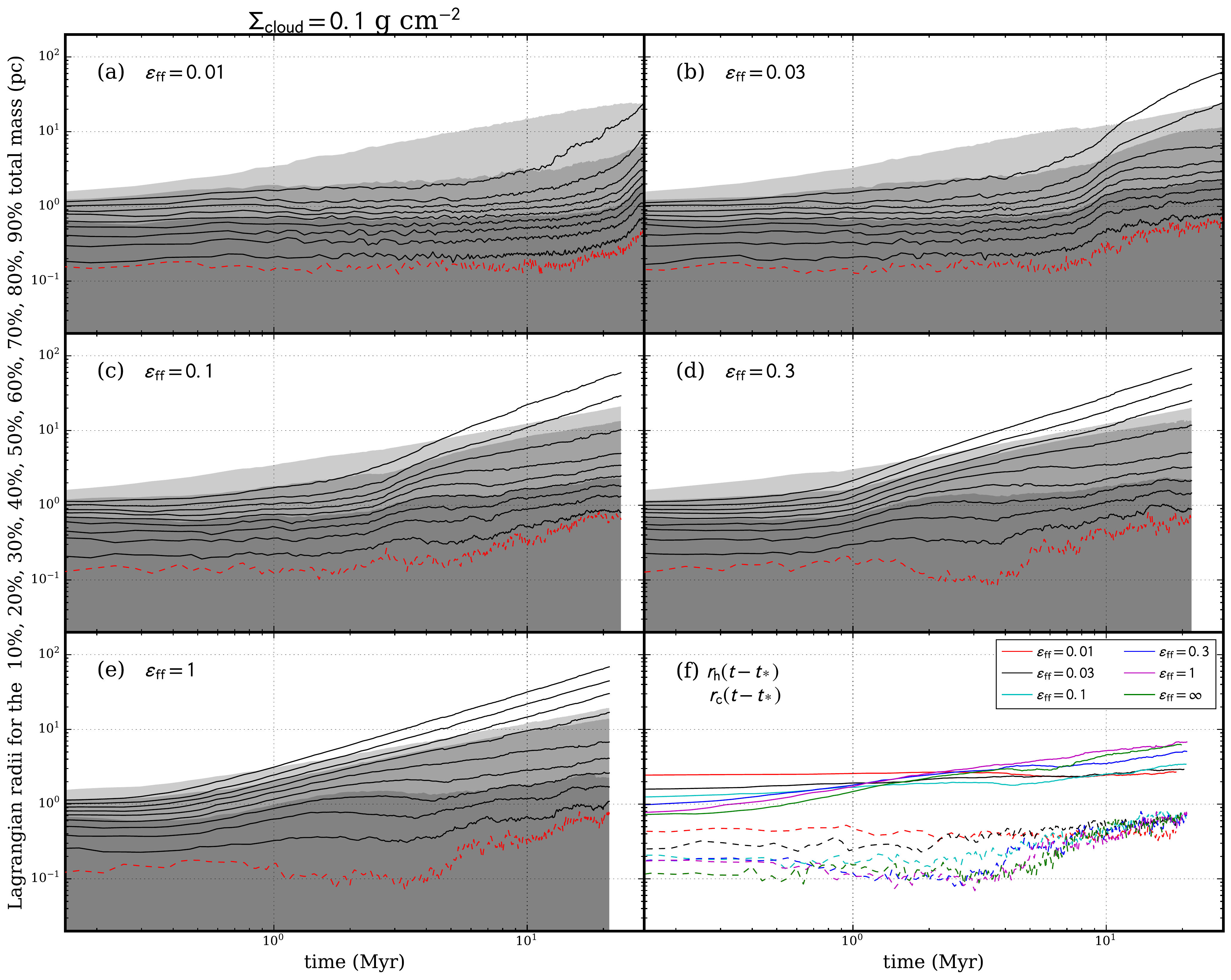}
\caption{
Time evolution of average Lagrangian radii for star clusters born from
a parent clump with $\Sigmacl=0.1\:{\rm g\:cm^{-2}}$ and for various
values of $\epsilonff$, as labelled in panels (a) to (e).  We
show the Lagrangian radii for the 10\%, 20\%, 30\%, 40\%, 50\%, 60\%,
70\%, 80\% and 90\% masses (black solid lines). Red dashed lines are
the core radii defined in \protect\cite{Aarseth2003}.  Gray shaded areas
represent the regions below the 50\%, 95\% and 100\% mass radius of
the bound cluster. Panel (f) shows the half-mass and core radii of all
the clusters together, including the $\epsilonff=\infty$
case.}
\label{fig:lrad1}
\end{figure*}

\begin{figure*}        
        \includegraphics[width=\textwidth]{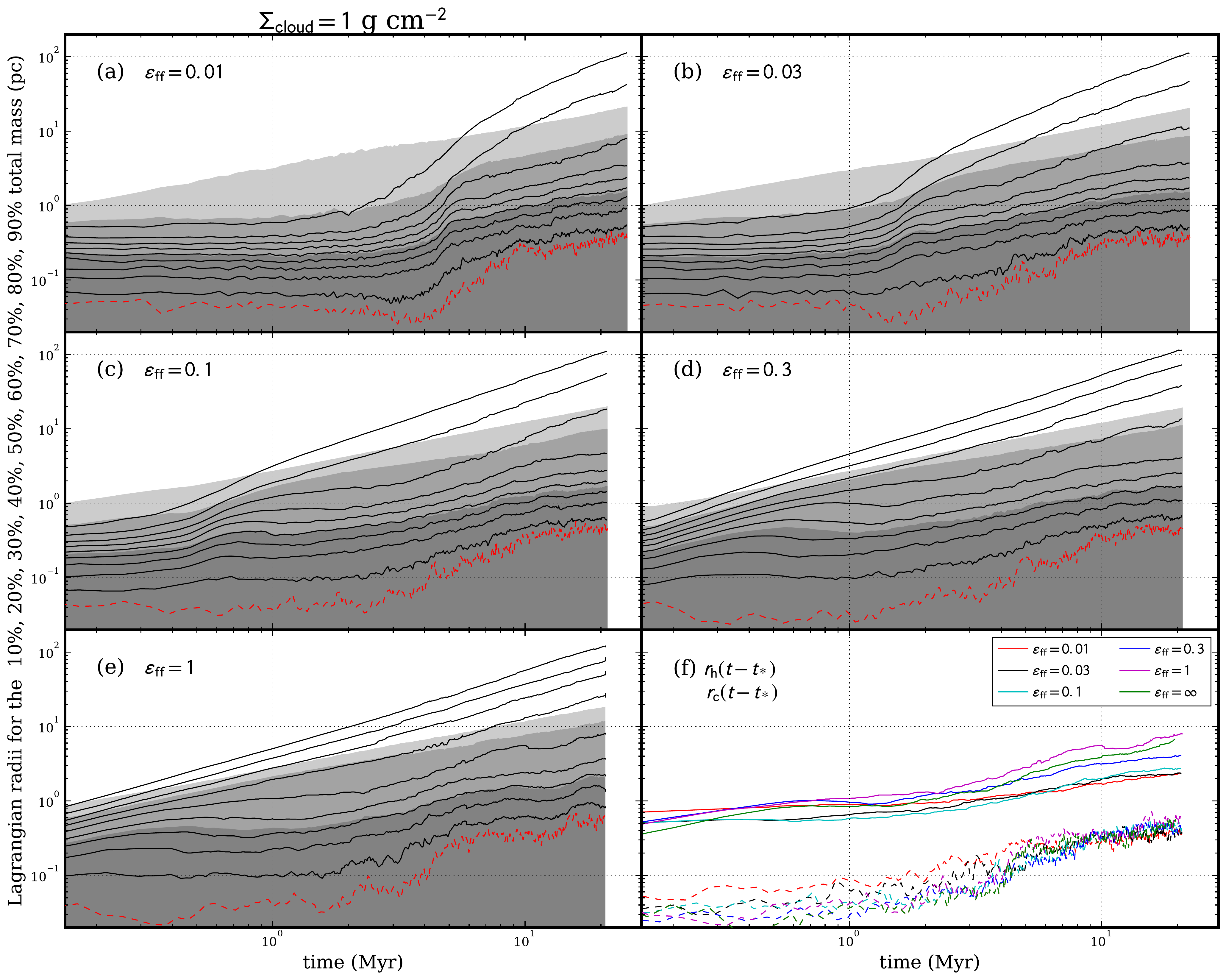}
        \caption{%
Same as Figure \ref{fig:lrad1} for star clusters born from a clump with
$\Sigmacl=1\:{\rm g\:cm^{-2}}$.
        }\label{fig:lrad2}
\end{figure*}

Figure~\ref{fig:bounds} shows the different values of bound fraction
($f_{\rm bound}$) at $t=t_*$. While there are some differences between
the models, that we discuss below, in general there is only modest
variation of $f_{\rm bound}$ for different values of $\epsilonff$.
The different values of $Q$ obtained at the end of the
formation stage, between $0.5< Q <0.8$, would suggest different bound
fractions \citep[see e.g.][]{Farias2018,Lee2016}, however the
differences in this narrow range are quite small. Second order
differences are a combination of dynamical effects such as early
dynamical ejections and the assumptions of the formation process,
e.g., the different rates of decrease of the gas mass.

Let us first ignore early dynamical effects such as virialization and
the rearrangement of the stellar masses. In this regime, the
sequential formation of stars does not affect the final bound
fraction. As each star is formed the escape velocity decreases in the
same amount instantly.  Therefore, how fast or slow stars are created
becomes irrelevant because of the instant residual gas loss assumed in
this work.
Gray dashed lines on Figure~\ref{fig:bounds} show an analytical
estimation considering the radial and time dependent escape velocity
and clump velocity distribution based on the cloud model described in
\S\ref{sec:gasmodel} (see Appendix~\ref{sec:bound_model} for
details). Thin lines with open symbols in Figure~\ref{fig:bounds} show
the resulting bound fractions when only removing stars that are born
unbound, i.e., ignoring dynamical ejections, and we can see that the
analytic estimations and numerical experiments agree well.

Then, the actual bound fraction is shown by the solid lines with solid
circles in Figure~\ref{fig:bounds}. Values of $\epsilonff =
0.1$ and 0.3 (low $\Sigma$) show slightly lower bound fractions,
probably due to the combination of the higher virial ratios and
dynamical ejections that happen during formation. Then, at low values
of $\epsilonff<0.1$ there are also slightly smaller bound
fractions: in this case, dynamical ejections play an important role
during formation, since it happens on several ($\gtrsim 10$) crossing
times. It is at $\epsilonff=0.1$ where both effects appear to
be minimized.

\subsection{Global evolution}
\label{sec:evol}

\begin{figure}
\centering
\includegraphics[width=\columnwidth]{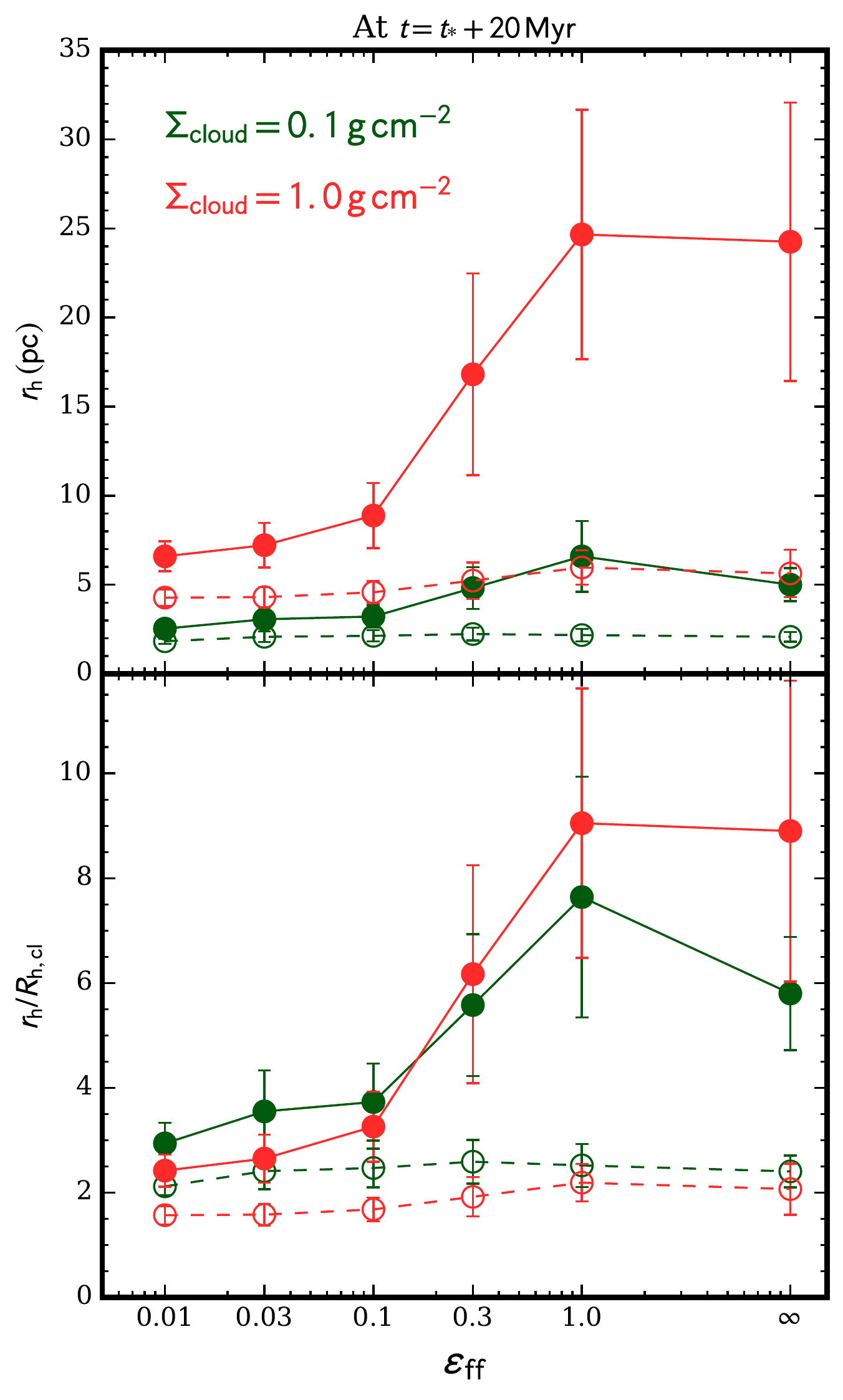}
\caption{
Size of star clusters at
20~Myr after star formation stops. The top panel shows the half-mass
radius in parsecs. The bottom panel shows the half-mass radius
relative to the initial clump size ($R_{\rm cl}$). Filled circles are
the values measured with all stars in the clusters, while open circles
are values measured only with the bound stars. The low (green) and
high (red) $\Sigma_{\rm cl}$ cases are indicated (see legend).
}
\label{fig:sizes}
\end{figure}

\begin{figure*}
\includegraphics[width=0.98\textwidth]{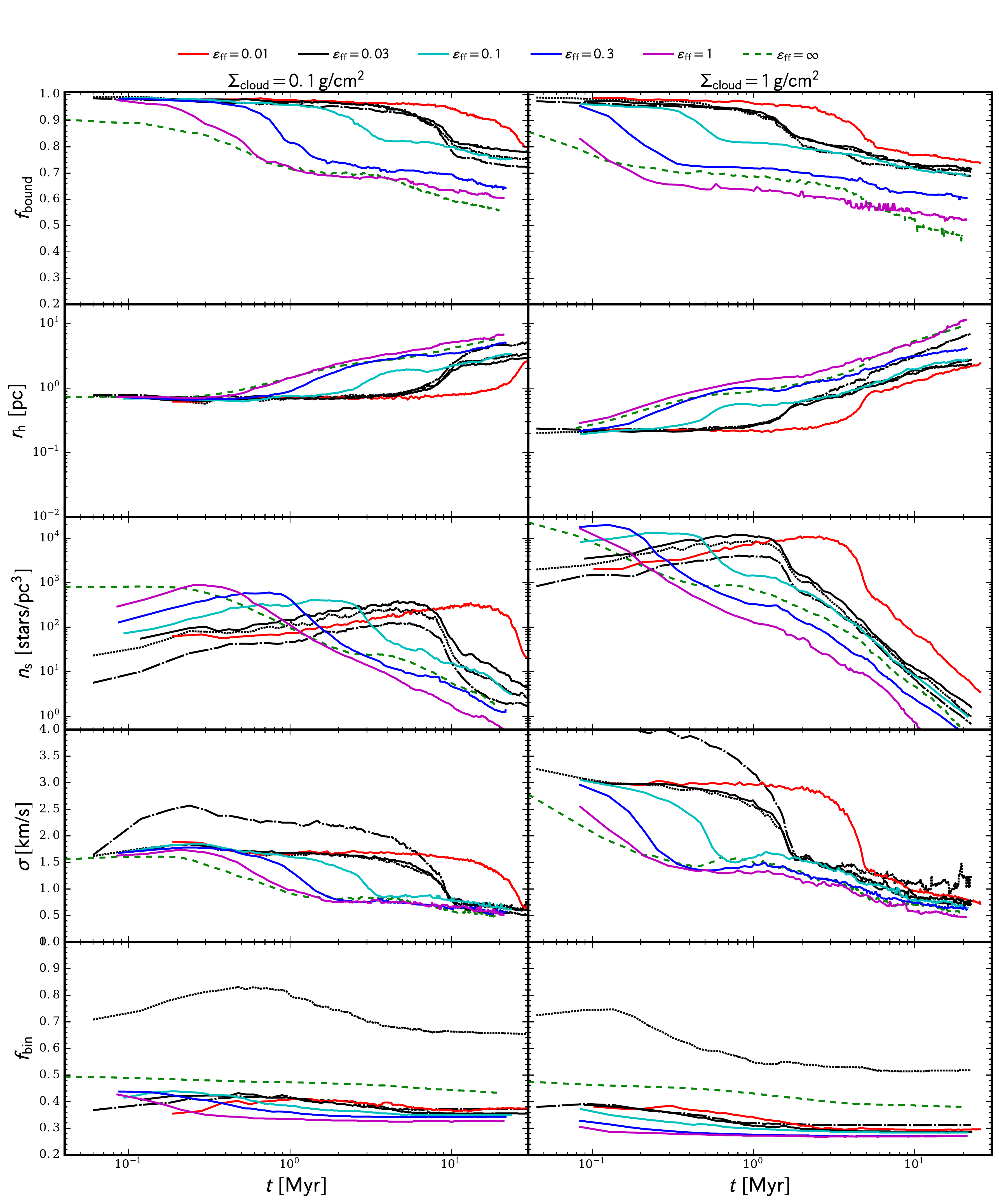}
\caption{
Time evolution of properties for our fiducial clusters formed with
SFE=50\% and a range of $\epsilonff$ values in different colors
(see legend). Solid lines show results for the \fiducial set, dotted
lines for the set \fullbinaries, dot-dashed lines for \segregated and
dashed lines the results of the fast formation limit described in
\citetalias{F17}. Left column shows clusters forming from a
$\Sigmacl=0.1$ g cm$^{-2}$ environment; right column from a
$\Sigmacl=1$ g cm$^{-2}$ environment. The lines in each panel show
median values calculated from the 20 simulations performed for each
set. Top row shows the fraction of bound mass in the cluster relative
to the instantaneous stellar mass, where in this figure unbound stars
inside the 95\% mass radius of the bound cluster are kept to show the
timescale of their escape. Second row shows the evolution of the half
mass radius $r_{\rm h}$ for all the stars in the simulation. Third row
shows the average number density of systems ($n_{s}$), where by
systems we refer to singles and binaries, measured inside the volume
defined by $r_{\rm h}$.  Fourth row show the evolution of the velocity
dispersion measured inside $r_{\rm h}$, and bottom row shows the
evolution of the global binary fraction.  }
\label{fig:evol}
\end{figure*}

We have seen that one of the key effects of varying $\epsilon_{\rm
  ff}$ is the ability of the star cluster to reach equilibrium before
gas is exhausted/ejected.  While the difference in their ability to
survive gas expulsion is not raised considerably, the condition on
which they face their early gas-free evolution is very different.  We
have evolved every model for 20 Myr after all stars are formed. The
global evolution under each model's parameters can be appreciated
through the Lagrangian radii shown in Figure~\ref{fig:lrad1} for the
low $\Sigma$ case and in Figure~\ref{fig:lrad2} for the high $\Sigma$
case.  Comparing between these mass surface densities, the differences
are similar as found in \citetalias{F17}: star clusters born with high
density expand at a faster rate than star clusters born in a less
dense state. However, the expansion is also regulated by the star
formation rate.

Panel (f) of Figures~\ref{fig:lrad1} and \ref{fig:lrad2} shows the
half mass and core radii for all the values of $\epsilonff$
with a time offset so that the starting point is the moment when they
start their gas free phase, i.e., $ t - t_*$. High values of
$\epsilonff$ result in star clusters that expand much faster
than the ones born with low $\epsilonff$. As we can extract
from Figure~\ref{fig:sizes}, where we compare the half mass radii at
20 Myr after formation, the extreme cases are when star clusters are
born with low density and low $\epsilonff$ and do not expand at
all following the expulsion of gas (they only expand initially a
factor 2 during the formation stage, see Figure~\ref{fig:qevol}).  On
the other hand, a star clusters born in one free fall time
($\epsilonff=1$) expands by a factor 9 during the first 20
Myr.

A summary of the evolution of different parameters is shown in
Figure~\ref{fig:evol}.  Including the evolution of the extremely mass
segregated case (\segregated set, in dot dashed lines), and the set
with 100\% binaries (\fullbinaries, shown in dotted lines), withe the
low $\Sigma$ case in the left panels and high-$\Sigma$ case in the
right.

Top panels show the evolution of the bound fraction. As discussed in
\S\ref{sec:fbounds}, at $t_*$ the fraction of bound stars is similar
at all $\epsilonff$. However, $t_*$ is different for each set
and since $f_{\rm bound}$ evolves due to dynamical ejections, at a
fixed time, e.g., $t_{\rm m} =10$\,Myr, the bound stellar fraction is
very different in each case. If we take a fixed physical time as a
reference, then we can say that star clusters born slowly better
survive gas exhaustion than star clusters that form fast.  However,
this is because the formation time ($t_{\rm *}$) is different for each
set and a fixed physical time does not represent the star clusters in
the same state.  Most of the stellar mass loss is during the star
formation phase due to dynamical ejections in a high density state and
stars being born unbound from the cluster (see
Section~\ref{sec:unbound}), but after the gas is gone, dynamical
ejections are still happening, and therefore, if we compare at a fixed
physical time ($t_{\rm m}$), we will count the mass loss during the
formation plus the mass loss during the time $t_{\rm m} - t_{*}$.  If
we compare the bound fraction at the end of star formation in each
case as shown in Figure~\ref{fig:bounds}, we see that the difference
between the models is not significant. Extreme cases such as
\segregated and \fullbinaries do not show significant deviations.

Second row panels show the evolution of stellar half mass radius in
physical time. We showed in \S\ref{sec:formation}, slow-forming star
clusters expand more during formation but after gas is gone they are
more stable against further expansion. However, since their formation
stage takes longer, slow-forming clusters are more compact at any time
in their evolution when comparing to star clusters formed fast at the
same time. While the \fullbinaries set expands at the same rate as the
\fiducial set, the \segregated set expands faster since the velocity
dispersions remain high due to the large amounts of close massive
stars. A direct consequence of the evolution of $r_{\rm h}$ is the
evolution of the system number densities ($n_{\rm s}$) within $r_{\rm
  h}$, shown in the third row. By system we refer to single and binary
stars, therefore the \fullbinaries set have slightly smaller number
densities than the \fiducial set. Also, since in the set \segregated
more massive stars are in the center, then there are fewer stars born
in the center in order to follow the density profile of the parent
clump. The difference in $n_{\rm h}$ due to these effects remains
during the whole evolution of the cluster. Overall, $n_{\rm h}$ is
high during the formation of stars and increases as new stars are born,
then when the gas is gone and star clusters start to expands $n_{\rm
  h}$ fall several orders of magnitude in all sets.

The fourth row of panels shows the evolution of the velocity
dispersion within $r_{\rm h}$, $\sigma_{\rm h}$.  Velocity dispersions
are high during the formation of stars, since the background gas is
present, and after gas is gone it falls until an equilibrium value is
established that is similar for all the models. Later decrease of
$\sigma_{\rm h}$ is relatively small and would be very difficult to
distinguish between the models. Note, $\sigma_{\rm h}$ depends on the
total density of the cluster, but its variation is small when the
initial mass of the system, $M_{\rm cl}$ falls by a factor of 0.5,
i.e., $\epsilon$. In the set \segregated, $\sigma_{\rm cl}$ rises
rapidly as all massive stars are close to each others, however, after
the rearrangement of the stellar distribution, the final value of
$\sigma_{\rm cl}$ is the same as the other models.

The last row of panels shows the evolution of the binary
fraction. Within the models, the evolution of the binary fraction is
similar, however in all cases studied here, there is more disruption
of binaries than in the case studied in \citetalias{F17}. This is
because in the cases here, the number densities stay higher for
longer, and most binaries have a reasonable chance to interact with
the other stars. However in the fast formation limit, the cluster
starts its life in expansion, not allowing much interaction between
them and the rest of stars. Binary stars are introduced uniformly over
time, however, binary stars are disrupted very early and the binary
fraction evolves to a smaller number than that given by the initial
conditions. We discuss the binary population in the next section.

\subsection{Binary population} 
\label{sec:binaries}

\begin{figure*}
        \centering
       $\begin{array}{rl}
       \includegraphics[width=0.5\textwidth]{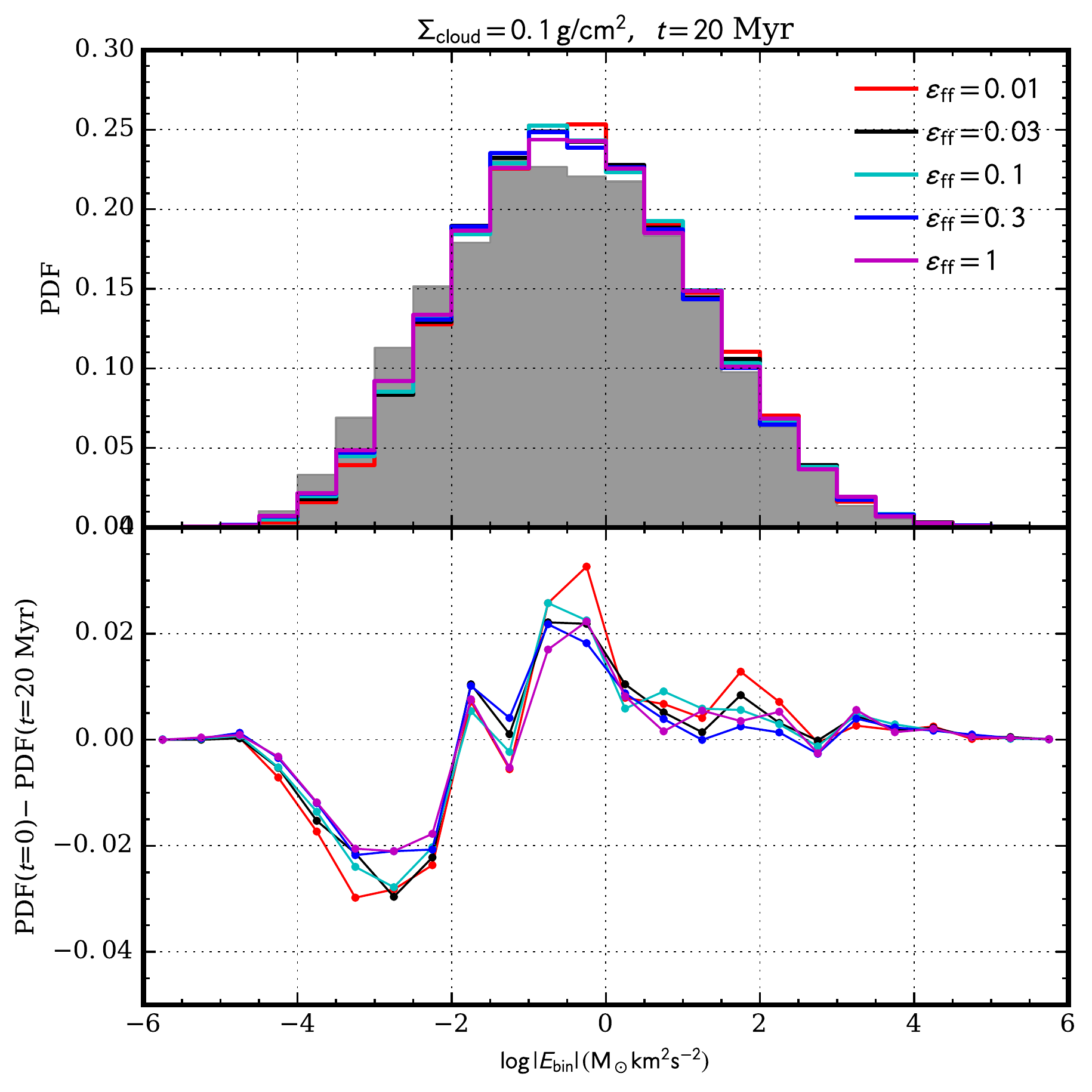}&
       \includegraphics[width=0.5\textwidth]{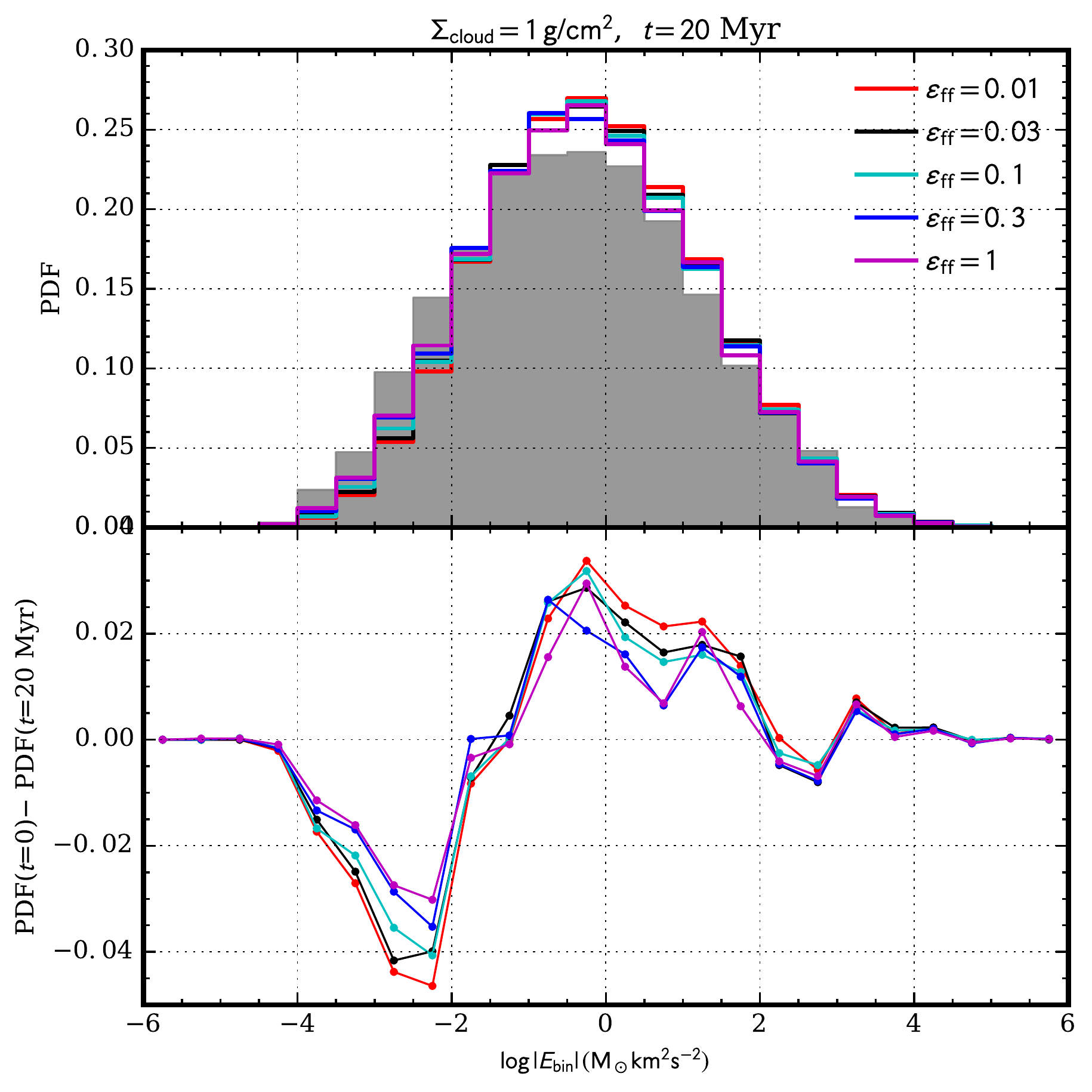}
        \end{array}$
        \caption{
The probability distribution functions (PDF) of binary binding
energies. Top panels shows the PDFs measured at 20 Myr with the shaded
area representing the distribution of the initial conditions given to
binaries when they form. Bottom panels show the differences in the
PDFs in comparison with the initial population. The left column shows
the result for clusters in the low mass surface density environment
($\Sigma_{\rm cl}=0.1\:{\rm g\:cm^{-2}}$); the right column shows the
high mass surface density environment ($\Sigma_{\rm cl}=1\:{\rm
  g\:cm^{-2}}$).}
        \label{fig:ebin}
\end{figure*}
\begin{figure*}
        \centering
       $\begin{array}{rl}
       \includegraphics[width=0.5\textwidth]{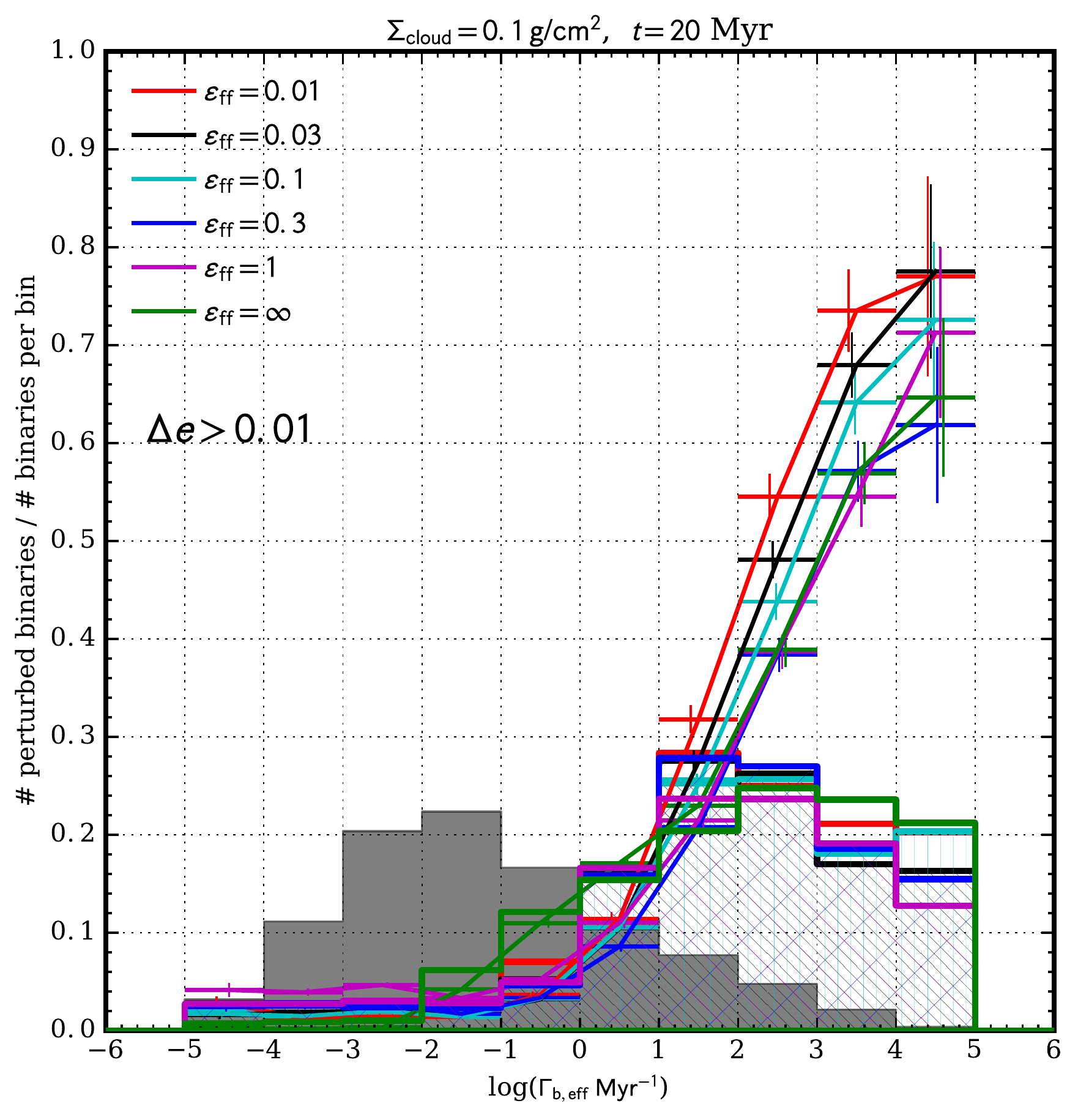}&
       \includegraphics[width=0.5\textwidth]{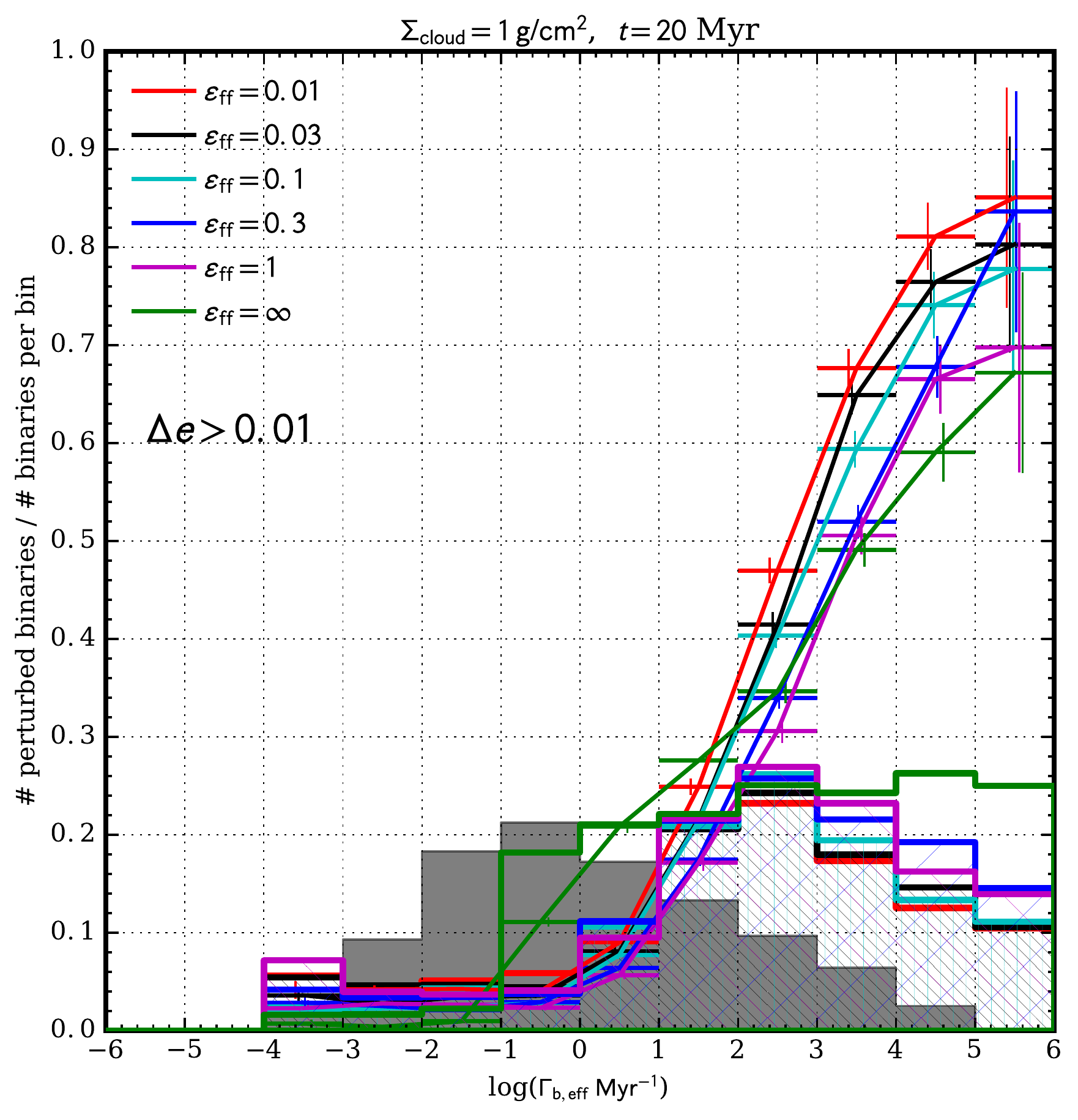}
        \end{array}$
        \caption{
Fraction of binaries in each $\Gamma_{\rm b,eff}$ bin that were
disrupted (lines with errorbars) and for which the initial
eccentricity changed by $\Delta e>0.1$ (hatched histograms).
The grey filled histogram shows the actual fraction of binaries in
each bin. Most binaries with high $\Gamma_{\rm b,eff}$ are disrupted,
however note that there are only a small fraction of binaries in this
regime. Thus, overall the changes in the binary populations are
relatively modest, but at high $\Gamma_{\rm b,eff}$ the evolution is
significant, including some dependence on $\epsilonff$.}
\label{fig:ecchist}
\end{figure*}

In the models presented here we have included a relatively high (50\%)
percentage of primordial binaries (though note this value is
compatible with observational estimates).  Binary population
parameters in this work are the same as in \citetalias{F17}. In
\citetalias{F17} we did not find that there was significant dynamical
processing of primordial binaries. Now in the models with gradual star
formation, as star cluster formation last longer, the chance that a
binary system has to interact with others increases since number
densities stay higher for longer. Therefore we would expect that the
degree of dynamical processing increases with lower values of
$\epsilonff$.  Indeed, this is what we see from
Figure~\ref{fig:ebin}, where we compare the initial distribution of
binding energies (shown by grey filled histograms) with the same
distribution measured at 20 Myr for the different values of
$\epsilonff$.
The differences between initial and evolved distributions are shown in
the bottom panels. In general the Heggie-Hut rule is followed as hard
binaries become harder and soft binaries become softer with time,
causing the peak of the PDF to move to the left.  In the high
$\Sigma_{\rm cl}$ case (right panels) the differences between models
is greater, where low $\epsilonff$ values results in larger
modifications of the initial PDF.  In the low $\Sigma_{\rm cl}$ cases
the differences are less prominent, even though the same trend is
still present.

However, the above results will depend on the adopted initial binary
population. In \citetalias{F17}, we calculated the interaction rates
of interactions with cross sections on the order of the separation of
the binary stars as:
\begin{eqnarray}
        \label{eq:gamma}
        \Gamma_{\rm b, eff} &=& \Gamma_{\rm b} \cal{F} 
\end{eqnarray}
where $\Gamma_{\rm b}$ is the interaction rate factor given by:
\begin{eqnarray}
        \Gamma_{\rm b}(a,\mu) & \lesssim& 
        9.67\times10^{-3} \left( \frac{n_{s}}{10^4\:{\rm pc}^{-3}} \right) 
        \left(         \frac{\sigma}{2~{\rm km\,s^{-1}}} \right) \nonumber \\
        \times & &  \left( \frac{a}{40~{\rm AU}} \right)^2 \frac{\langle m_s
        \rangle^2}{\mu^2}\:{\rm Myr}^{-1} \quad,
        \label{eq:gamma2}
\end{eqnarray}
where $a$ is the binary semi-major axis, $\mu = m_1m_2/(m_1+m_2)$ is
the reduced mass of the binary with components of masses $m_1$ and
$m_2$. And $\langle m_{\rm s}\rangle$ is the average system mass,
which in a system with a binary fraction $f_{\rm bin}$ and average
mass $\langle m_i \rangle$ this is $\langle m_{\rm s}\rangle \approx
(1+f_{\rm bin})\langle m_i \rangle $, ignoring the presence of higher
order multiple systems.  The inequality comes from the fact that the
derivation of $\Gamma_{\rm b}$ assumes that if the perturber is a
binary, then its semimajor axis is smaller or similar to the binary in
question, which normally holds for soft binaries but not usually for
the hard ones.  Gravitational focusing effects are accounted by the
factor $\cal{F}$ in Eq.~\ref{eq:gamma} given by:
\begin{eqnarray}
        \cal{F}&\approx& 1 + \left[ 5.55 
        \left(\frac{\mu}{{M_\odot}} \right)
        \left( \frac{a}{40\:{\rm AU}} \right)^{-1}
        \left( \frac{\sigma}{2\:{\rm km/s}}\right)^{-2} \right. \nonumber \\  
        & &\times \left.  \left( \frac{m_1+m_2}{m_s} +1 \right)^{\vphantom{1}} \right].
\end{eqnarray}

While these interaction rates change as the cluster evolves and the
internal properties of the binary systems change with each
interaction, it serves as a good estimator of which stars are most
likely to be perturbed during the simulations. The grey histograms of
Figure~\ref{fig:ecchist} show the fraction of binary stars in each
logarithmic $\Gamma_{\rm b, eff}$ bin, given the chosen primordial
binary population properties.  In \citetalias{F17} we explored how
initial eccentricities changed as a function of $\Gamma_{\rm b,
  ff}$. First we considered the fraction of binaries that have their
eccentricities modified by a 1\% change or greater. This fraction
increases with $\Gamma_{\rm b,eff}$ until reaching a level of about
20\% by $\Gamma_{\rm b,eff} = 1\,{\rm Myr}^{-1}$, after which it
remains flat.

Now in the models of gradual cluster formation, with longer formation
timescales we expect a larger fraction of modified eccentricities for
$\Gamma_{\rm b,eff}>1$~Myr$^{-1}$.  However, as we can see from the
hatched histograms of Figure~\ref{fig:ecchist} this does not happen,
instead the fraction of modified eccentricities decreases with larger
$\Gamma_{\rm b,eff}$. The explanation is that so far we did this
analysis considering only binary stars that remained bound after
20~Myr. However, if we analyse the fraction of binaries that were
disrupted in each $\Gamma_{\rm b,eff}$ bin (see solid lines with
Poisson errorbars in Figure~\ref{fig:ecchist}) we can see that this
increases very strongly with $\Gamma_{\rm b,eff}$. In fact, at the
highest values of $\Gamma_{\rm b,eff}$, i.e., $\gtrsim4\:{\rm
  Myr}^{-1}$, the majority of binaries are disrupted and those that
are not almost all have had their eccentricities modified by at least
1\%.

However, even though the fraction of perturbed binaries is high at
large values of $\Gamma_{\rm b,eff}$, only a small fraction of binary
systems reach these high interaction rates. It remains to be explored
to which extent the binary population can be dynamically processed in
these models, if we choose different primordial binary populations
with different formation timescales.

\subsection{Radial structure}
\label{sec:profiles}

\begin{figure*}
        $\begin{array}{c}
         \includegraphics[width=\textwidth]{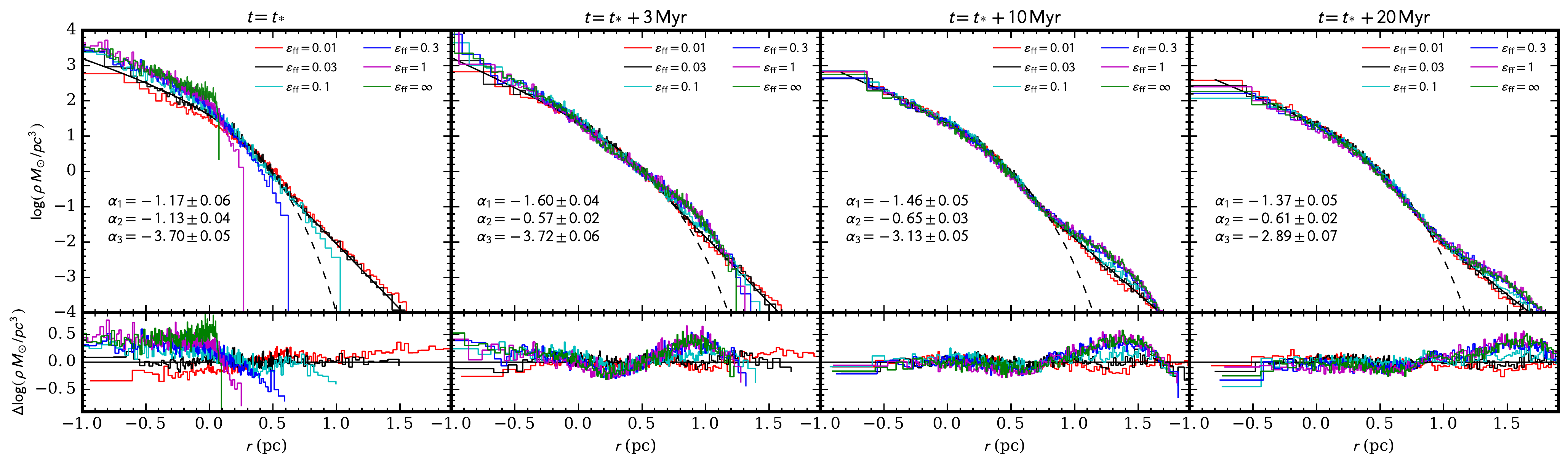}\\
         \includegraphics[width=\textwidth]{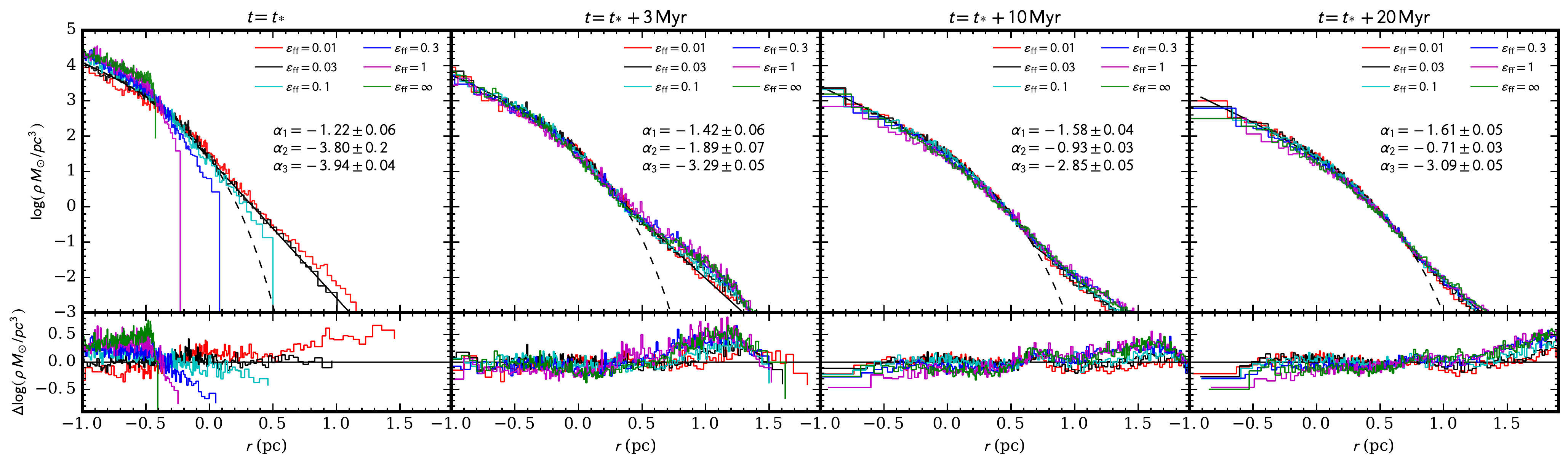}
        \end{array}$
  \caption{
Evolution of the average radial stellar density profile of the
clusters forming in the low (top panels) and high (bottom panels)
$\Sigma_{\rm cl}$ environments, with the different values of
$\epsilonff$ indicated in the legends. 
Evolution is given after the star formation ends, i.e., setting $t_{*}$
as the starting point.  Step functions show the measured profiles while
the solid black line shows a fitting function to the fiducial
$\epsilonff=0.03$ case that we use for comparisons.  The fitting
function is a two part function (see text) and we show the
continuation of the first part with a black dashed line. Auxiliary
bottom panels shows the residuals of every model with respect to the
$\epsilonff=0.03$ fitting function in order to better see the
differences between the models. Best fitting parameters are shown in
each panel.
}
        \label{fig:density}
\end{figure*}
\begin{figure*}
        $\begin{array}{c}
         \includegraphics[width=\textwidth]{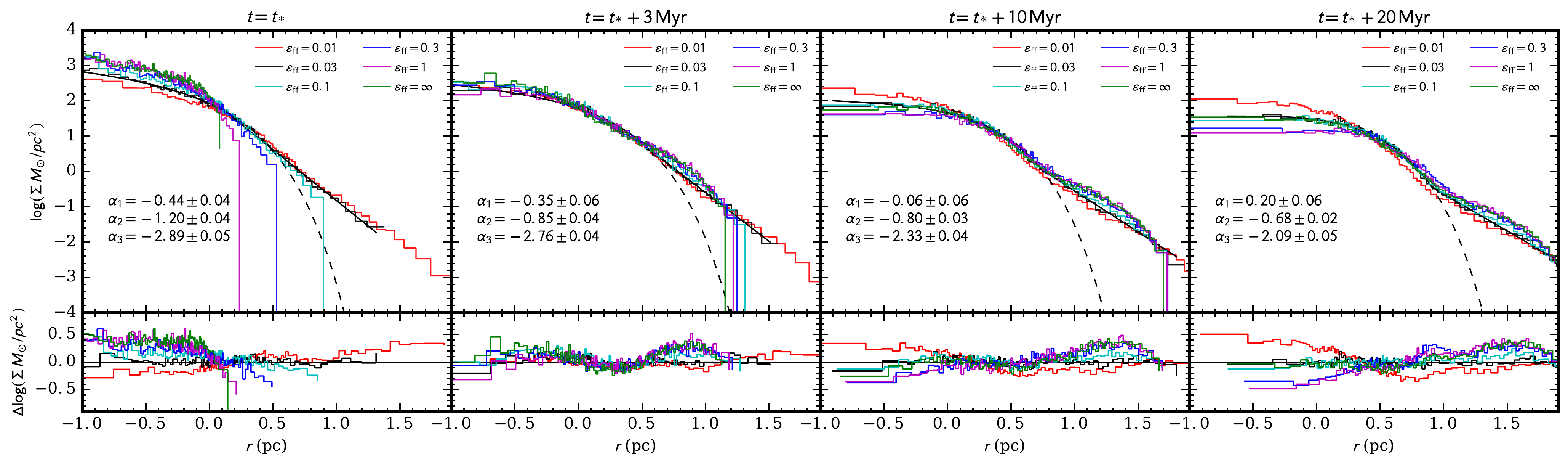}\\
         \includegraphics[width=\textwidth]{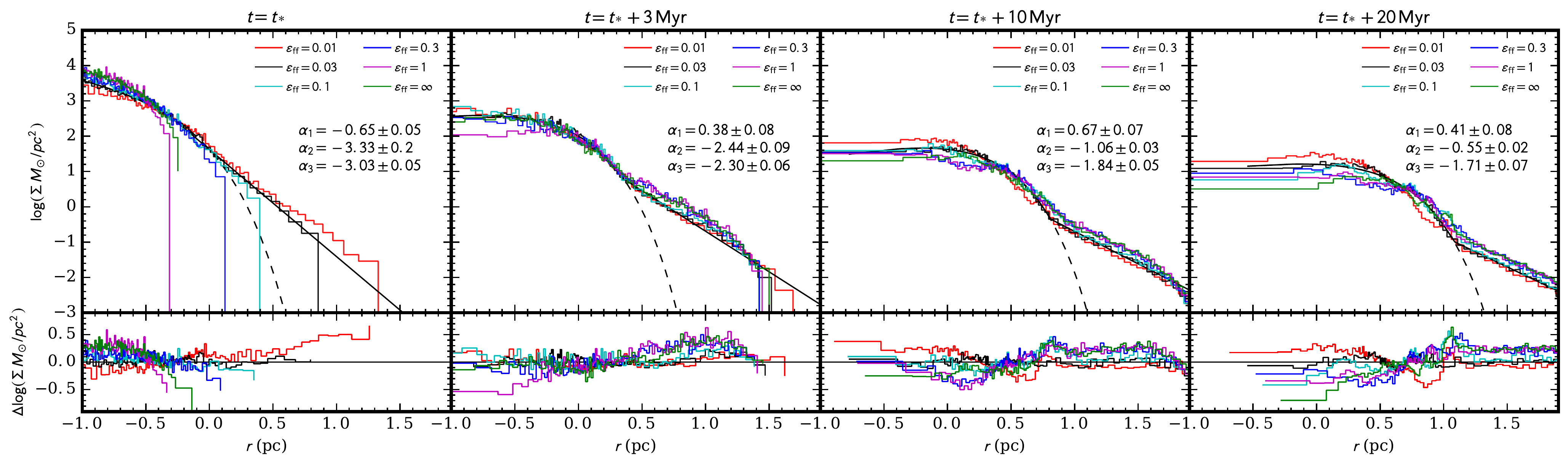}
        \end{array}$
        \caption{ Same as Figure~\ref{fig:density} but for the surface density profile.}
        \label{fig:surface_density}
\end{figure*}
\begin{figure*}
        $\begin{array}{c}
         \includegraphics[width=\textwidth]{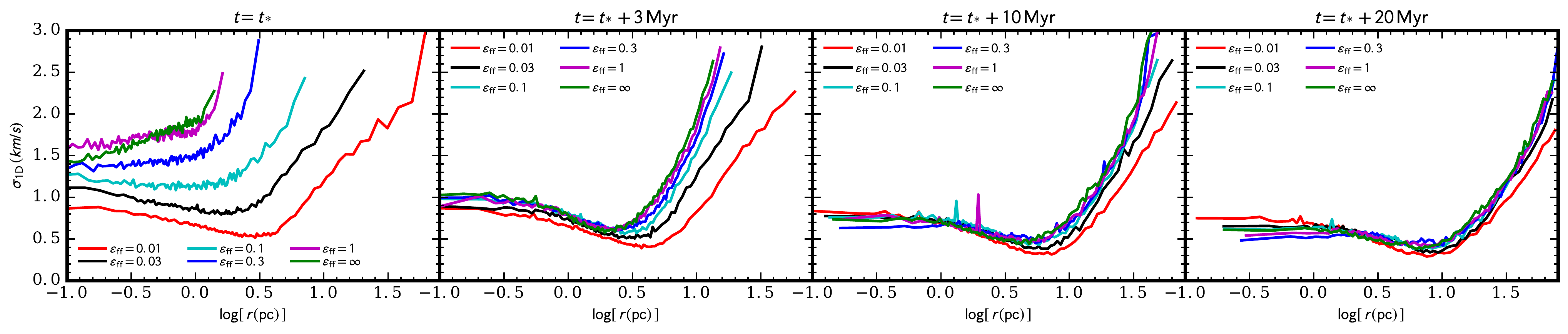}\\
         \includegraphics[width=\textwidth]{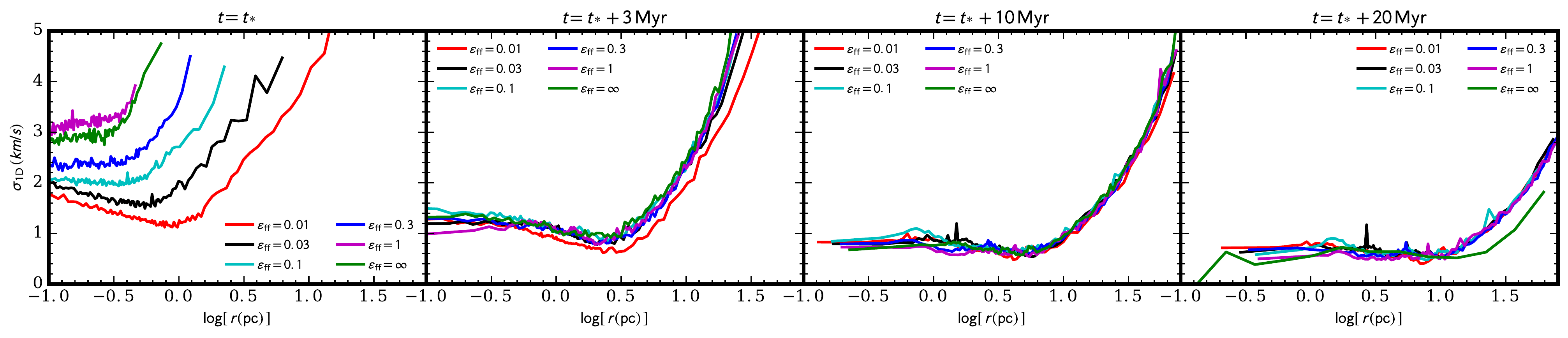}
        \end{array}$
        \caption{ 
Evolution of the one dimensional velocity dispersion profile after
star formation has ended for star clusters forming in the low (top
panels) and high (bottom panels) $\Sigma_{\rm cl}$ cases. The values
of the corresponding $\epsilonff$ are shown in the legends. Velocity
dispersions are measured treating stellar binaries as unresolved
systems, in order to avoid contamination from the high velocities
coming from their binary orbits.
        }
        \label{fig:velocity_dispersion}
\end{figure*}

The radial structure of star clusters, in their projected forms, are
one of the most direct observables of real star clusters. We have
first measured the average radial profiles for the 3D stellar mass
density (Figure~\ref{fig:density}) and then its 2D projection, i.e.,
the stellar mass surface density profile
(Figure~\ref{fig:surface_density}), for both high and low-$\Sigma_{\rm
  cl}$ cases and at different times after star formation ended, i.e.,
at $t - t_{\rm *}=$ 0, 3, 10 and 20 Myr.

In order to better see the differences in the density profiles when
varying $\epsilonff$ we compare the models using $\epsilonff=0.03$ as
a baseline case using a fitting function to describe it.  Density
profiles obtained here show a different behaviour in the inner zones,
dominated by the bound cluster, than in the outskirts, dominated by
the unbound population.  Then, the obtained profiles can be well
described by a two part function of the form: 
\begin{eqnarray}
       \label{eq:densityfit}
        f(r) &\propto& \left\{  
        \begin{array}{lr}
                r^{\alpha_1}\exp ( \alpha_2 r) &, r \leq r_{\rm crit} \\
                & \\
                r^{\alpha_3} &, r > r_{\rm crit}\\
        \end{array} 
        \right.
\end{eqnarray}
with $\alpha_1$, $\alpha_2$ and $\alpha_3$ fitting constants, $r$ in
units of parsecs and $r_{\rm crit}$ the critical radius at which the
$\alpha_3$ power law becomes a better description.  Bottom auxiliary
panels of Figures~\ref{fig:density} and~\ref{fig:surface_density}
show the residuals with respect to Eq.~\ref{eq:densityfit} fitted to
the $\epsilonff=0.03$ profiles.

One of the most obvious effects shown in these profiles is that just
after star formation has finished the clusters that took the longest
to form, i.e., low values of $\epsilonff$ have more extended
radial distributions. Note these include stars that were born unbound
from the clump and those that become dynamically ejected.

We showed in \citetalias{F17} that star clusters that form
instantaneously exhibit a density excess in a surrounding ``halo'',
which is caused by a significant fraction of stars that were born
unbound because of the initial high turbulent velocities.
Since all of these stars were lost at the same time, this excess is
quite prominent since they are at similar distances from the center.
We can see in Figures~\ref{fig:density} and \ref{fig:surface_density}
that as star cluster formation lasts longer, this excess becomes less
prominent. In all simulations presented here there are stars that are
born unbound from their natal regions, however, as they are born
gradually they also leave the system gradually being more spread out
in radial distance. They can also become mixed with stars that were
dynamically ejected in the way. Therefore, one consequence of the
models presented here is that star clusters forming slowly, should not
have ``bumps'' in their density profiles.

We also have measured the evolution of the 1D velocity dispersion
$\sigma_{\rm 1D}$ after star formation ended (see
Figure~\ref{fig:velocity_dispersion}). The first panel ($t=t_{*}$)
clearly shows the consequence of a longer star formation timescale
(low $\epsilonff$) on the dynamical stability of the newly
formed star cluster. At the end of the formation time, star cluster
that formed quickly have a velocity profile of the parent cloud, i.e.,
increasing with radius (see green lines on
Figure~\ref{fig:velocity_dispersion}), while star clusters forming
more slowly have a velocity dispersion profile that decreases with
radius (see red lines). This is valid only for the bound part of the
cluster and here any initial signature of the velocity dispersion
profile is quickly erased after gas is expelled and the differences
become very small. At large radii, where there are unbound stars, the
velocity dispersion rises again because faster stars are farther from
the cluster. Thus the velocity dispersion profile clearly shows the
boundary where the bound cluster ends and the unbound population
starts, i.e., where $\sigma$ has a minimum.

\subsection{Age gradients}
\label{sec:age}

\begin{figure*}
        $\begin{array}{c}
         \includegraphics[width=\textwidth]{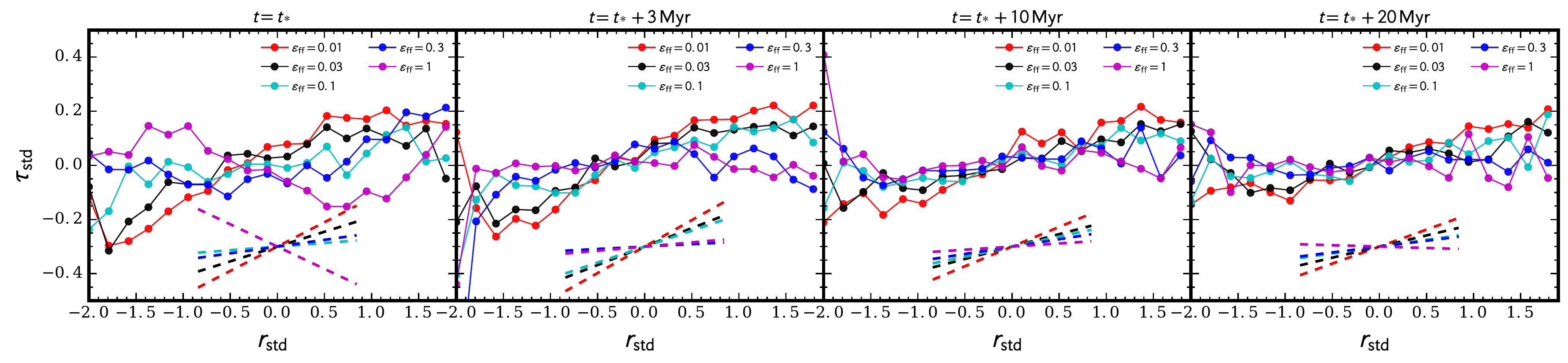}\\
         \includegraphics[width=\textwidth]{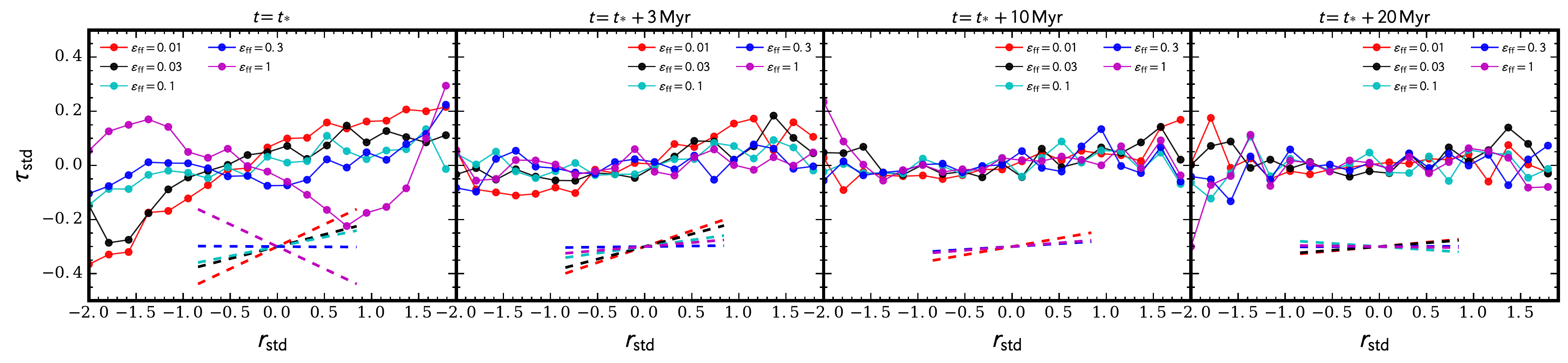}\\
        \begin{array}{rl}
         \includegraphics[width=0.4\textwidth]{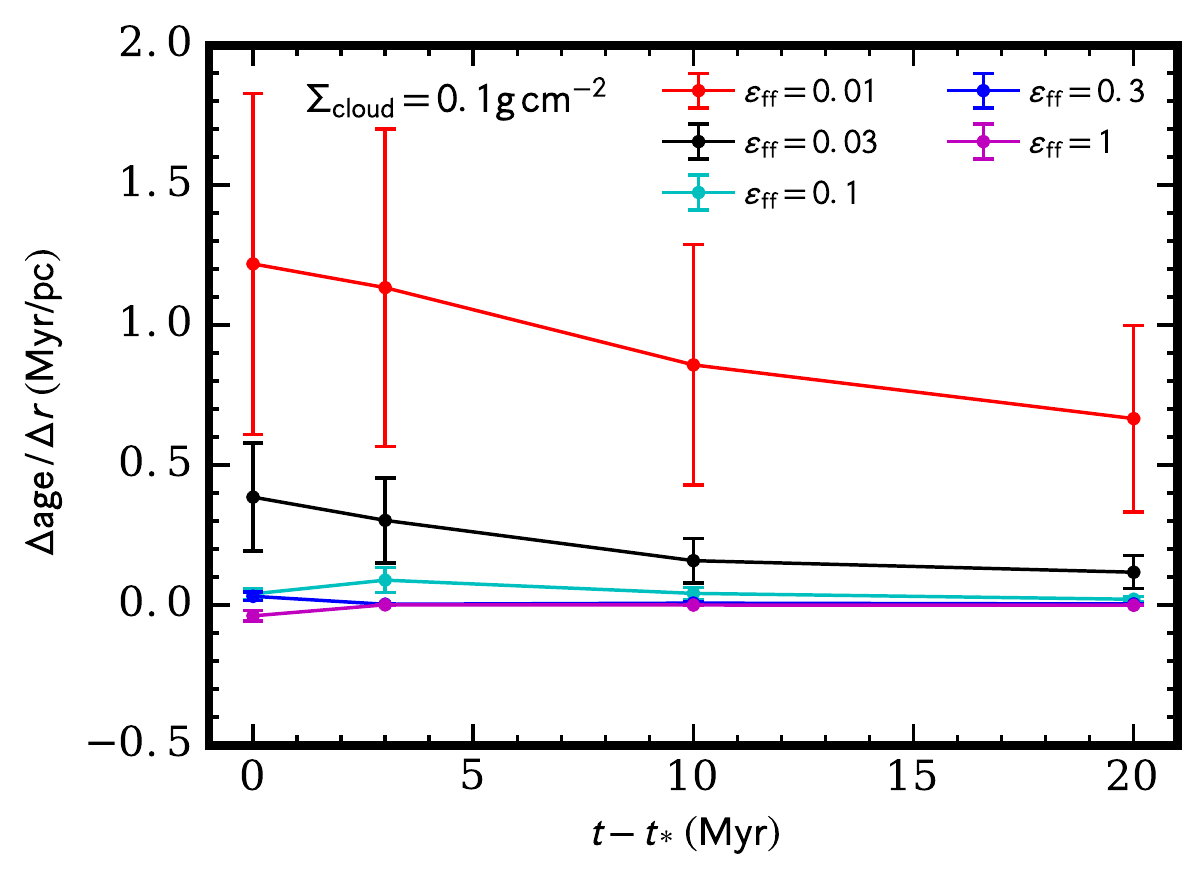}&
         \includegraphics[width=0.4\textwidth]{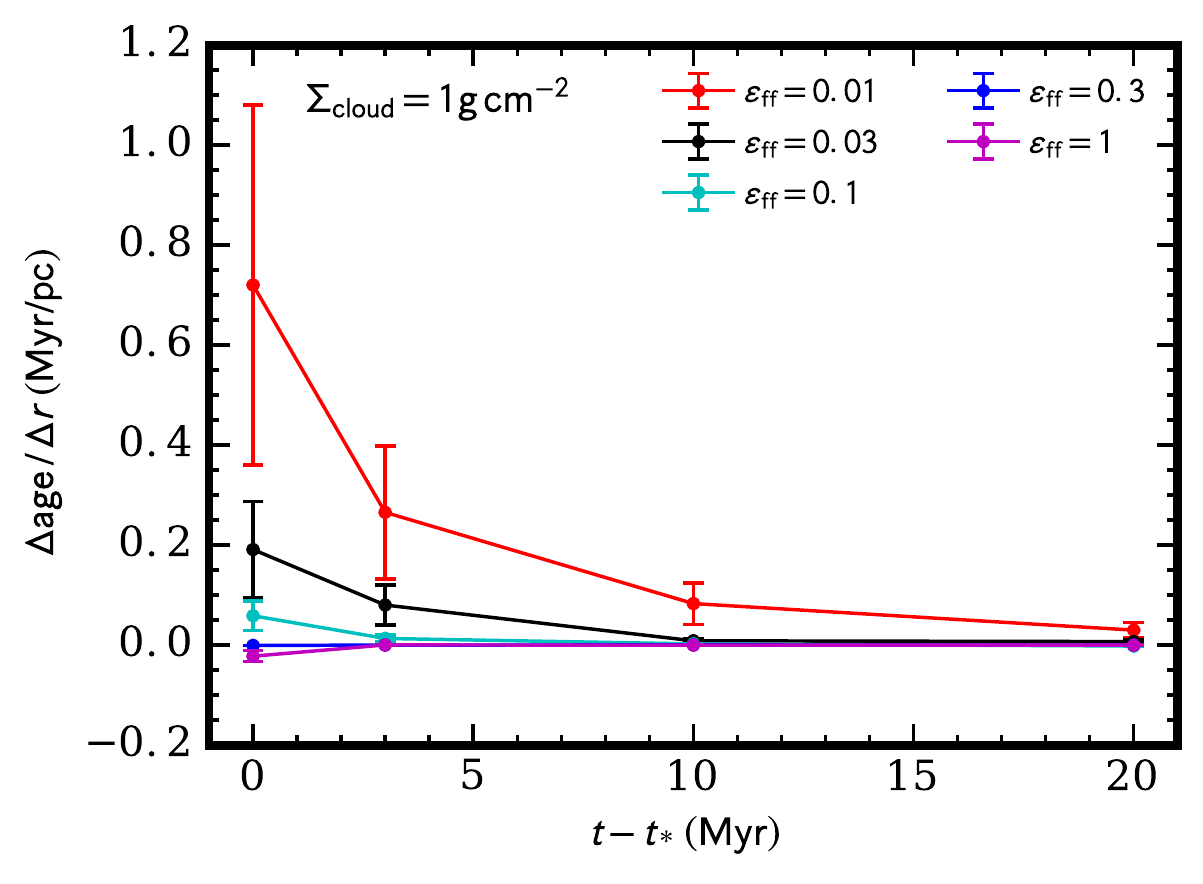}
        \end{array}\\
        \end{array}$
\caption{
Evolution of the average age radial gradients for the low (top panels)
and high (middle panels) $\Sigma_{\rm cl}$ cases. Ages and radial
distances are standarized, i.e., the mean value is subtracted and the
result is divided by the standard deviation. Only stars within
$3r_{\rm h}$ are considered in the analysis. A linear fit has been
performed in the ranges $-1 <r_{\rm std} < 1 $ and the resulting
gradients are shown below each curve.  Bottom panels show the
evolution of these gradients in physical units.  }
        \label{fig:ages}
\end{figure*}

One important feature of this work is that stars are born gradually
during the simulations. This means that stars have different ages with
an age spread equal to $t_{*}$. Stars are born uniformly in time and
also in space but following the power law density profile of the
initial clump gas. Therefore, in this work, we are able to look for
radial age gradients that may form dynamically. In order to be able to
compare with recent works on this matter, we have adopted the method
used by \cite{Getman2018} who measured and compared age gradients in
several young star clusters using standarized values for age and
distance to the center. A given value, e.g., of age ($\tau$), is
``standarized'' by subtracting the mean of the sample and dividing the
result by the standard deviation, i.e., for the $i$-th star:
\begin{eqnarray}
        \tau_{i,\rm std} &=&   \frac{\tau_{i} - \langle \tau \rangle}{\sigma_{\tau}} \\
        r_{i,\rm std} &=&  \frac{r_{i} - \langle r \rangle}{\sigma_{r}}.
\end{eqnarray}

In order to avoid biasing results by stars that are very far from the
cluster, e.g., because of dynamical ejections, that would increase the
mean value of distance and standard deviation, we only consider stars
within $3r_{\rm h}$ of the center of each cluster. This is also
motivated by the fact that such stars are very difficult to associate
observationally with a given cluster and would be normally not
included in such studies.

The resulting radial profiles of average age are shown in
Figure~\ref{fig:ages}. In order to show the different gradients we
perform linear fits in the standarized distances between -1 and 1. The
resulting gradients are shown below the curves for clarity. At the end
of star formation (left panels) there is a clear trend in which slower
forming clusters show a steeper radial gradient of mean age, i.e.,
with younger stars in the center and older stars in the outskirts.
For faster forming clusters (in fewer dynamical times) the gradient
becomes flatter and even turns around for the case of $\epsilon_{\rm
  ff}= 1$. The reason for these gradients is the expansion of the
cluster during the formation stage. While the star-forming region is
constrained to be within $R_{\rm cl}$ by the model assumptions, slow
forming clusters expand during the formation phase as they move
towards equilibrium, so older stars tend to be found in the outer
regions. Such effects will generally occur in real systems if the
star-forming region can be described as a localized dense gas clump.

As the cluster evolves without gas, this signature starts to decrease
as stars (in the bound cluster) are mixed again and the gradient
disappears. However, the gradients can remain strong for a
considerable period of time. In the low $\Sigma_{\rm cl}$ case it
is possible to still see the effects at 10 Myr, though for the high
$\Sigma_{\rm cl}$ case, the signature disappear sooner because of the
faster dynamical evolution implied by the high density.

The recent work of \cite{Getman2018} has concluded that age gradients
are generally present in young star clusters and the majority
($\sim80\%$) of these gradients are positive, i.e., younger stars in
the core compared to the outskirts \citep[see also][]{Getman2014}.  In
Figure~\ref{fig:ages} we show that slow forming star clusters tend to
form more positive and steeper age gradients relative to fast forming
clusters.  In terms of physical units, \citep{Getman2018} obtained
gradients between 0.75-1.5 Myr pc$^{-1}$ depending on the model used
to estimate stellar ages.  In this work, we have been able to reach
such values only with slow forming star clusters ($\epsilonff<0.03$,
as can be seen in the bottom panels of Figure~\ref{fig:ages}.  When
transforming to physical units the difference between the models are
amplified, given that slow forming clusters have larger age spreads by
construction, and therefore can naturally form greater gradients.
However, we note that the fundamental difference also appears in
standarized units and use of physical units only amplifies the
differences.  Then, high $\Sigmacl$ cases show smaller initial age
gradients because $t_{*}$ is smaller. However, the rapid dynamical
evolution of these systems also remove the gradients more quickly.

\subsection{Ejected stars and kinematic structure}
\label{sec:unbound}

\begin{figure*}
        $\begin{array}{c}
                \includegraphics[width=\textwidth]{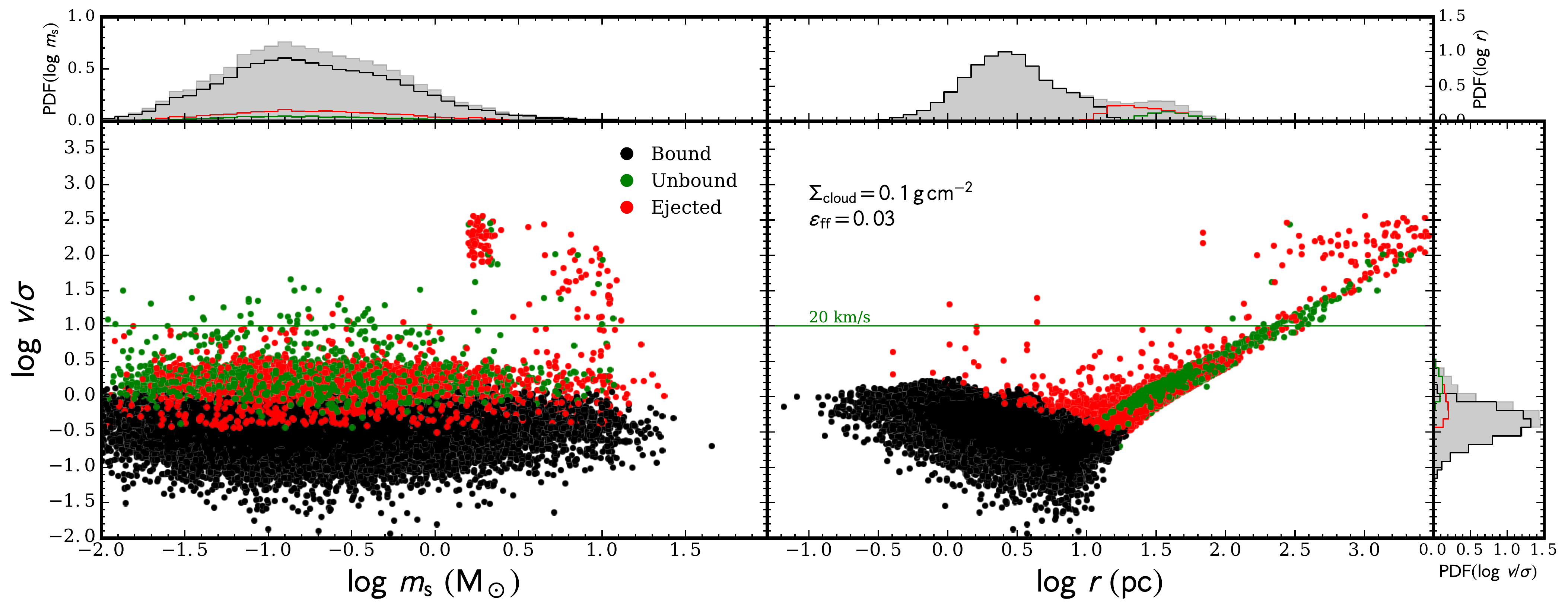}\\
                \includegraphics[width=\textwidth]{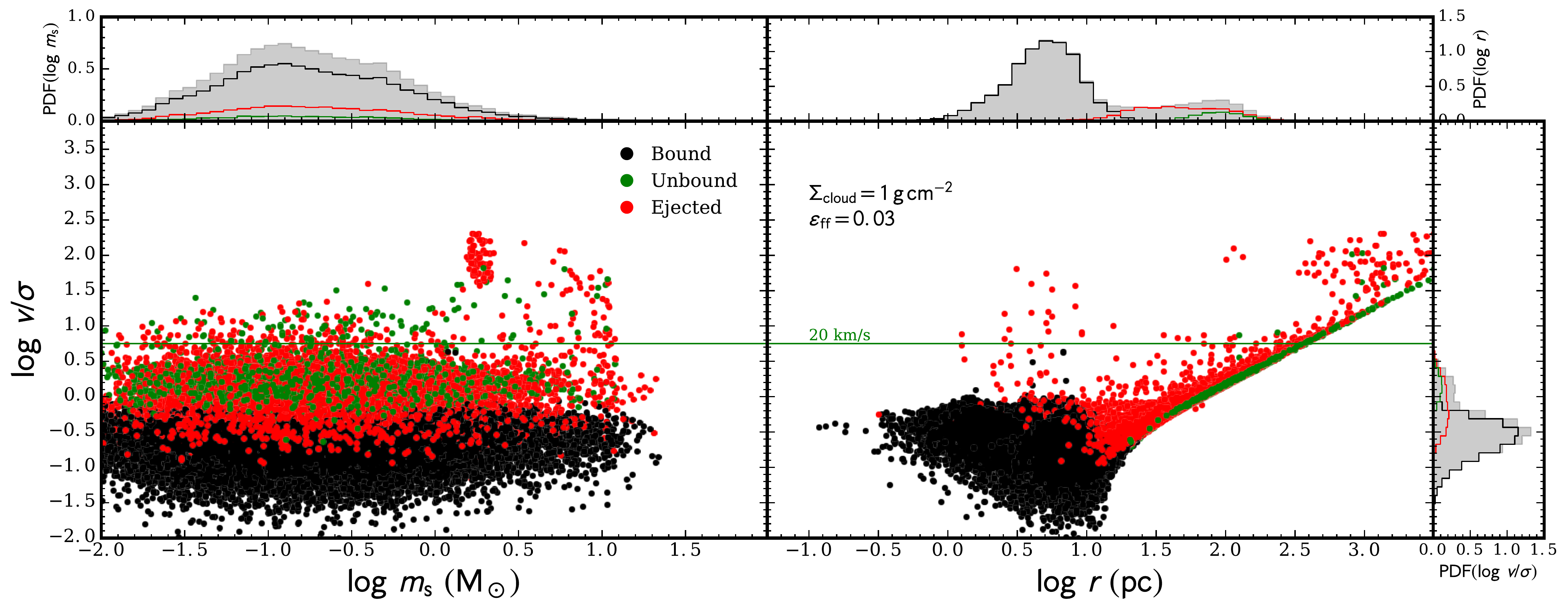}
        \end{array}$
  \caption{
Normalized velocity versus stellar mass (left panels) and versus
distance from cluster center (right panels) for all stars in the
fiducial case with $\epsilonff=0.03$ for simulations with
$\Sigmacl=0.1\:{\rm g\:cm}^{-2}$ (top) and $\Sigmacl=1\:{\rm
  g\:cm}^{-2}$ (bottom), measured at 20 Myr. Stars are separated in
three groups: bound stars (black); stars born unbound (green); and
ejected stars (red).  Velocity values are normalized by the
mass-averaged velocity dispersion of the parent clump (see
\ref{tab:ic}). Small top and side panels show the PDFs considering all
stars in the set (gray shaded area), and the fraction of the PDFs that
correspond to each group of stars (lines). An example velocity
threshold of 20~km/s is shown by a green line.
}
        \label{fig:mv}
\end{figure*}

Following our analysis of \citetalias{F17}, we identify bound members
in the simulations snapshot by snapshot, classifying stars in three
groups: (1) bound stars, which are the ones that remain bound until
the end of the simulation; (2) unbound stars, which were unbound from
the first snapshot that they appear, i.e., they are born
unbound\footnote{However, note that we are constrained by the time
  resolution of the simulations. This will cause us to classify as
  unbound some stars that were dynamically ejected just after being
  bound and before the next output time, however we expect that the
  number of such cases is very small.}; (3) Ejected stars, which are
stars that appear bound in one snapshot but later they do not. With
the ejected stars, we also subdivide them in: (A) Strong dynamical
ejections, i.e., stars that show $\Delta T_i \ge 2\Delta \Omega_i$,
when ejected; (B) Supernovae ejections, including stars receiving
velocity kicks during their supernovae phase, and stars ejected
because they were binary companions of a star that exploded as
supernovae; and (C) gentle ejections, stars that become unbound
because of the evolving potential of the cluster, which we distinguish
by $\Delta T_i < 2\Delta \Omega_i$. In this work, we refer to runaway
stars, as dynamically ejected stars with velocities above 20~km/s, but
note that this is an arbitrary threshold for such a definition.

Following the format of \citetalias{F17}, we show the structure formed
by these three groups of stars in the velocity-mass and
velocity-distance diagrams in Figure \ref{fig:mv}, all measured at
20~Myr. Velocities are normalized by $\sigma_{\rm cl}$ of the parent
clump (see~\ref{tab:clumps} and the 20~km/s threshold is marked as a
green horizontal line for reference. The corresponding PDFs are shown
in the respective side panels.  The structures formed in the
velocity-distance diagrams are similar to those in \citetalias{F17},
however the number of stars that born unbound (green) has decreased
considerably and these are a smaller fraction of the corresponding
final PDFs. The population of ejected stars (red) is stronger in these
models, however they are now spread over a wider range of distances
and velocities. This is a consequence of the large and slow decrease
on the escape velocity of the system and the spread in the formation
times of the stars.

\begin{figure*}
       \centering
       $\begin{array}{rl}
       \includegraphics[width=0.5\textwidth]{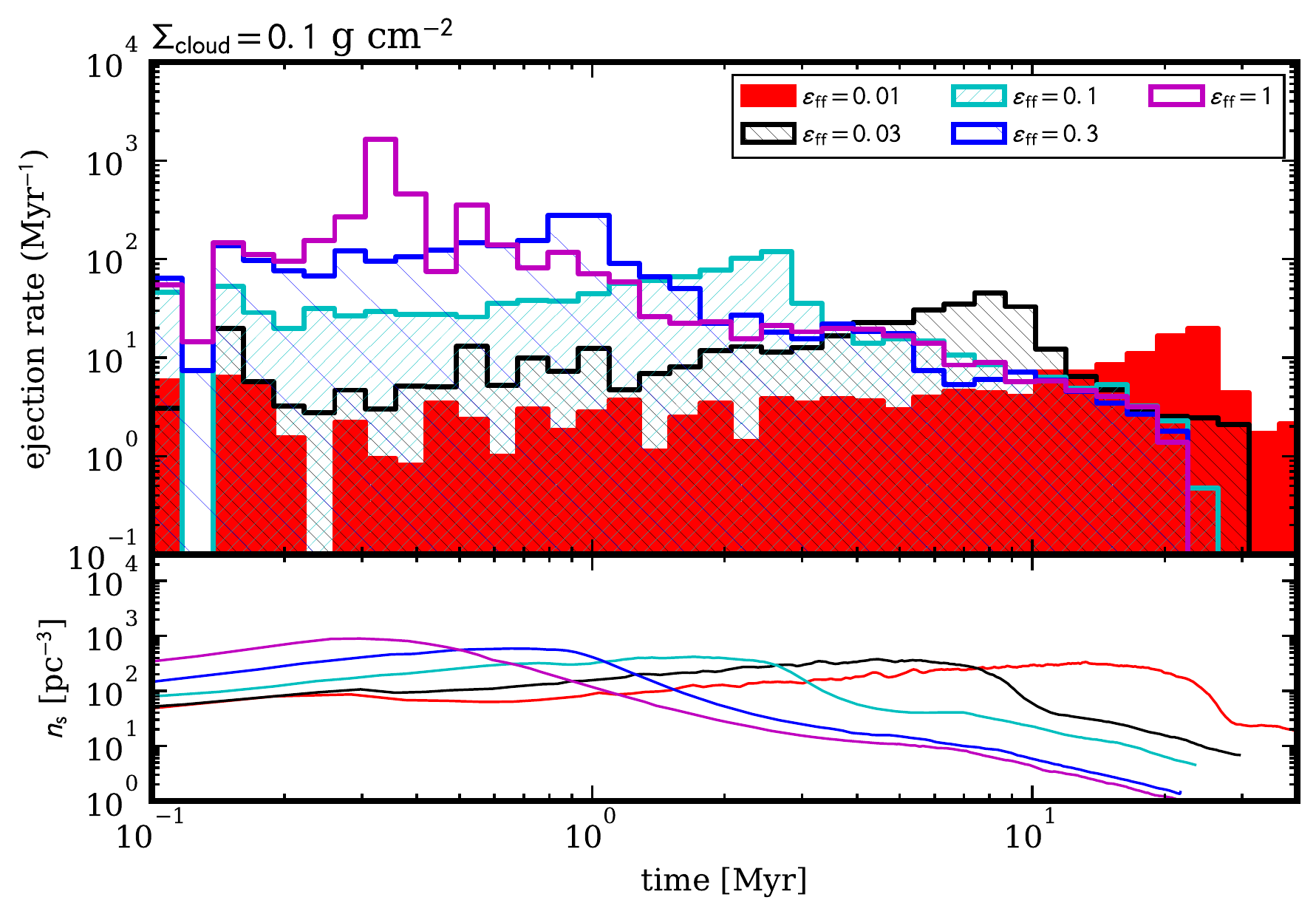}&
       \includegraphics[width=0.5\textwidth]{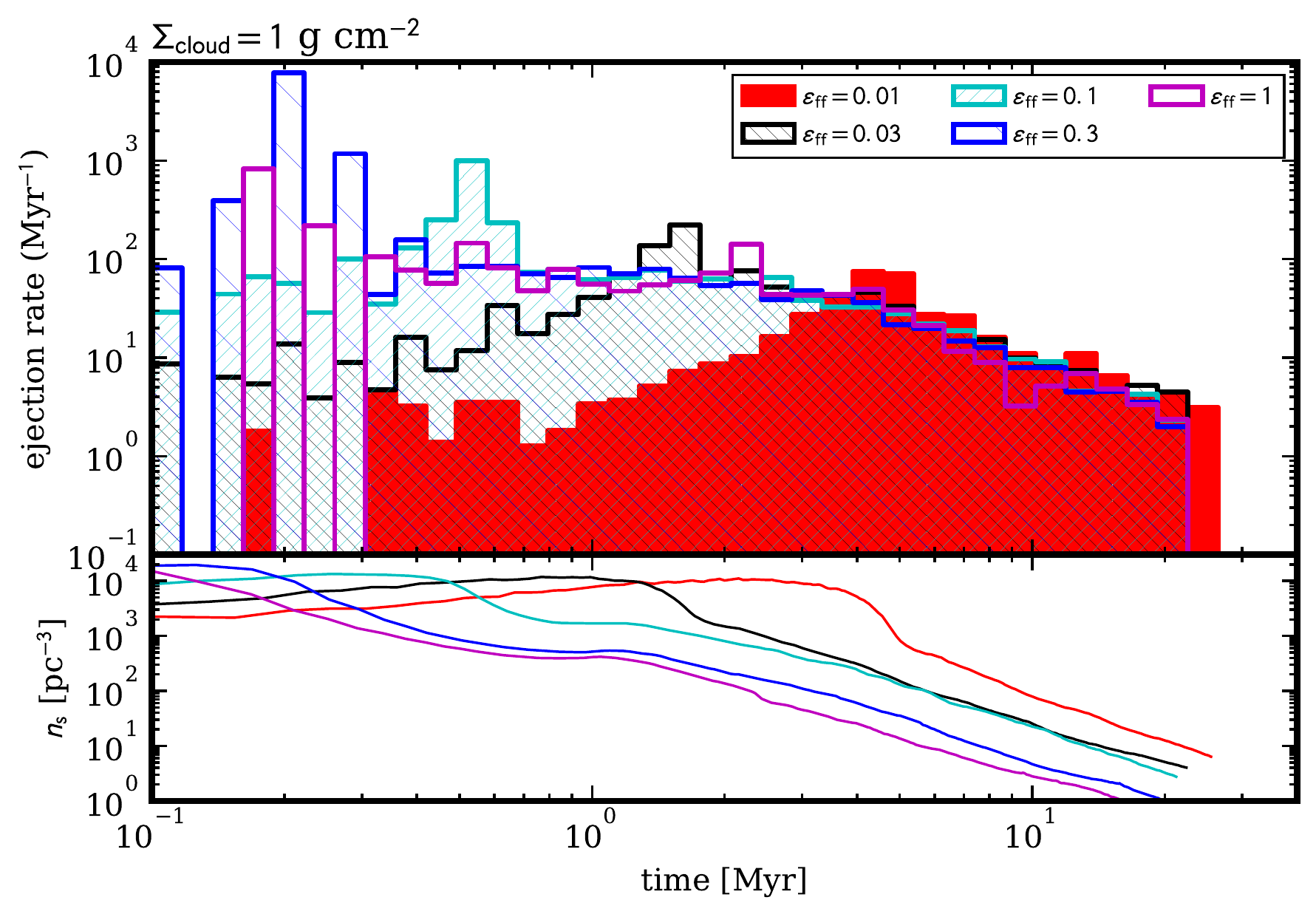}
        \end{array}$
       \caption{
Average dynamical ejection rates for star clusters in the \fiducial
set with different values of $\epsilonff$. As a reference of the
cluster structure, bottom panels show the time evolution of the number
density inside the half mass radius of the cluster.}
       \label{fig:rwevol}
\end{figure*}

We also recorded the time when strong ejections happen and we measured
the dynamical ejection rate, shown in Figure~\ref{fig:rwevol} for the
low (left) and high (right) $\Sigmacl$ cases. During the formation of
the clusters, the ejection rate remains higher and is closely
correlated to the number densities (shown in the bottom panels). As
stars are formed, the central number densities grow, along with the
ejection rates.  The peak density in the different models, which is
reached at the end of the formation stage, is smaller at low
$\epsilonff$.  As we saw in \S\ref{sec:evol}, star clusters expand
more during formation when $\epsilonff$ is small, and therefore the
peak number density is also smaller than in the high $\epsilonff$
regime. This difference causes the peak in ejection rate to be smaller
at low $\epsilonff$, but significantly broader. After gas is gone and
the cluster expands, ejection rates fall
at a similar rate in all models.

\begin{figure*}
        \includegraphics[width=\textwidth]{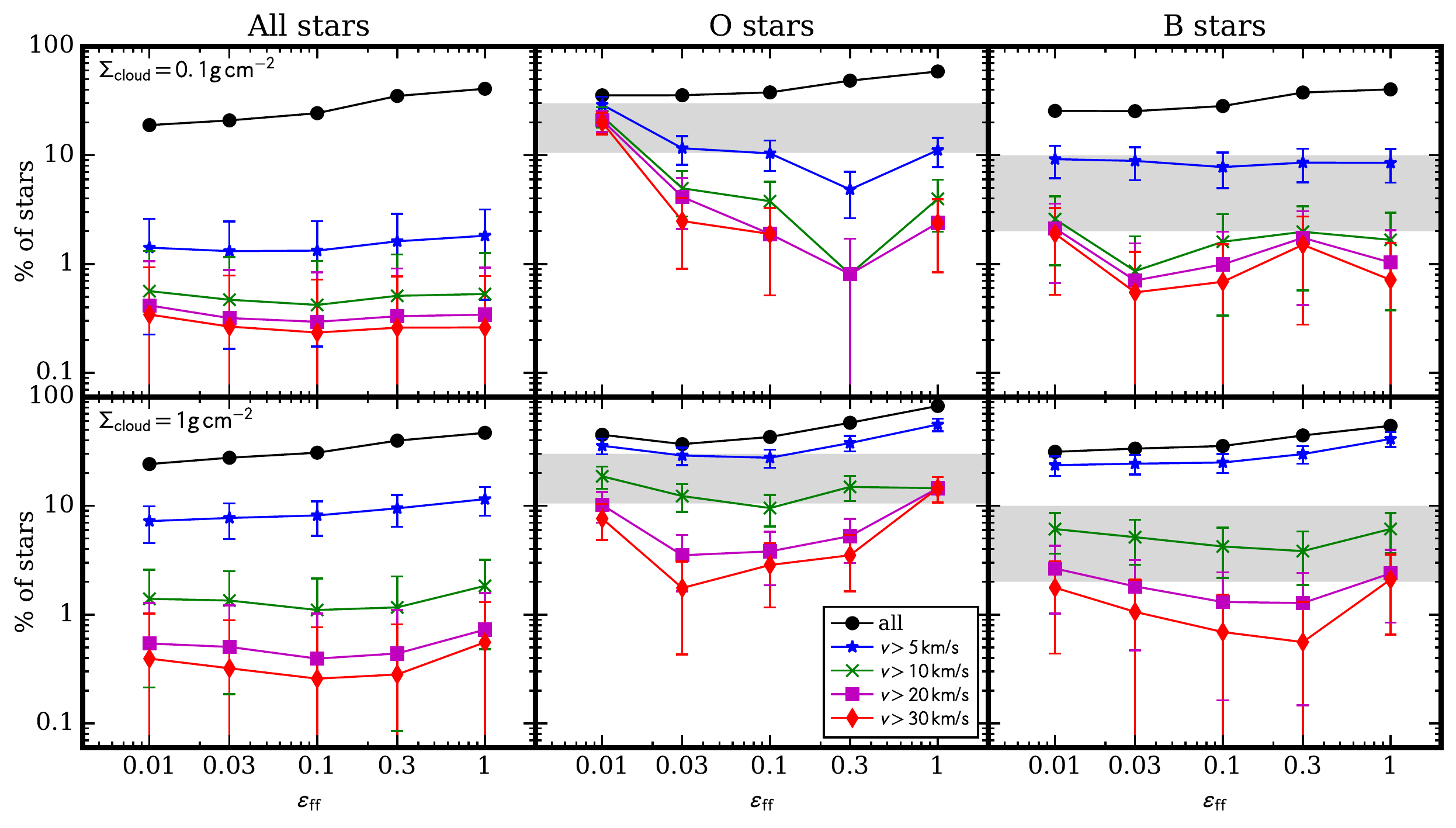}
        \caption{
Percentage of ejected stars relative to all stars in each mass range
for results for the low (top panels) and high (bottom panels)
$\Sigmacl$ cases.  The first column shows the results when
using all stars in the set, center panels when considering only O
stars and left panels for B stars.  Different colors shows different
velocity thresholds to define ``ejected'' and gray shaded areas the
range of values found in the literature for runaway stars under
various definitions (see text).
}
        \label{fig:runaways}
\end{figure*}

\begin{figure*}
       \centering
       \includegraphics[width=\textwidth]{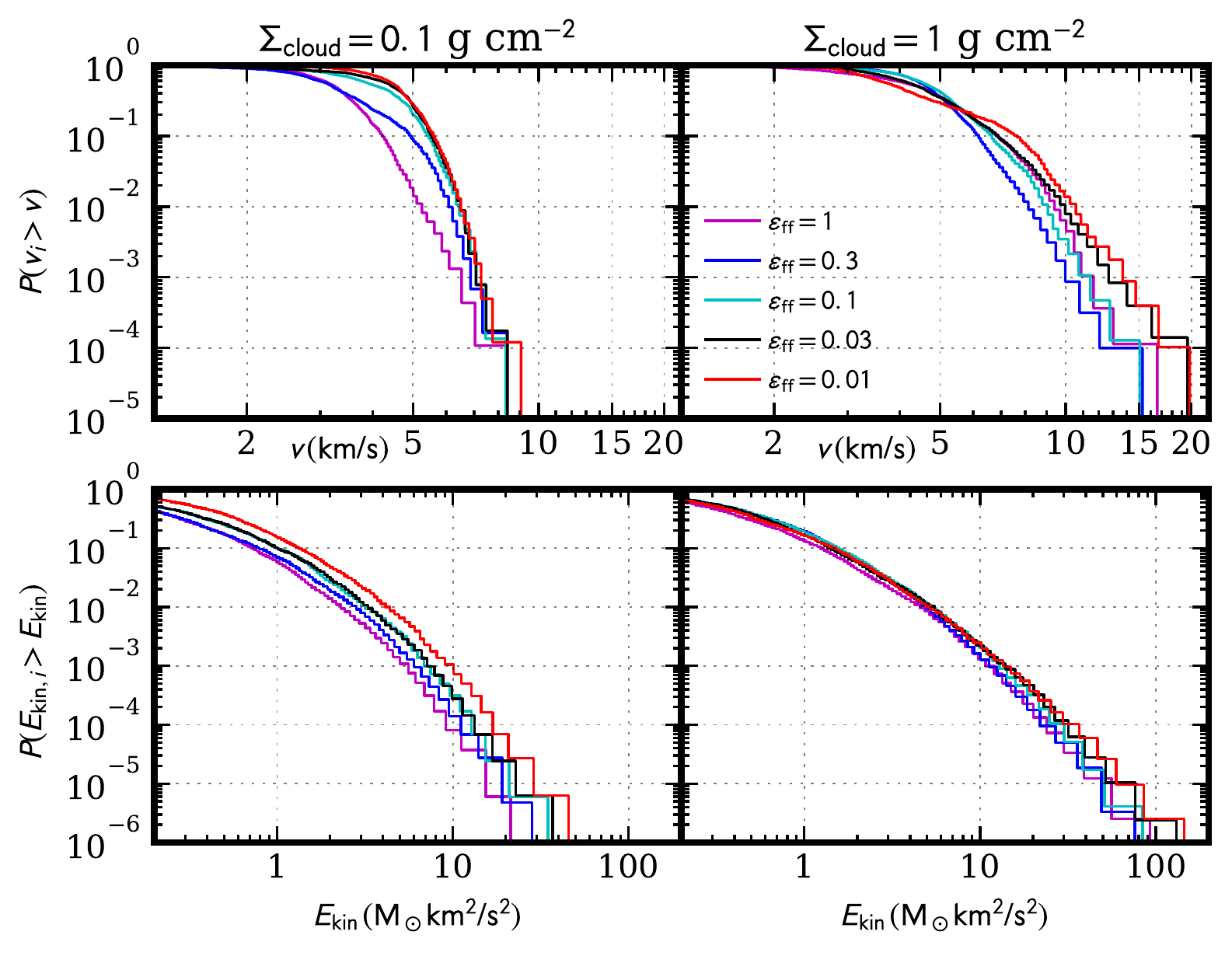}
       \caption{
Velocity (top panels) and kinetic energy (bottom panels) distributions
of ejected stars, shown as the probability of finding an ejected star
with velocity or energy above a given value.  }
       \label{fig:kenergy}
\end{figure*}

Figure~\ref{fig:runaways} (left column) shows the collected percentage of
ejected stars in each simulation set, including for various threshold
velocities that can be used to define ``runaways''. Then the middle
and right columns show the same results, but for O and B stars,
respectively. Note, these are only dynamically ejected stars via
$3+$-body interactions and not those ejected due to binary supernova
explosions. Filled black circles show the total percentage of
dynamical ejections, especially including ``gentle'' ejections, from
the 20 simulations of each set. The percentage of ejected stars
increases modestly as $\epsilonff$ increases.
This is because in the faster formation regime, the escape velocity
falls faster, and there are more gentle ejections because stars do not
have time to relax to a state of dynamical equilibrium.

When considering ejections at speeds $>5\:$km/s the effect of mass
surface density is evident, since in the high $\Sigmacl$ case the
velocity dispersion and escape speed of the initial clump is quite
close to this threshold. In the low mass surface density case there
are greater fractions of runaway O stars in the slowest formation
limit, which we attribute to there being more time for mass
segregation and the cluster remaining in a denser state for longer. In
the low $\Sigmacl$ regime, these trends continue as we consider larger
velocity thresholds to define the runaway population. In the high
$\Sigmacl$ these trends are weaker and appear to be roughly constant
with $\epsilonff$.

However, we note that even though the fraction of massive runaway
stars appears to be relatively constant with $\epsilonff$, the number
of ejections are larger in the fast formation regime. Therefore, the
fraction of massive runaway stars is higher at low $\epsilonff$ if we
only consider ejected stars. This indicates that dynamical ejections
in slowly forming star clusters are more energetic than in the fast
formation regime, but less numerous.

Indeed, we can see that this is the case if we examine the kinetic
energy ($E_{\rm kin}$) distribution of the ejected population. Figure
~\ref{fig:kenergy} shows the cumulative distributions of velocities
(top panels) and kinetic energies (bottom panels) for ejected stars,
for the low and high mass surface density cases. More precisely, it
shows the probability of finding an ejected star having a velocity or
kinetic energy higher than a given value. If we look at the kinetic
energy distributions, we see that in the low-$\Sigmacl$ case,
probabilities of finding high energy ejected stars in the $\epsilonff
= 0.01$ case are around one one order of magnitude higher than in the
fast $\epsilonff=1$ scenario. This trend is similar in the
high-$\Sigmacl$ case, however differences are smaller given that
higher densities are a more important factor when increasing the
energetics of interactions, not only because interaction rates are
higher, but also because binaries involved in these interactions are
harder than in the low sigma case, and therefore they can provide more
energy. In this sense, density is always a first order factor
increasing numbers and energies of runaway stars. Star cluster
formation timescales enter as a second order factor because
interaction rates have more time to act and the more energetic
ejections are a consequence of having more time for less likely
interactions to happen, i.e., the more energetic ones.

Observations of runaways have inferred that about 10-30\% of O stars
and 2-10\% of B stars are observed to be runaways
\citep{Gies1987,Stone1991,DeWit2005}, ranges marked as gray areas on
Figure~\ref{fig:runaways}. These ranges are somewhat uncertain given
the different definitions on the literature and completeness
\citep[see][]{Eldridge2011}. At this point, most of our results appear
to be below the observational limits, except at low $\epsilon_{\rm
  ff}$. However, this result is highly dependent on the velocity
cutoff that we choose, as well as those used to define the
observational samples.  These results suggest that the timescale of
star formation could play an important role at reproducing observed
numbers of runaways. However, it may still be important to include
more realistic assumptions in future work, potentially including
primordial substructure and various degrees of primordial mass
segregation.

\section{Discussion}
\label{sec:discussion}

We have presented the second step in the modeling of star clusters
born from the turbulent clump model of \cite{mt03}. In this paper, we
have relaxed the assumption of instantaneous star formation used in
\citetalias{F17} and in most previous $N$-body studies, and we have
focused on exploring the effects of the star cluster formation
timescale on the evolution of the clusters, parameterized by the star
formation efficiency per free fall time
\citep[$\epsilonff$,][]{Krumholz2005}.  By exploring this parameter,
we have studied the dynamics of the formation of star clusters
including the gradual formation of stars. This stage has previously
only been modeled by means of (magneto-)hydrodynamical simulations of
star formation, typically using sink particles that have limited
ability to follow accurate binary orbits. We note that the exception
is the $N$-body study of \cite{Proszkow2009}. However, this focused on
relatively low densities and without including binaries or stellar
evolution.

Thus, our work is the first attempt to study comprehensively the
dynamics of the stars during star cluster formation.  Several
simplifying assumptions have been made in our study. We have used an
uniform and fixed star formation rate, defined by the initial
properties of the clump.  However, as the density of the clump
decreases, the free fall time does as well and the formation timescale
would then be even longer than presented in this work. Also, given
that the free fall time is density dependent, the central area of the
clump will have shorter physical timescales and star formation may
happen faster in the central (denser) region of the clump rather than
uniformly \citep[see, e.g.,][]{Parmentier2013}.  However,
observationally is very difficult to estimate if this effect is
actually occuring \citep[see, e.g.,][]{Dario2014}, and there are many
other potential complications, e.g., local variations of
feedback. Thus in this first study with gradual star formation we have
preferred to investigate the constant star formation rate case and
defer more complex star formation histories to future studies.

In this study, the gaseous clump was modeled in a simplistic way, only
decreasing due to the mass of the stars formed and gas ejected. The
clump, however, should evolve in much more complex ways as stars are
formed.  We have assumed that any residual gas is removed instantly as
stars form, however this will not be the case in real star-forming
clumps.  Protostellar outflows and other feedback not only will drive
the turbulence that potentially can support the cloud, but also will
be the agent that removes some gas from the clump.

Several authors have shown that slowly removing gas results in
clusters that survive gas expulsion better than rapid gas removal
\citep[e.g.,][]{Baumgardt2007,Smith2013b}. Our work also finds this
basic trend in that our models with gradual star formation have higher
bound fractions than the instantaneous cases of Paper I. However, we
see little variation in the bound fraction as the star formation rate
decreases, with these values all being $>0.8$. A broader range of
initial conditions and effects of some basic model assumptions, such
as keeping a fixed velocity disperison of new born stars and the
detailed treatment of gas explusion, need to be explored for their
effects on the bound fraction.

An important property that we must include when advancing the realism
of our modeling is the inclusion of substructure, since young stars in
clusters exhibit substructure \citep[e.g.][]{Gutermuth2008,Dario2014}
and it has been shown that substructure influences several important
dynamical processes such as mass segregation
\citep{Allison2009b,Dominguez2018} and pre-gas expulsion dynamical
equilibrium \citep[e.g.][]{Farias2015}.  Still, our approach has been
to first present the case without substructure (except for a radial
gradient of initial clump density), and we defer the exploration of
effects of sub-structure to a future paper in this series.

\section{Conclusions}
\label{sec:conclusions}

In this modeling of star cluster formation with a pure stellar
dynamical study, we have made simple assumptions for the formation of
the stars and then explored the effects of different rates of star
formation to form a cluster. We summarize our results as follows:
\begin{enumerate}
        \item The critical difference between models with different $\epsilonff$
                is that star clusters forming slowly ($\epsilonff\lesssim0.1$) have
                several dynamical timescales to reach equilibrium before the gas is
                exhausted/ejected. Thus they start their gas-free evolution in
                a more stable configuration.

        \item High values of $\epsilonff$ result in star clusters that expand much
                faster than the ones with low $\epsilonff$, which is a consequence
                of the first result.  The expansion rate, however, depends strongly on
                the initial density of the parent clump, and to a second degree on the
                level of initial mass segregation and multiplicity.

        \item The various dynamical states obtained through the
          different formation timescales, are not reflected in the
          bound fractions that star cluster have after formation, with
          these all being relatively high, i.e., $f_{\rm
            bound}\simeq0.8$ to 0.9. In this aspect, the treatment of
          the expulsion of residual gas by star formation feedback,
          along with assumptions of the initial velocity dispersion
          that stars are born with, may be key features of the
          modeling that need to be explored further to assess the
          reliability of this result.

        \item The level of mass segregation developed during the
          formation of the cluster, is highly dependent on
          $\epsilonff$, with slow-forming clusters having higher
          levels of mass segregation at the end of their formation.
          However, many crossing times are needed to develop high
          levels of mass segregation in these spherical models, in
          agreement with previous studies.

        \item Star clusters that form slowly tend to produce more
          energetic dynamical ejections than fast forming star
          clusters, a consequence of the longer times that the cluster
          can remain in a dense state. However, $\Sigmacl$ is the main
          parameter that rules the energetics of dynamical ejections
          with $\epsilonff$ as a second order factor. We have shown
          that slow-forming star clusters produce more runaway stars
          than fast-forming ones under the same density environments,
          especially in the low-$\Sigmacl$ scenario.

        \item We have found that gradual formation of stars naturally
          causes age gradients within star clusters as the stellar
          component expands in order to reach equilibrium. The
          steepness of the age gradient is highly dependent on
          $\epsilonff$ and the difference between models remains
          visible for several Myr. In the framework of this work, we
          were able to reproduce recent observational values from
          \cite{Getman2018} only if $\epsilonff<0.03$.

\end{enumerate}

\section*{Acknowledgements}

The authors would like to than Sverre Aarseth  for useful advice at the beginning of this
project and for making \texttt{Nbody6} publicly available. We also thank Maxwell Cai for
useful discussions.




\bibliographystyle{mnras}
\bibliography{bibfile} 

\begin{thebibliography}{}
\makeatletter
\relax
\def\mn@urlcharsother{\let\do\@makeother \do\$\do\&\do\#\do\^\do\_\do\%\do\~}
\def\mn@doi{\begingroup\mn@urlcharsother \@ifnextchar [ {\mn@doi@}
  {\mn@doi@[]}}
\def\mn@doi@[#1]#2{\def\@tempa{#1}\ifx\@tempa\@empty \href
  {http://dx.doi.org/#2} {doi:#2}\else \href {http://dx.doi.org/#2} {#1}\fi
  \endgroup}
\def\mn@eprint#1#2{\mn@eprint@#1:#2::\@nil}
\def\mn@eprint@arXiv#1{\href {http://arxiv.org/abs/#1} {{\tt arXiv:#1}}}
\def\mn@eprint@dblp#1{\href {http://dblp.uni-trier.de/rec/bibtex/#1.xml}
  {dblp:#1}}
\def\mn@eprint@#1:#2:#3:#4\@nil{\def\@tempa {#1}\def\@tempb {#2}\def\@tempc
  {#3}\ifx \@tempc \@empty \let \@tempc \@tempb \let \@tempb \@tempa \fi \ifx
  \@tempb \@empty \def\@tempb {arXiv}\fi \@ifundefined
  {mn@eprint@\@tempb}{\@tempb:\@tempc}{\expandafter \expandafter \csname
  mn@eprint@\@tempb\endcsname \expandafter{\@tempc}}}

\bibitem[\protect\citeauthoryear{{Aarseth}}{{Aarseth}}{2003}]{Aarseth2003}
{Aarseth} S.~J.,  2003, {Gravitational N-Body Simulations}

\bibitem[\protect\citeauthoryear{{Allison}, {Goodwin}, {Parker}, {Portegies
  Zwart}, {de Grijs}  \& {Kouwenhoven}}{{Allison} et~al.}{2009a}]{Allison2009}
{Allison} R.~J.,  {Goodwin} S.~P.,  {Parker} R.~J.,  {Portegies Zwart} S.~F.,
  {de Grijs} R.,   {Kouwenhoven} M.~B.~N.,  2009a, \mn@doi [\mnras]
  {10.1111/j.1365-2966.2009.14508.x}, \href
  {http://adsabs.harvard.edu/abs/2009MNRAS.395.1449A} {395, 1449}

\bibitem[\protect\citeauthoryear{{Allison}, {Goodwin}, {Parker}, {de Grijs},
  {Portegies Zwart}  \& {Kouwenhoven}}{{Allison} et~al.}{2009b}]{Allison2009b}
{Allison} R.~J.,  {Goodwin} S.~P.,  {Parker} R.~J.,  {de Grijs} R.,  {Portegies
  Zwart} S.~F.,   {Kouwenhoven} M.~B.~N.,  2009b, \mn@doi [\apjl]
  {10.1088/0004-637X/700/2/L99}, \href
  {http://adsabs.harvard.edu/abs/2009ApJ...700L..99A} {700, L99}

\bibitem[\protect\citeauthoryear{{Bastian} \& {Goodwin}}{{Bastian} \&
  {Goodwin}}{2006}]{Bastian2006}
{Bastian} N.,  {Goodwin} S.~P.,  2006, \mn@doi [\mnras]
  {10.1111/j.1745-3933.2006.00162.x}, \href
  {http://adsabs.harvard.edu/abs/2006MNRAS.369L...9B} {369, L9}

\bibitem[\protect\citeauthoryear{{Baumgardt} \& {Kroupa}}{{Baumgardt} \&
  {Kroupa}}{2007}]{Baumgardt2007}
{Baumgardt} H.,  {Kroupa} P.,  2007, \mn@doi [\mnras]
  {10.1111/j.1365-2966.2007.12209.x}, \href
  {http://adsabs.harvard.edu/abs/2007MNRAS.380.1589B} {380, 1589}

\bibitem[\protect\citeauthoryear{{Da Rio}, {Tan}  \& {Jaehnig}}{{Da Rio}
  et~al.}{2014}]{Dario2014}
{Da Rio} N.,  {Tan} J.~C.,   {Jaehnig} K.,  2014, \mn@doi [\apj]
  {10.1088/0004-637X/795/1/55}, \href
  {https://ui.adsabs.harvard.edu/#abs/2014ApJ...795...55D} {795}

\bibitem[\protect\citeauthoryear{De~Wit, Testi, Palla  \& Zinnecker}{De~Wit
  et~al.}{2005}]{DeWit2005}
De~Wit W.,  Testi L.,  Palla F.,   Zinnecker H.,  2005, Astronomy \&
  Astrophysics, 437, 247

\bibitem[\protect\citeauthoryear{{Dom{\'{\i}}nguez}, {Fellhauer}, {Bla{\~n}a},
  {Farias}  \& {Dabringhausen}}{{Dom{\'{\i}}nguez}
  et~al.}{2017}]{Dominguez2018}
{Dom{\'{\i}}nguez} R.,  {Fellhauer} M.,  {Bla{\~n}a} M.,  {Farias} J.~P.,
  {Dabringhausen} J.,  2017, \mn@doi [\mnras] {10.1093/mnras/stx1883}, \href
  {http://adsabs.harvard.edu/abs/2017MNRAS.472..465D} {472, 465}

\bibitem[\protect\citeauthoryear{{Eldridge}, {Langer}  \& {Tout}}{{Eldridge}
  et~al.}{2011}]{Eldridge2011}
{Eldridge} J.~J.,  {Langer} N.,   {Tout} C.~A.,  2011, \mn@doi [\mnras]
  {10.1111/j.1365-2966.2011.18650.x}, \href
  {http://adsabs.harvard.edu/abs/2011MNRAS.414.3501E} {414, 3501}

\bibitem[\protect\citeauthoryear{{Elmegreen}}{{Elmegreen}}{2000}]{Elmegreen2000}
{Elmegreen} B.~G.,  2000, \mn@doi [\apj] {10.1086/308361}, \href
  {http://adsabs.harvard.edu/abs/2000ApJ...530..277E} {530, 277}

\bibitem[\protect\citeauthoryear{{Elmegreen}}{{Elmegreen}}{2007}]{Elmegreen2007}
{Elmegreen} B.~G.,  2007, \mn@doi [\apj] {10.1086/521327}, \href
  {http://adsabs.harvard.edu/abs/2007ApJ...668.1064E} {668, 1064}

\bibitem[\protect\citeauthoryear{{Farias}, {Smith}, {Fellhauer}, {Goodwin},
  {Candlish}, {Bla{\~n}a}  \& {Dominguez}}{{Farias} et~al.}{2015}]{Farias2015}
{Farias} J.~P.,  {Smith} R.,  {Fellhauer} M.,  {Goodwin} S.,  {Candlish} G.~N.,
   {Bla{\~n}a} M.,   {Dominguez} R.,  2015, \mn@doi [\mnras]
  {10.1093/mnras/stv790}, \href
  {http://adsabs.harvard.edu/abs/2015MNRAS.450.2451F} {450, 2451}

\bibitem[\protect\citeauthoryear{{Farias}, {Tan}  \& {Chatterjee}}{{Farias}
  et~al.}{2017}]{F17}
{Farias} J.~P.,  {Tan} J.~C.,   {Chatterjee} S.,  2017, \mn@doi [\apj]
  {10.3847/1538-4357/aa63f6}, \href
  {http://adsabs.harvard.edu/abs/2017ApJ...838..116F} {838, 116}

\bibitem[\protect\citeauthoryear{{Farias}, {Fellhauer}, {Smith},
  {Dom{\'{\i}}nguez}  \& {Dabringhausen}}{{Farias} et~al.}{2018}]{Farias2018}
{Farias} J.~P.,  {Fellhauer} M.,  {Smith} R.,  {Dom{\'{\i}}nguez} R.,
  {Dabringhausen} J.,  2018, \mn@doi [\mnras] {10.1093/mnras/sty597}, \href
  {http://adsabs.harvard.edu/abs/2018MNRAS.476.5341F} {476, 5341}

\bibitem[\protect\citeauthoryear{{Getman}, {Feigelson}  \& {Kuhn}}{{Getman}
  et~al.}{2014}]{Getman2014}
{Getman} K.~V.,  {Feigelson} E.~D.,   {Kuhn} M.~A.,  2014, \mn@doi [\apj]
  {10.1088/0004-637X/787/2/109}, \href
  {http://adsabs.harvard.edu/abs/2014ApJ...787..109G} {787, 109}

\bibitem[\protect\citeauthoryear{{Getman}, {Feigelson}, {Kuhn}, {Bate}, {Broos}
   \& {Garmire}}{{Getman} et~al.}{2018}]{Getman2018}
{Getman} K.~V.,  {Feigelson} E.~D.,  {Kuhn} M.~A.,  {Bate} M.~R.,  {Broos}
  P.~S.,   {Garmire} G.~P.,  2018, \mn@doi [\mnras] {10.1093/mnras/sty302},
  \href {http://adsabs.harvard.edu/abs/2018MNRAS.476.1213G} {476, 1213}

\bibitem[\protect\citeauthoryear{{Gies}}{{Gies}}{1987}]{Gies1987}
{Gies} D.~R.,  1987, \mn@doi [The Astrophysical Journal Supplement Series]
  {10.1086/191208}, \href
  {https://ui.adsabs.harvard.edu/#abs/1987ApJS...64..545G} {64, 545}

\bibitem[\protect\citeauthoryear{{Gutermuth} et~al.,}{{Gutermuth}
  et~al.}{2008}]{Gutermuth2008}
{Gutermuth} R.~A.,  et~al., 2008, \mn@doi [\apj] {10.1086/524722}, \href
  {http://adsabs.harvard.edu/abs/2008ApJ...674..336G} {674, 336}

\bibitem[\protect\citeauthoryear{{Gutermuth}, {Megeath}, {Myers}, {Allen},
  {Pipher}  \& {Fazio}}{{Gutermuth} et~al.}{2009}]{Gutermut2009}
{Gutermuth} R.~A.,  {Megeath} S.~T.,  {Myers} P.~C.,  {Allen} L.~E.,  {Pipher}
  J.~L.,   {Fazio} G.~G.,  2009, \mn@doi [\apjs] {10.1088/0067-0049/184/1/18},
  \href {http://adsabs.harvard.edu/abs/2009ApJS..184...18G} {184, 18}

\bibitem[\protect\citeauthoryear{{Hartmann} \& {Burkert}}{{Hartmann} \&
  {Burkert}}{2007}]{Hartmann2007}
{Hartmann} L.,  {Burkert} A.,  2007, \mn@doi [\apj] {10.1086/509321}, \href
  {http://adsabs.harvard.edu/abs/2007ApJ...654..988H} {654, 988}

\bibitem[\protect\citeauthoryear{{Hobbs}, {Lorimer}, {Lyne}  \&
  {Kramer}}{{Hobbs} et~al.}{2005}]{Hobbs2005}
{Hobbs} G.,  {Lorimer} D.~R.,  {Lyne} A.~G.,   {Kramer} M.,  2005, \mn@doi
  [\mnras] {10.1111/j.1365-2966.2005.09087.x}, \href
  {http://adsabs.harvard.edu/abs/2005MNRAS.360..974H} {360, 974}

\bibitem[\protect\citeauthoryear{{Hurley}, {Pols}  \& {Tout}}{{Hurley}
  et~al.}{2000}]{Hurley2000}
{Hurley} J.~R.,  {Pols} O.~R.,   {Tout} C.~A.,  2000, \mn@doi [\mnras]
  {10.1046/j.1365-8711.2000.03426.x}, \href
  {http://adsabs.harvard.edu/abs/2000MNRAS.315..543H} {315, 543}

\bibitem[\protect\citeauthoryear{{Hurley}, {Tout}  \& {Pols}}{{Hurley}
  et~al.}{2002}]{Hurley2002}
{Hurley} J.~R.,  {Tout} C.~A.,   {Pols} O.~R.,  2002, \mn@doi [\mnras]
  {10.1046/j.1365-8711.2002.05038.x}, \href
  {http://adsabs.harvard.edu/abs/2002MNRAS.329..897H} {329, 897}

\bibitem[\protect\citeauthoryear{{Kroupa}}{{Kroupa}}{2001}]{Kroupa2001}
{Kroupa} P.,  2001, \mn@doi [\mnras] {10.1046/j.1365-8711.2001.04022.x}, \href
  {http://adsabs.harvard.edu/abs/2001MNRAS.322..231K} {322, 231}

\bibitem[\protect\citeauthoryear{{Krumholz} \& {McKee}}{{Krumholz} \&
  {McKee}}{2005}]{Krumholz2005}
{Krumholz} M.~R.,  {McKee} C.~F.,  2005, \mn@doi [\apj] {10.1086/431734}, \href
  {http://adsabs.harvard.edu/abs/2005ApJ...630..250K} {630, 250}

\bibitem[\protect\citeauthoryear{{Lada} \& {Lada}}{{Lada} \&
  {Lada}}{2003}]{Lada2003}
{Lada} C.~J.,  {Lada} E.~A.,  2003, \mn@doi [\araa]
  {10.1146/annurev.astro.41.011802.094844}, \href
  {http://adsabs.harvard.edu/abs/2003ARA%26A..41...57L} {41, 57}

\bibitem[\protect\citeauthoryear{{Lee} \& {Goodwin}}{{Lee} \&
  {Goodwin}}{2016}]{Lee2016}
{Lee} P.~L.,  {Goodwin} S.~P.,  2016, \mn@doi [\mnras] {10.1093/mnras/stw988},
  \href {http://adsabs.harvard.edu/abs/2016MNRAS.460.2997L} {460, 2997}

\bibitem[\protect\citeauthoryear{{McKee} \& {Ostriker}}{{McKee} \&
  {Ostriker}}{2007}]{Mckee2007}
{McKee} C.~F.,  {Ostriker} E.~C.,  2007, \mn@doi [\araa]
  {10.1146/annurev.astro.45.051806.110602}, \href
  {http://adsabs.harvard.edu/abs/2007ARA%26A..45..565M} {45, 565}

\bibitem[\protect\citeauthoryear{{McKee} \& {Tan}}{{McKee} \&
  {Tan}}{2003}]{mt03}
{McKee} C.~F.,  {Tan} J.~C.,  2003, \mn@doi [\apj] {10.1086/346149}, \href
  {http://adsabs.harvard.edu/abs/2003ApJ...585..850M} {585, 850}

\bibitem[\protect\citeauthoryear{{Nakamura} \& {Li}}{{Nakamura} \&
  {Li}}{2007}]{Nakamura2007}
{Nakamura} F.,  {Li} Z.-Y.,  2007, \mn@doi [\apj] {10.1086/517515}, \href
  {http://adsabs.harvard.edu/abs/2007ApJ...662..395N} {662, 395}

\bibitem[\protect\citeauthoryear{{Nakamura} \& {Li}}{{Nakamura} \&
  {Li}}{2014}]{Nakamura2014}
{Nakamura} F.,  {Li} Z.-Y.,  2014, \mn@doi [\apj]
  {10.1088/0004-637X/783/2/115}, \href
  {http://adsabs.harvard.edu/abs/2014ApJ...783..115N} {783, 115}

\bibitem[\protect\citeauthoryear{{Parker}, {Wright}, {Goodwin}  \&
  {Meyer}}{{Parker} et~al.}{2014}]{Parker2014a}
{Parker} R.~J.,  {Wright} N.~J.,  {Goodwin} S.~P.,   {Meyer} M.~R.,  2014,
  \mn@doi [\mnras] {10.1093/mnras/stt2231}, \href
  {http://adsabs.harvard.edu/abs/2014MNRAS.438..620P} {438, 620}

\bibitem[\protect\citeauthoryear{{Parmentier} \& {Pfalzner}}{{Parmentier} \&
  {Pfalzner}}{2013}]{Parmentier2013}
{Parmentier} G.,  {Pfalzner} S.,  2013, \mn@doi [\aap]
  {10.1051/0004-6361/201219648}, \href
  {http://adsabs.harvard.edu/abs/2013A%26A...549A.132P} {549, A132}

\bibitem[\protect\citeauthoryear{{Pavl{\'{\i}}k}, {Je{\v r}{\'a}bkov{\'a}},
  {Kroupa}  \& {Baumgardt}}{{Pavl{\'{\i}}k} et~al.}{2018}]{Pavlik2018}
{Pavl{\'{\i}}k} V.,  {Je{\v r}{\'a}bkov{\'a}} T.,  {Kroupa} P.,   {Baumgardt}
  H.,  2018, preprint, \href
  {http://adsabs.harvard.edu/abs/2018arXiv180605192P} {} (\mn@eprint {arXiv}
  {1806.05192})

\bibitem[\protect\citeauthoryear{{Pfalzner}, {Vincke}  \& {Xiang}}{{Pfalzner}
  et~al.}{2015}]{Pfalzner2015}
{Pfalzner} S.,  {Vincke} K.,   {Xiang} M.,  2015, \mn@doi [\aap]
  {10.1051/0004-6361/201425100}, \href
  {http://adsabs.harvard.edu/abs/2015A%26A...576A..28P} {576, A28}

\bibitem[\protect\citeauthoryear{{Proszkow} \& {Adams}}{{Proszkow} \&
  {Adams}}{2009}]{Proszkow2009}
{Proszkow} E.-M.,  {Adams} F.~C.,  2009, \mn@doi [\apjs]
  {10.1088/0067-0049/185/2/486}, \href
  {http://adsabs.harvard.edu/abs/2009ApJS..185..486P} {185, 486}

\bibitem[\protect\citeauthoryear{{Raghavan} et~al.,}{{Raghavan}
  et~al.}{2010}]{Raghavan2010}
{Raghavan} D.,  et~al., 2010, \mn@doi [\apjs] {10.1088/0067-0049/190/1/1},
  \href {http://adsabs.harvard.edu/abs/2010ApJS..190....1R} {190, 1}

\bibitem[\protect\citeauthoryear{{Reggiani} \& {Meyer}}{{Reggiani} \&
  {Meyer}}{2011}]{Reggiani2011}
{Reggiani} M.~M.,  {Meyer} M.~R.,  2011, \mn@doi [\apj]
  {10.1088/0004-637X/738/1/60}, \href
  {http://adsabs.harvard.edu/abs/2011ApJ...738...60R} {738, 60}

\bibitem[\protect\citeauthoryear{{Smith}, {Goodwin}, {Fellhauer}  \&
  {Assmann}}{{Smith} et~al.}{2013}]{Smith2013b}
{Smith} R.,  {Goodwin} S.,  {Fellhauer} M.,   {Assmann} P.,  2013, \mn@doi
  [\mnras] {10.1093/mnras/sts106}, \href
  {http://adsabs.harvard.edu/abs/2013MNRAS.428.1303S} {428, 1303}

\bibitem[\protect\citeauthoryear{{Stone}}{{Stone}}{1991}]{Stone1991}
{Stone} R.~C.,  1991, \mn@doi [\aj] {10.1086/115880}, \href
  {https://ui.adsabs.harvard.edu/#abs/1991AJ....102..333S} {102, 333}

\bibitem[\protect\citeauthoryear{{Tan}, {Krumholz}  \& {McKee}}{{Tan}
  et~al.}{2006}]{Tan2006}
{Tan} J.~C.,  {Krumholz} M.~R.,   {McKee} C.~F.,  2006, \mn@doi [\apjl]
  {10.1086/504150}, \href {http://adsabs.harvard.edu/abs/2006ApJ...641L.121T}
  {641, L121}

\bibitem[\protect\citeauthoryear{{Tanaka}, {Tan}  \& {Zhang}}{{Tanaka}
  et~al.}{2017}]{Tanaka2017}
{Tanaka} K.~E.~I.,  {Tan} J.~C.,   {Zhang} Y.,  2017, \mn@doi [\apj]
  {10.3847/1538-4357/835/1/32}, \href
  {http://adsabs.harvard.edu/abs/2017ApJ...835...32T} {835, 32}

\bibitem[\protect\citeauthoryear{{Wang}, {Spurzem}, {Aarseth}, {Nitadori},
  {Berczik}, {Kouwenhoven}  \& {Naab}}{{Wang} et~al.}{2015}]{Wang2015}
{Wang} L.,  {Spurzem} R.,  {Aarseth} S.,  {Nitadori} K.,  {Berczik} P.,
  {Kouwenhoven} M.~B.~N.,   {Naab} T.,  2015, \mn@doi [\mnras]
  {10.1093/mnras/stv817}, \href
  {http://adsabs.harvard.edu/abs/2015MNRAS.450.4070W} {450, 4070}

\bibitem[\protect\citeauthoryear{Wang, Kroupa  \& Jerabkova}{Wang
  et~al.}{2018}]{Wang2018}
Wang L.,  Kroupa P.,   Jerabkova T.,  2018, \mn@doi [Monthly Notices of the
  Royal Astronomical Society] {10.1093/mnras/sty2232}, p. sty2232

\makeatother
\end{thebibliography}

\appendix




\appendix

\onecolumn
\section{Analytical estimation of bound fractions}
\label{sec:bound_model}

Given that all properties of the newly formed stars are given by an
analytical prescription, it is possible to estimate the amount of
stellar mass that should keep bound at birth.  Stars are born with
one-dimensional component velocity dispersions following a Normal
distribution with $\sigma_x=\sigma_y=\sigma_z$ given by Equation
\ref{eq:sigmar}. Then, at a given radius, stars are born with a
Maxwell-Boltzman velocity distribution with scale parameter
$a(r)=\sqrt{3}\sigma_{\rm cl}(r)$.  Therefore, we can use the
cumulative distribution function (CDF) to find the fraction of stars
that have velocities below the escape velocity at a given radius.

The escape velocity at a given radius $r$ is given by $v_{\rm
  esc}=\sqrt{-2\Phi(r,t)}$, where $\Phi(r,t)$ is given by
Eq.~\ref{eq:pot}. For convenience, we form the expression
$v_{\sigma}(r,t) =v_{\rm esc}(r,t)/\sigma_{\rm cl}(r)$, that reduces
to:
\begin{eqnarray}
        v_{\sigma}(r,t) &=& 
        \sqrt{
        \frac{ 4(\krho -1)}
        {3(\krho-2)} 
        \phi_{\rm B}
        \left[ 1 - (3-\krho) \left(
        \frac{r}{R_{\rm cl}} 
        \right)^{ \krho -2} \right] \frac{M_{\rm total}(t)}{M_{\rm cl,0}} },
\end{eqnarray}
where $\phi_{\rm B}$ a factor that accounts for the magnetic field
support in the cloud (see \citetalias{F17} and \citetalias{mt03}), and
the factor $M_{\rm total}(t)/M_{\rm cl,0}$ is the total mass of the
system (stars and gas) at a given time.

Then, using the Maxwell-Boltzman CDF, the fraction of stars that that
born bound at a given radius $r$, is:
\begin{eqnarray}
        \label{eq:fb1}
        f_{\rm bound}(r) &=&  
        {\rm erf}\left( \frac{v_{\sigma}(r) }{\sqrt{2}} \right) 
        - \sqrt{\frac{2}{\pi}}v_{\sigma}(r) \exp\left( -\frac{v_{\sigma}^2(r)}{2} \right) .
\end{eqnarray}
The total fraction of mass is different at each radius, therefore the
total bound fraction is Eq.~\ref{eq:fb1} mass averaged over the clump,
which reduces to:
\begin{eqnarray}
        \label{eq:fbtot}
        f_{\rm bound,tot} &=&  \frac{3-\krho}{R_{\rm cl}} \int_0^{R_{\rm cl}} f_{\rm bound}(r)
        \left(\frac{r}{R_{\rm cl}}\right)^{2-\krho} dr,
\end{eqnarray}
which can only be solved numerically.
Equation \ref{eq:fbtot} can be evaluated at $t=0$ assuming that
$M_{\rm total} = M_{\rm cl,0}\epsilon$ assuming all gas is lost
instantly to obtain the bound fraction of
$\epsilonff=\infty$. However, in order to obtain the values for other
formation rates, it needs to be integrated over time using a numerical
method.

\section{Ancillary results for the full set of simulations}
\begin{figure*}
        $\begin{array}{rl}
        \includegraphics[width=0.49\textwidth]{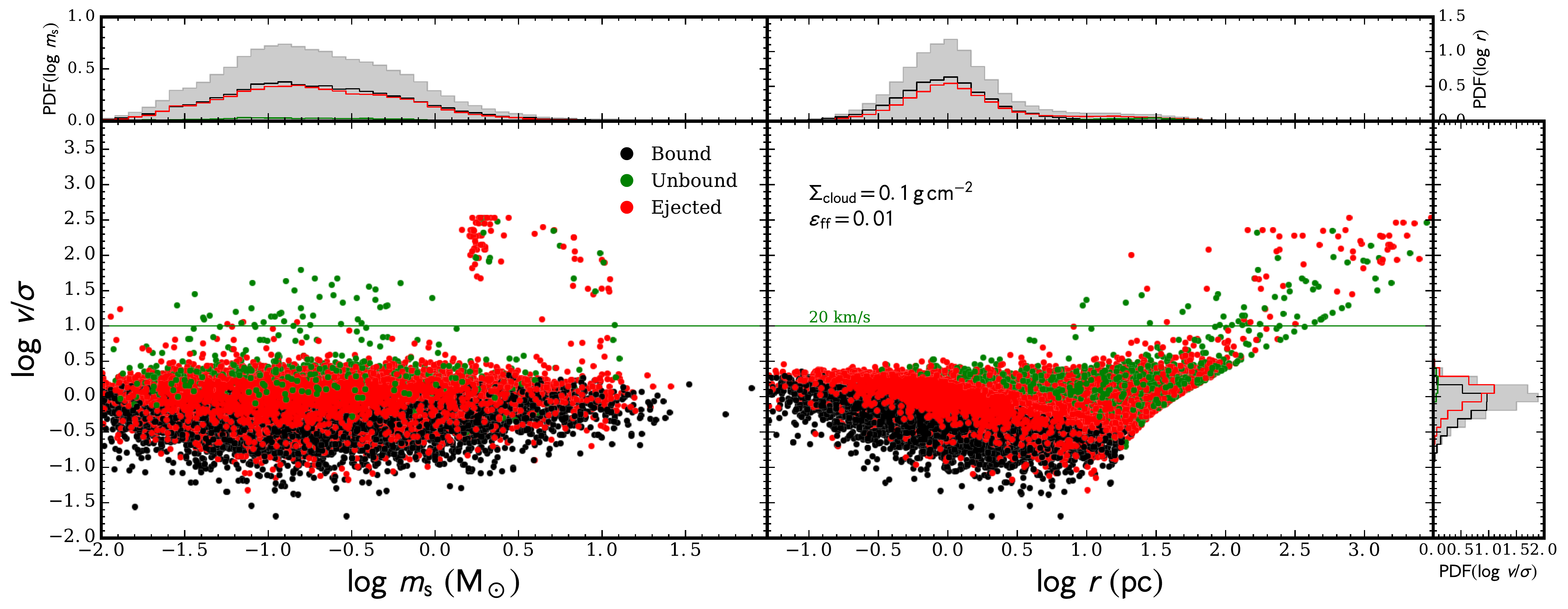} &
        \includegraphics[width=0.49\textwidth]{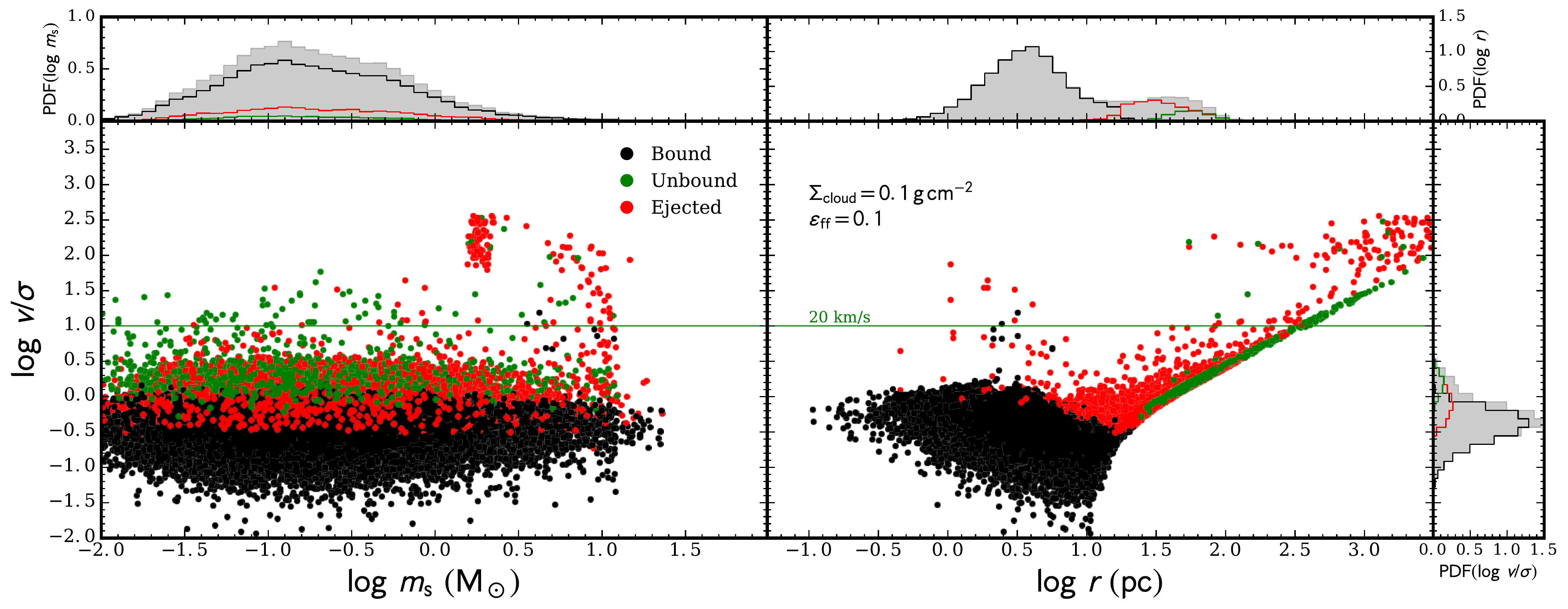} \\
        \includegraphics[width=0.49\textwidth]{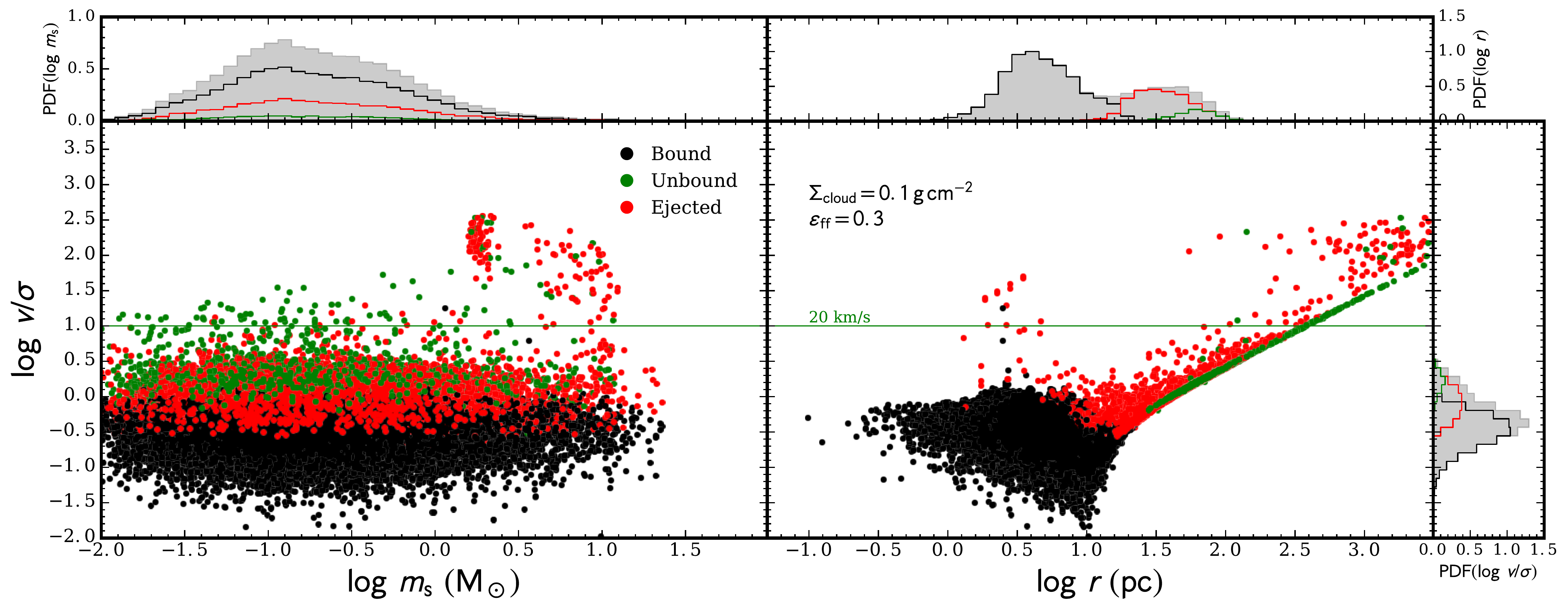} &
        \includegraphics[width=0.49\textwidth]{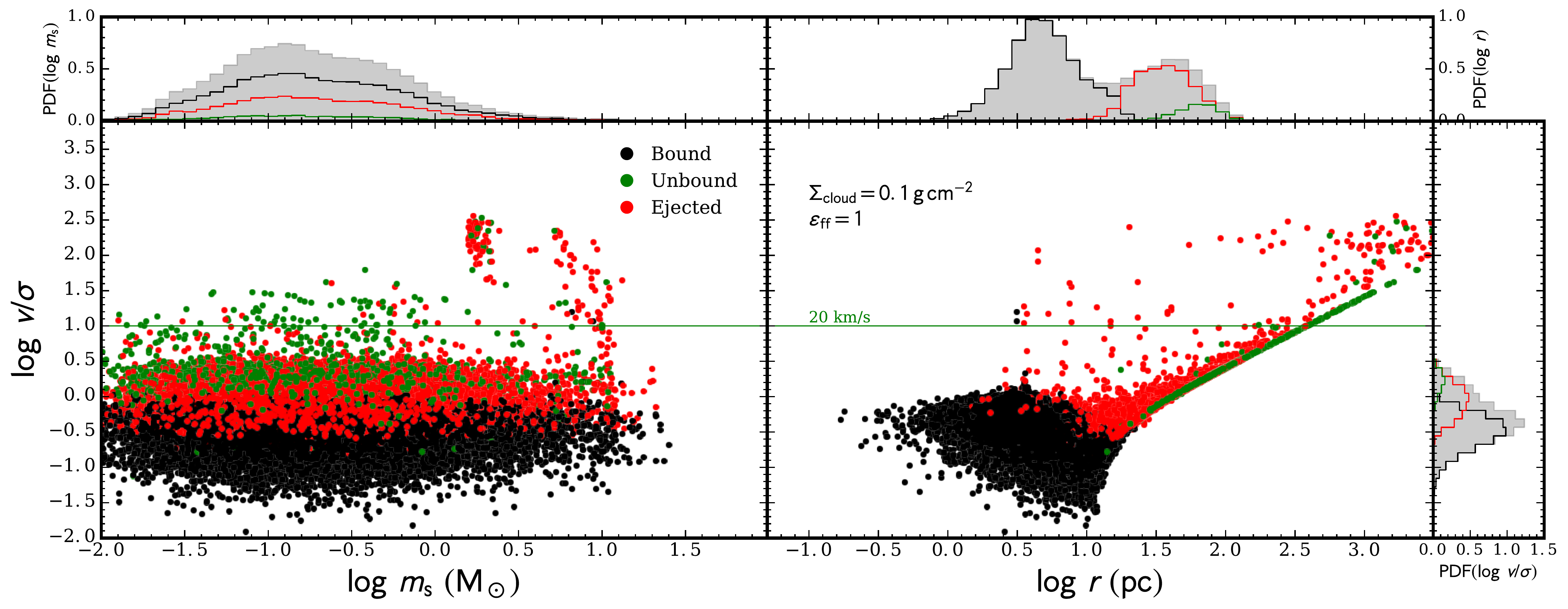} \\
        \end{array}$
        \caption{
Same as Figure \ref{fig:mv}, but now for simulations sets with $\epsilonff=0.01$,
0.1, 0.3, 1.0. 
}
\label{fig:mvall}
\end{figure*}

\begin{figure*}
        $\begin{array}{rl}
        \includegraphics[width=0.49\textwidth]{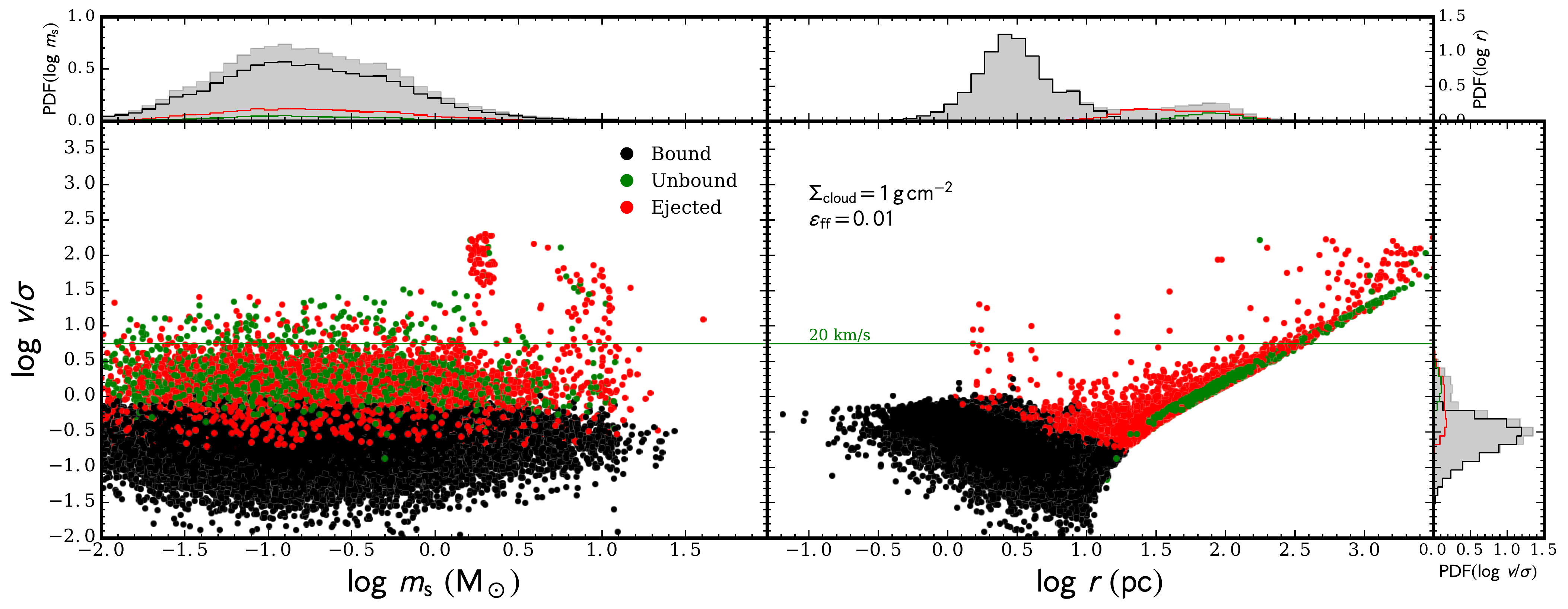} &
        \includegraphics[width=0.49\textwidth]{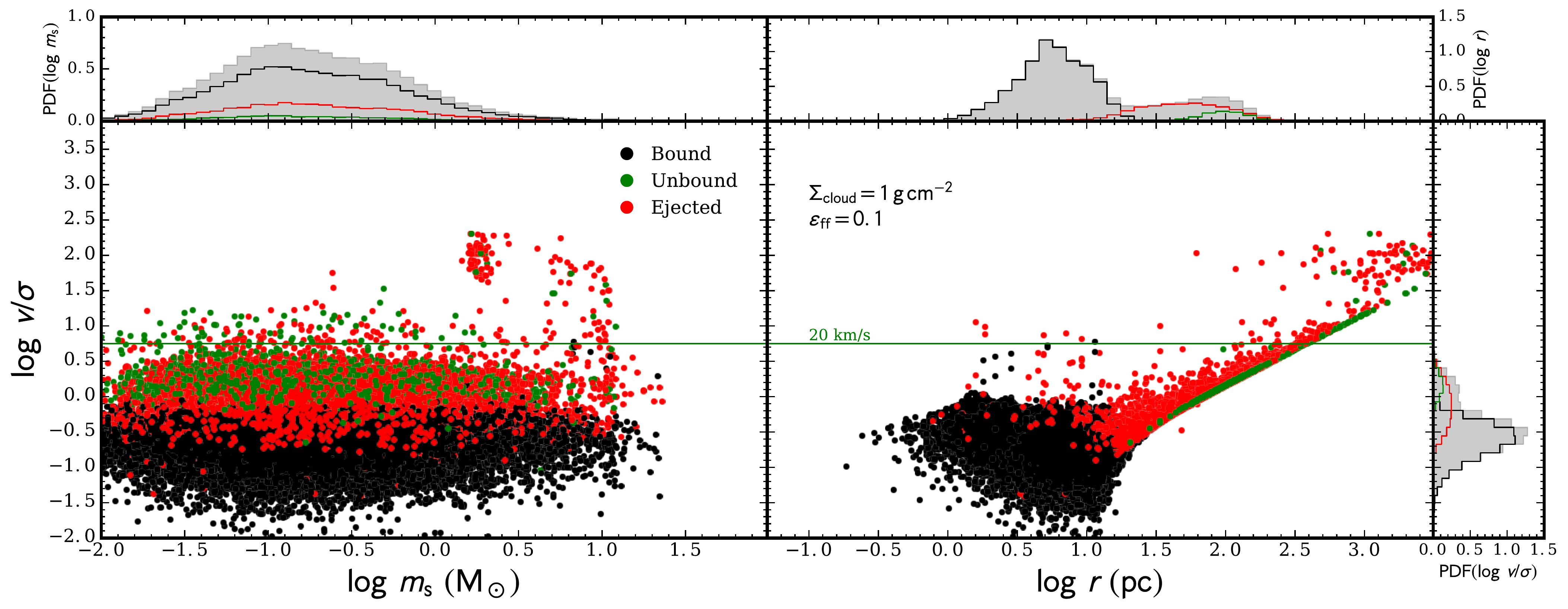} \\
        \includegraphics[width=0.49\textwidth]{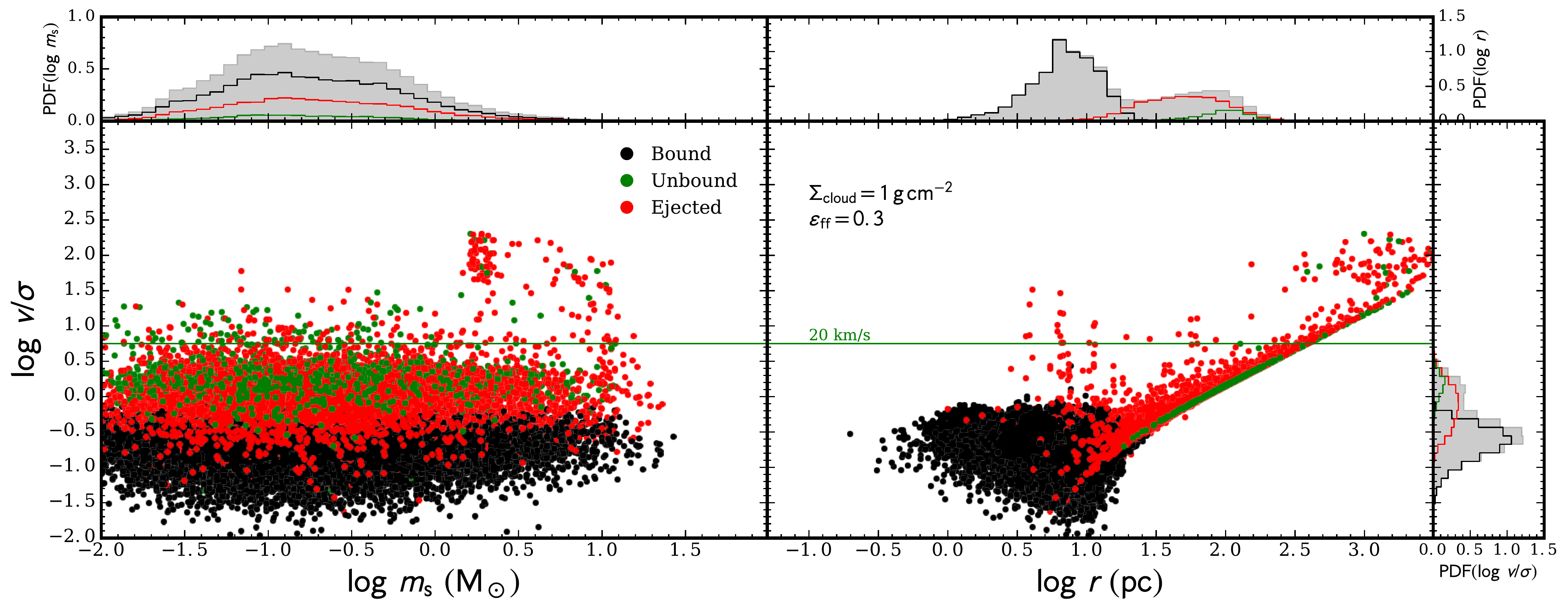} &
        \includegraphics[width=0.49\textwidth]{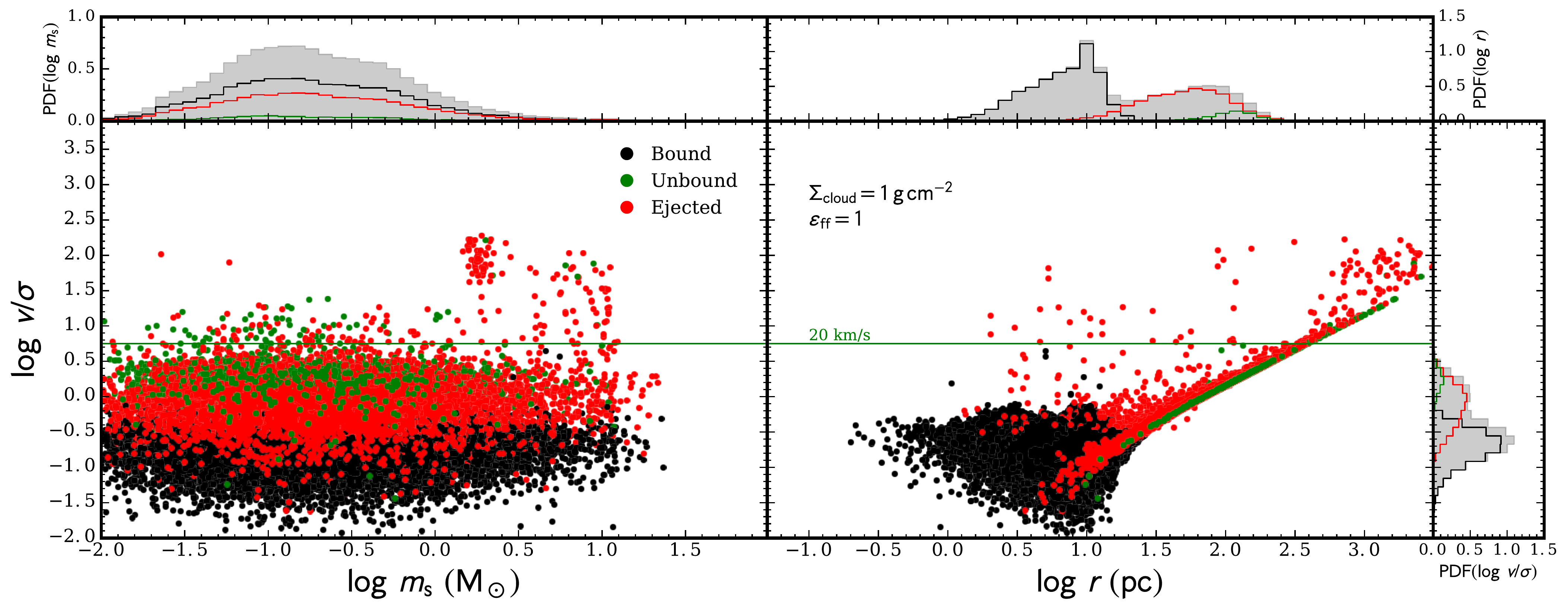} \\
        \end{array}$
        \caption{
Same as Figure \ref{fig:mvall}, but for $\Sigmacl=1\,$g cm$^{-2}$.}
\label{fig:mvallss1}
\end{figure*}


\bsp	
\label{lastpage}
\end{document}